\documentclass[prd,preprint,superscriptaddress,amsmath,amssymb,nofootinbib]{revtex4}
\usepackage{here}
\usepackage[dvipdfmx]{graphicx}
\usepackage{graphicx, color}
\usepackage{wrapfig}
\usepackage{graphicx}
\usepackage{dcolumn}
\usepackage{bm}
\usepackage{amssymb}
\usepackage{amsmath}
\usepackage{epsfig}    
\usepackage{color}
\usepackage{slashed}
\usepackage{hhline}
\usepackage{color}

\def\be{\begin{equation}}
\def\ee{\end{equation}}
\newcommand{\bea}{\begin{eqnarray}}
\newcommand{\eea}{\end{eqnarray}}
\newcommand{\nn}{\nonumber}

\begin{document}

{\begin{flushright}{APCTP Pre2020-017}
\end{flushright}}

\title{Modular $A_4$ symmetry and light dark matter with gauged $U(1)_{B-L}$} 
\author{Keiko I. Nagao}
\email{nagao@dap.ous.ac.jp}
\affiliation{Okayama University of Science, Faculty of Science,  Department of Applied Physics, Ridaicho 1-1, Okayama, 700-0005, Japan}
\author{Hiroshi Okada}
\email{hiroshi.okada@apctp.org}
\affiliation{Asia Pacific Center for Theoretical Physics (APCTP) - Headquarters San 31, Hyoja-dong,
Nam-gu, Pohang 790-784, Korea}
\affiliation{Department of Physics, Pohang University of Science and Technology, Pohang 37673, Republic of Korea}

\date{\today}

\begin{abstract}
 We propose a radiative seesaw model with a light dark matter candidate (DM) under modular $A_4$ and gauged $U(1)_{B-L}$ symmetries, in which neutrino masses are generated via one-loop level, and we have a bosonic DM candidate $\sim 0.01-50$ GeV.
DM is stabilized by nonzero modular weight and a remnant $Z_2$ symmetry of $U(1)_{B-L}$ symmetry.
The naturalness of tiny DM mass is indirectly realized by the radiatively induced mass of neutral fermions that interact with our DM candidate. Thus, DM mass does not have to exceed the neutral fermion masses, whose scale is assumed to be 50 GeV.
 Taking several benchmark points of DM mass in DM analysis, we show the numerical analysis for the neutrino oscillation data in the normal and inverted hierarchy cases and find several features for both cases. 
  \end{abstract}
\maketitle

\section{Introduction}
A radiative seesaw model is one of the sophisticated ideas to connect massive neutrinos and dark matter candidate~\cite{Ma:2006km}, both of which are considered beyond the standard model (SM). Moreover, this model can typically be within a low energy scale$\sim$TeV, since relatively large Yukawa couplings are expected.
Neutrino masses are generated only via interactions of DM at loop levels. This is why we can naturally explain that the neutrino mass is tiny enough without the requirement of small Yukawa couplings.
In order to realize this kind of model, we often introduce symmetry in order to stabilize DM in addition to gauge symmetries in SM, and Abelian discrete symmetries (such as $Z_2$) are frequently applied to this kind of model.
 Also, flavor physics raises our serious target from this model due to not so small Yukawa couplings.
Thus, it would be a natural consequence that stabilized symmetry of DM is extended to a flavor symmetry in order to explain DM and flavor physics at the same time.

Recently, attractive flavor symmetries are proposed by papers~\cite{Feruglio:2017spp,
deAdelhartToorop:2011re}, in which they applied modular non-Abelian discrete flavor
symmetries to quark and lepton sectors.
One remarkable advantage is that any dimensionless couplings can also be transformed as non-trivial representations under those symmetries. Therefore, we do not need so many scalars to find a predictive mass matrix. Another advantage is that we have a modular weight from the modular origin that can play a role in stabilizing DM when appropriate charge assignments are distributed to each of the fields of models.
Along the line of this idea, a vast reference has recently appeared in the literature, {\it e.g.},  $A_4$~\cite{Feruglio:2017spp, Criado:2018thu, Kobayashi:2018scp, Okada:2018yrn, Nomura:2019jxj, Okada:2019uoy, deAnda:2018ecu, Novichkov:2018yse, Nomura:2019yft, Okada:2019mjf,Ding:2019zxk, Nomura:2019lnr,Kobayashi:2019xvz,Asaka:2019vev,Zhang:2019ngf, Gui-JunDing:2019wap,Kobayashi:2019gtp,Nomura:2019xsb, Wang:2019xbo,Okada:2020dmb,Okada:2020rjb, Behera:2020lpd, Behera:2020sfe, Nomura:2020opk, Nomura:2020cog, Asaka:2020tmo, Okada:2020ukr, Nagao:2020snm, Okada:2020brs, Yao:2020qyy, Chen:2021zty, Kashav:2021zir, Okada:2021qdf, deMedeirosVarzielas:2021pug, Nomura:2021yjb, Hutauruk:2020xtk},
$S_3$ \cite{Kobayashi:2018vbk, Kobayashi:2018wkl, Kobayashi:2019rzp, Okada:2019xqk, Mishra:2020gxg, Du:2020ylx},
$S_4$ \cite{Penedo:2018nmg, Novichkov:2018ovf, Kobayashi:2019mna, King:2019vhv, Okada:2019lzv, Criado:2019tzk,
Wang:2019ovr, Zhao:2021jxg, King:2021fhl, Ding:2021zbg, Zhang:2021olk, gui-jun},
$A_5$~\cite{Novichkov:2018nkm, Ding:2019xna,Criado:2019tzk}, double covering of $A_5$~\cite{Wang:2020lxk, Yao:2020zml, Wang:2021mkw, Behera:2021eut}, larger groups~\cite{Baur:2019kwi}, multiple modular symmetries~\cite{deMedeirosVarzielas:2019cyj}, and double covering of $A_4$~\cite{Liu:2019khw, Chen:2020udk, Li:2021buv}, $S_4$~\cite{Novichkov:2020eep, Liu:2020akv}, and the other types of groups \cite{Kikuchi:2020nxn, Almumin:2021fbk, Ding:2021iqp, Feruglio:2021dte, Kikuchi:2021ogn, Novichkov:2021evw} in which masses, mixing, and CP phases for the quark and/or lepton have been predicted~\footnote{For interest readers, we provide some literature reviews, which are useful to understand the non-Abelian group and its applications to flavor structure~\cite{Altarelli:2010gt, Ishimori:2010au, Ishimori:2012zz, Hernandez:2012ra, King:2013eh, King:2014nza, King:2017guk, Petcov:2017ggy}.}.
Moreover, a systematic approach to understanding the origin of CP transformations has been discussed in Ref.~\cite{Baur:2019iai}, 
and CP/flavor violation in models with modular symmetry was discussed in Refs.~\cite{Kobayashi:2019uyt,Novichkov:2019sqv}, 
and a possible correction from K\"ahler potential was discussed in Ref.~\cite{Chen:2019ewa}. Furthermore,
systematic analysis of the fixed points (stabilizers) has been discussed in Ref.~\cite{deMedeirosVarzielas:2020kji}.

In this paper, we propose a radiative seesaw model with a modular $A_4$ and gauged $U(1)_{B-L}$ symmetries, in which the neutrino masses are generated via one-loop level, and we have a relatively light bosonic DM candidate $\sim 0.01-50$ GeV.
DM is stabilized by nonzero modular weight and a remnant $Z_2$ symmetry of $U(1)_{B-L}$ symmetry.
The naturalness of tiny mass of DM is indirectly realized by the radiatively induced mass of neutral fermions that interact with our DM candidate. Thus, DM mass must not exceed the mass of the neutral fermion masses, whose scale is supposed to be 50 GeV. We numerically analyze our DM candidate and then search for the allowed region of neutrino oscillation data.

This letter is organized as follows.
In Sec. II, we review our model, formulating our renormalizable Lagrangian, Higgs potential, and discuss DM including a numerical search for the allowed mass region.
Then, we formulate the neutrino sector and show our numerical analysis to satisfy the neutrino oscillation data in cases of normal and inverted hierarchies.
In Sec. III, we devote the summary to our results and the conclusion.
In the appendix, we briefly review the modular symmetry.

\section{Model setup}
\begin{table}[t!]
\begin{tabular}{|c||c|c|c|c|c||c|c|c|c|}\hline\hline  
& 
~$L_{L_i}$~& ~$e_{R_i}$~& ~$N_{R_a}$~& ~$S_{L_a}$~& ~$S_{R_a}$~& ~$H$~& ~$\eta$~& ~$\varphi$~& ~$\chi$~\\\hline\hline 
$SU(2)_L$ & $\bm{2}$  & $\bm{1}$  & $\bm{1}$  & $\bm{1}$ & $\bm{1}$   & $\bm{2}$ & $\bm{2}$ & $\bm{1}$ & $\bm{1}$     \\\hline 
$U(1)_Y$  
&  $-\frac12$  & $-1$ & $0$  & $0$  & $0$ & $\frac12$  & $\frac12$ & $0$  & $0$  \\\hline
$U(1)_{B-L}$ & $-1$  & $-1$  & $-1$  & $-1/2$  & $-1/2$ & $0$  & $0$  & $1$ & $-1/2$\\\hline
$A_{4}$   
& $\{\bf 1\}$  & $\{\bf 1\}$  & $\bf3$  & $\bf3$ & $\bf3$ & $\bf1$   & $\bf1$ & $\bf1$ & $\bf1$\\\hline
$-k$   
& $0$  & $0$  & ${-1}^*$  & $-1$  & $0$ & $0$  & $-3$   & $0$  & $-2$ \\\hline
\end{tabular}
\caption{ 
Charge assignments of our fields
under $SU(2)_L\otimes U(1)_Y\otimes U(1)_{B-L}\otimes A_4\otimes (-k)$, where $\{\bf 1\}\equiv (1,1',1'')$, $SU(3)_C$ is the same as the SM and singlet for all the new fields, and $-1^*$ indicates complex conjugate of $\tau$ should be applied under the modular transformation. $B-L$ charge assignment for the quark sector is $-1/3$ to cancel the chiral anomaly.
The lower indices $(i,a)$ are the number of families that runs over $1-3$.}
\label{tab:1}
\end{table}

In this section, we explain our model.

\subsection{Renormalizable Yukawa Lagrangian and Higgs potential}
As for the fermion sector, we introduce three families of right-handed neutral fermions $N_{R}$ to cancel anomaly coming from gauged $U(1)_{B-L}$ symmetry, where we impose $N_R$ into $(3,-1^*)$ under $(A_4,-k)$.
Here, $-1^*$ denotes complex conjugate of $\tau$ should be applied under the modular transformation.
In addition, we introduce three families of vector-like neutral fermions $S_L$ and $S_R$ with $(-1/2,3)$ under $(U(1)_{B-L},A_4)$ that implies $U(1)_{B-L}$ anomaly is automatically cancelled among them.
{\it $S_{L,R}$ plays a role in generating the mass of $N_R$ at one loop level.}
While we impose $(-1,0)$ charge for $(S_L,S_R)$ under modular weight $-k$.
In the quark sector of the SM fermions, we impose $(Q_L,\bar d_R,\bar u_R)=(3,\{1\},\{1\})$ under $A_4$ symmetry and $(-2,0,-4)$ under the modular weight of $-k$, where $-1/3$ charge is imposed in order to cancel the anomaly of gauged $U(1)_{B-L}$ and $\{\bf 1\}\equiv (1,1',1'')$. Even though we will not discuss masses and mixings on quark sector, 
there exists allowed region nearby $\tau=i$ or $\tau=i\times \infty$ in ref.~\cite{Okada:2019uoy}. 
These fixed points are known that there are residual symmetries $Z_2:\{1,S\}$ and $Z_3:\{1,T,T^2\}$, respectively.
\footnote{More systematic analysis for quark and lepton has been don by ref.~\cite{Yao:2020qyy}.}
In the lepton sector of the SM, both $L_L$ and $e_R$ have three types of singlets $\{1\}$ under $A_4$ with zero modular weight charges. Thanks to their assignments of $A_4$ symmetry, we find diagonal charged-lepton masses.
Therefore, we do not need to diagonalize the mass matrix of charged-leptons at leading order. 
$-1$ charge for $L_L$ and $e_R$ is imposed in order to cancel the anomaly of the $U(1)_{B-L}$ that is the same as the typical gauged $U(1)_{B-L}$ symmetry. 

As for boson sector, we add an isospin doublet inert boson $\eta\equiv [\eta^+,(\eta_R + i\eta_I)/\sqrt2]^T$, two isospin singlets $\varphi\equiv (v'+z'+i r)/\sqrt2$ and $\chi\equiv (\chi_R+i \chi_I)/\sqrt2$
with $(0,1,-3)$, $(1,1,0)$, and $(-1/2,1,-2)$ under $(U(1)_{B-L},A_4,-k)$, respectively, in addition to the SM Higgs $H\equiv [h^+,(v_H+h + iz)/\sqrt2]^T$, where $H$ is totally neutral under $U(1)_{B-L}\otimes A_4\otimes (-k)$.
$h^+$ and $z,z'$ are absorbed by the charged gauge boson $W^+$ and the two neutral gauge bosons $Z,Z'$ to get masses, respectively, after spontaneous symmetry breaking of electroweak and $U(1)_{B-L}$ symmetry.  
$\varphi$ plays a role in breaking the gauged $U(1)_{B-L}$ symmetry spontaneously,
while $\eta$ is expected to be an inert boson that contributes to the active neutrino mass matrix together with $\chi$ and neutral heavier fermions.
Field contents and their assignments are summarized in Table~\ref{tab:1}.

Under these symmetries, 
the valid Higgs potential is given by 
\begin{align}
{\it V}&= 
\frac{Y^{(4)}_{1}}2 \mu \varphi \chi^2 
+\frac{a}4 Y^{(6)}_{1}(H^\dag \eta)^2 
+{\rm h.c.}.
\label{Eq:pot}
\end{align}
Due to the above terms, we have mass differences between the real part and the imaginary one for the neutral components of $\eta$ and $\chi$, respectively.
Once we define these mass eigenstates to be $m_{\eta_R}$, $m_{\eta_I}$, $m_{\chi_R}$, and $m_{\chi_I}$,
the mass differences are respectively given by
\begin{align}
\delta m_\chi^2 \equiv m_{\chi_R}^2 - m_{\chi_I}^2 
 = Y^{(4)}_1 \mu v'/\sqrt2,\quad
\delta m_\eta^2 \equiv m_{\eta_R}^2-m_{\eta_I}^2= a(Y^{(6)}_1) v_H^2/4. 
\label{eq:massdiff}
\end{align}
Notice here that these mass differences $\delta m_\chi^2$ and $\delta m_\eta^2$ are respectively crucial in the light DM mass and active neutrino mass matrix, since they are proportional to these mass differences. 
%

We show each of the allowed Lagrangian in the lepton sector that is invariant under these symmetries.
One of the Yukawa Lagrangian is given by
\begin{align}
&(Y^{(4)}_{\rm 3})^* \otimes \bar L_L \otimes  N_R \otimes \tilde\eta + {\rm h.c.}\nn\\
=&
\alpha_\eta \bar L_{L_1} \tilde\eta (y'_1  N_{R_1} +y'_2  N_{R_2} +y'_3  N_{R_3})
+
\beta_\eta\bar L_{L_2} \tilde\eta (y'_1  N_{R_2} +y'_3  N_{R_1} +y'_2  N_{R_3}) \nn\\
&+
\gamma_\eta \bar L_{L_3} \tilde\eta (y'_1  N_{R_3} +y'_3  N_{R_2} +y'_2  N_{R_1})
+ {\rm h.c.}
.\label{Eq:neut1}
\end{align}
Here, $\tilde \eta\equiv i\sigma_2 \eta^*$, being $\sigma_2$ second Pauli matrix, we define $Y^{(4)}_{3}\equiv [y'_1,y'_2,y'_3]^T$ and $\alpha_{\eta},\beta_{\eta},\gamma_{\eta}$ are real free parameters without loss of generality.
Then the Yukawa coupling is given by
\begin{align}
f =
\left[
\begin{array}{ccc}
\alpha_\eta &0 &  0  \\ 
0 &  \beta_\eta &  0 \\ 
0  & 0 & \gamma_\eta \\ 
\end{array}\right]
\left[
\begin{array}{ccc}
y'^*_1 & y'^*_2 &  y'^*_3  \\ 
 y'^*_3 &  y'^*_1 &  y'^*_2 \\ 
 y'^*_2  &  y'^*_3 &  y'^*_1 \\ 
\end{array}\right]
=\alpha_\eta
\left[
\begin{array}{ccc}
1 &0 &  0  \\ 
0 &  \tilde\beta_\eta &  0 \\ 
0  & 0 &\tilde \gamma_\eta \\ 
\end{array}\right]
\left[
\begin{array}{ccc}
y'^*_1 & y'^*_2 &  y'^*_3  \\ 
 y'^*_3 &  y'^*_1 &  y'^*_2 \\ 
 y'^*_2  &  y'^*_3 &  y'^*_1 \\ 
\end{array}\right]
=\alpha_\eta \tilde f,
\end{align}
where $\tilde\beta_\eta\equiv \beta_\eta/\alpha_\eta$ and  $\tilde\gamma_\eta\equiv \gamma_\eta/\alpha_\eta$.
This coupling plays a role in connecting the active neutrino masses and new fields.

The Second Yukawa Lagrangian is given by
\begin{align}
&Y^{(4)}_{\rm 3} \otimes \bar N_R \otimes  S_L \otimes \chi +
Y^{(4)}_{\rm \bar3} \otimes \bar N_R \otimes  S_L \otimes \chi +
Y^{(4)}_{\rm 1} \otimes \bar N_R \otimes S_L \otimes \chi  +
Y^{(4)}_{\rm 1'} \otimes \bar N_R \otimes S_L \otimes \chi  +
 {\rm h.c.}\nn\\
=&
\alpha_N 
[f'_1(2\bar N_{R_1}S_{L_1} -\bar N_{R_2}S_{L_2} -\bar N_{R_3}S_{L_3})
+
f'_2(2\bar N_{R_3}S_{L_2} -\bar N_{R_2}S_{L_1} -\bar N_{R_1}S_{L_3})\nn\\
&+
f'_3(2\bar N_{R_2}S_{L_3} -\bar N_{R_1}S_{L_2} -\bar N_{R_3}S_{L_1})] \chi\nn\\
&+
\beta_N 
[f'_1(\bar N_{R_3}S_{L_3} -\bar N_{R_2}S_{L_2})
+
f'_2(\bar N_{R_2}S_{L_1} -\bar N_{R_1}S_{L_3})
+
f'_3(\bar N_{R_1}S_{L_2} -\bar N_{R_3}S_{L_1})] \chi \nn\\
+&\gamma_NY^{(4)}_1(\bar N_{R_1}S_{L_1} +\bar N_{R_2}S_{L_2}+\bar N_{R_3}S_{L_3})] \chi
+\delta_NY^{(4)}_{1'}(\bar N_{R_1}S_{L_3} +\bar N_{R_2}S_{L_2}+\bar N_{R_2}S_{L_1})] \chi
+ {\rm h.c.},
\label{eq:ykw1}\end{align}
where $\alpha_N,\beta_N,\gamma_N,\delta_N$ are complex free parameters.
Then the corresponding Yukawa coupling is given by
\begin{align}
g =
\alpha_N
\left[\begin{array}{ccc}
2y'_1 & -y'_3 &  -y'_2  \\ 
-y'_2 &  -y'_1 &  2y'_3 \\ 
 -y'_3  & 2y'_2 & -y'_1 \\ 
\end{array}\right]+
\beta_N
\left[\begin{array}{ccc}
0 & y'_3 &  -y'_2  \\ 
y'_2 &  -y'_1 &  0 \\ 
-y'_3  & 0 & y'_1 \\ 
\end{array}\right]+
\gamma_N
\left[\begin{array}{ccc}
1 & 0 &  0  \\ 
0 &  1&  0 \\ 
0  & 0 & 1 \\ 
\end{array}\right]+
\delta_N
\left[\begin{array}{ccc}
0 & 0 &  1 \\ 
1 & 0 &  0 \\ 
0 & 1 &  0 \\ 
\end{array}\right]
.\label{eq:ykw2}
\end{align}

The remaining terms are masses for exotic neutral fermions,
and the first mass term is given by 
\begin{align}
& Y^{(2)}_3\otimes \bar S^C_L\otimes S_L\otimes \varphi+{\rm h.c.}\nn\\
&=\alpha_L
[f_1(2\bar S_{L_1}^C S_{L_1} -\bar S_{L_2}^CS_{L_3} -\bar S_{L_3}^CS_{L_2})
+
f_2(2\bar S_{L_2}^CS_{L_2} -\bar S_{L_3}^CS_{L_1} -\bar S_{L_1}^CS_{L_3})\nn\\
&+
f_3(2\bar S_{L_3}^CS_{L_3} -\bar S_{L_1}^CS_{L_2} -\bar S_{L_2}^CS_{L_1})]  \varphi,
\end{align}
where $Y^{(2)}_3\equiv [f_1,f_2,f_3]^T$.
After the spontaneous breaking of $\varphi$, we find the mass matrix for Majorana mass matrix for $S_L$ as follows:
\begin{align}
M_{S_L}= \frac{\alpha_L v'}{\sqrt2}
\left[\begin{array}{ccc}
2y_1 & -y_3 &  -y_2  \\ 
-y_3 &  2y_2 & -y_1 \\ 
 -y_2  &-y_1 & 2y_3 \\ 
\end{array}\right]
.\label{eq:m1}
\end{align}
$M_{S_L}$ with 3$\times$3 is diagonalized by a unitary matrix $V_L$ as $D_{S_L}\equiv V_L^T M_{S_L} V_L$ and $S_L=V_L\psi_L$, where $D_{L}$ is mass eigenvalue and $\psi_L$ is mass eigenstate.

The Majorana mass term on $S_R$ is directly given by 
\begin{align}
& M_{S_R} \bar S^C_R\otimes S_R+{\rm h.c.}
= M_{S_R} \bar S^C_R
\left[\begin{array}{ccc}
1 &0 & 0  \\ 
0 &  0 & 1 \\ 
0  &1 & 0 \\ 
\end{array}\right] S_R
.\label{eq:m1}
\end{align}
$M_{S_R}$ with 3$\times$3 is diagonalized by a unitary matrix $V_R$ as $D_{S_R}\equiv V_R^T M_{S_R} V_R$ and $S_R=V_R\psi_R$, where $D_R$ is mass eigenvalue and $\psi_R$ is mass eigenstate.
The concrete mass eigenstate and its mass eigenvalues are given by
\begin{align}
&V_R
=
\left[\begin{array}{ccc}
1 &0 & 0  \\ 
0 &  1/\sqrt2 &   -i/\sqrt2 \\ 
0  & 1/\sqrt2 &  i/\sqrt2 \\ 
\end{array}\right],
\quad
D_{S_R}=M_{S_R}{\rm Diag}[1,1,1].
\label{eq:m1}
\end{align}

Notice here that we do not have any other terms such as $\bar N^C_R N_R$, $\bar S_L S_R$, $\bar S_L N_R$,$\bar S^C_R N_R$ at tree level, but only the $\bar N^C_R N_R$ is generated at one-loop level. 
$\bar N^C_R N_R$ is found via the one-loop level through the following Yukawa term:
\begin{align}
&\bar N_{R_i} g_{ij} S_{L_j} (\chi_R+i\chi_I)/\sqrt2 +{\rm h.c.} 
= \bar N_{R_i} g_{ij} V_{L_{jk}}  \psi_{L_k} (\chi_R+i\chi_I)/\sqrt2 +{\rm h.c.}\nn\\
&= \bar N_{R_i} G_{ik}   \psi_{L_k} (\chi_R+i\chi_I) +{\rm h.c.},\label{eq:dm-int}
\end{align}
where $G_{ik}\equiv   \sum_{j=1}^3 g_{ij}  V_{L_{jk}}/\sqrt2$.
Then the mass matrix of $N_R$ is found as follows:
\begin{align}
M_{N_{ab}}&=\frac{1}{(4\pi)^2}\sum_{k=1}^3
G_{ak} D_{S_{L_k}} G^T_{kb}
\left[
\frac{m_{\chi_R}^2}{m_{\chi_R}^2-D_{S_{L_k}}^2} \ln\left(\frac{m_{\chi_R}^2}{D_{S_{L_k}}^2}\right)
-
\frac{m_{\chi_I}^2}{m_{\chi_I}^2-D_{S_{L_k}}^2} \ln\left(\frac{m_{\chi_I}^2}{D_{S_{L_k}}^2}\right)
\right].
\end{align}
$M_{N}$ with 3$\times$3 is diagonalized by a unitary matrix $V_N$ as $D_{N}\equiv V_N^T M_{N} V_N$ and $N_R=V_N\psi_N$, where $D_{N}$ is mass eigenvalue and $\psi_N$ is mass eigenstate.

\subsection{Dark matter}
What is the DM candidate in the model? In Table \ref{tab:1}, the bosons $\eta, \varphi, \chi$, and the fermion $S_{L_a}, S_{R_a},N_{R_a}$ are neutral components. 
 The lightest one of $S_{L_a}$ and $S_{R_a}$ would not be a natural DM candidate, since it has a mass at tree level.
Naively, $N_{R}$, i.e., $\psi_N$ in the mass eigenstate, is supposed to be relatively light because its mass is loop induced. However, their interactions are restricted by neutrino physics, as we will discuss later. As a consequence, its relic abundance via freeze-out mechanism is much larger than the Planck observation. The coupling is too small to produce the appropriate relic abundance by freeze-out mechanism, however, note that it is not feebly enough to accommodate the abundance via the freeze-in mechanism. Therefore other components than $\psi_N$ should be the DM in the model. Hence, we consider the lightest component of $\chi$ to be DM.
Let us suppose the parameter $Y_1^{(4)}\mu v'$ are positive for simplicity, then $\chi_I$ is the DM candidate. 

\subsubsection{Numerical analysis of $\chi$}
Even though $\chi$ has a mass at tree-level, we have an upper bound that DM mass should not exceed the mass of the lightest $\psi_N$. This is because $\chi$ directly interacts with $\psi_N$ as can be seen in Eq.(\ref{eq:dm-int}), and lighter one has to remain in the current Universe. Here, we fix the lightest mass of $\psi_N$ to be more than 50 GeV.

\begin{figure}[htbp]
  \includegraphics[width=70mm]{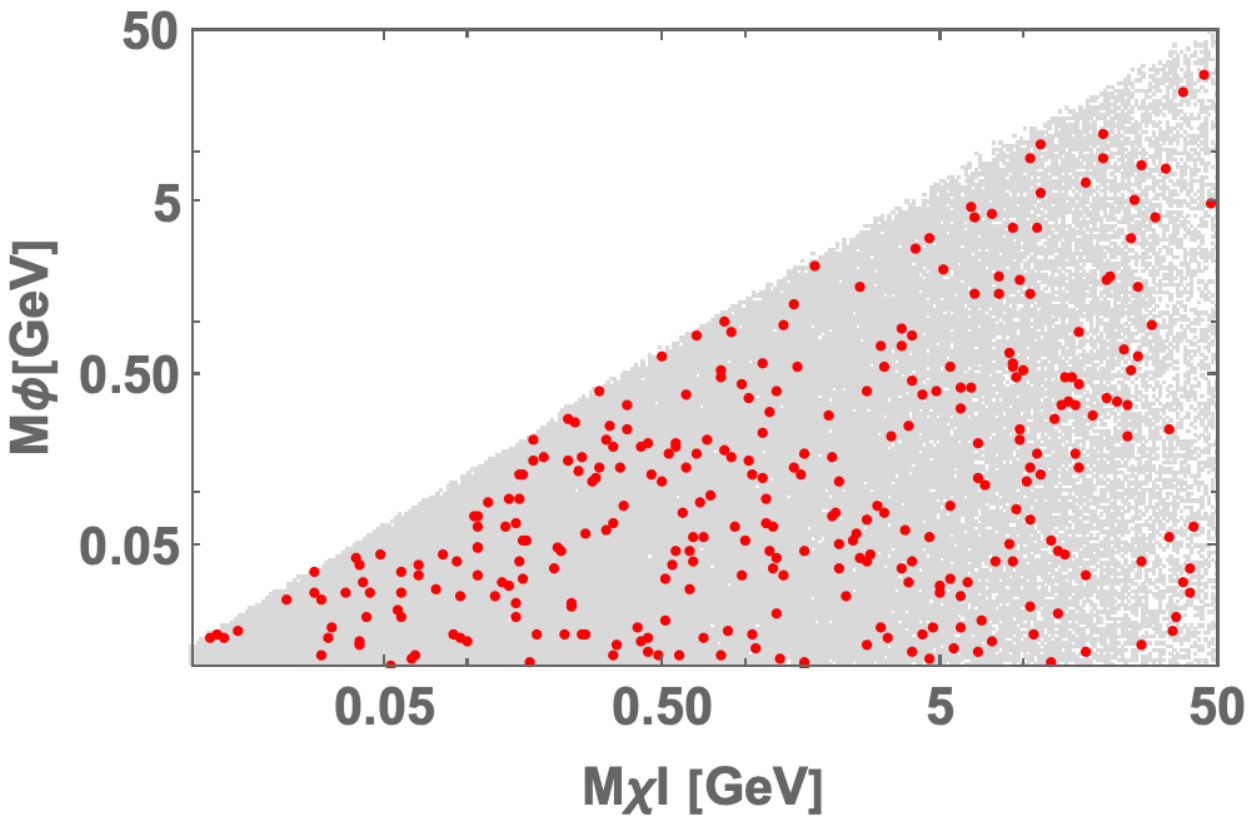}
   \caption{Relic abundance of DM $\chi_I$. Red and gray points represent the points where the relic abundance agrees with the observation and is smaller than that, respectively.}
   \label{fig:DMrelic}
\end{figure}

In Figure \ref{fig:DMrelic}, it is shown that the relic abundance of the DM candidate $\chi_I$ produced by the freeze-out mechanism is shown. The relic abundance and spin-independent cross section are calculated using MicrOmegas code \cite{Belanger:2020gnr} by inputting the Lagrangian in the model. 
The parameter space is scanned by random number parameters in the range: 
\begin{align}
&m_{\chi_I}=[0.01, 50]\ {\rm GeV},\
m_\phi=[0.01, 50]\ {\rm GeV},\\
&g_{\varphi}=[10^{-3}, 2\pi],\ 
Y_1^{(4)}=[10^{-4}, 10^{-2}],\\ 
&\mu v'=[10^{-5}, 10^4]\  {\rm GeV}^2,\ 
\sin{x}=[-0.1, 0.1], \label{eq:parameters}
\end{align}
where $x$ is mixing angle of neutral scalar $\varphi$ and Higgs boson, which would be examined by the direct detection constraint. 
Red and gray points represent the allowed region where the relic abundance agrees with the observed relic abundance $\Omega h^2 =0.120 \pm 0.001$ \cite{Planck:2018vyg} in 3$\sigma$, and is smaller than the constraint, respectively. In all the parameter regions, there are points where the relic abundance matches the observation. The associated main annihilation process is $\chi_I \chi_I \to \varphi \to \varphi \varphi$. Its thermally averaged annihilation cross section in the non-relativistic limit is
\begin{align}
\langle \sigma v_{\mathrm{M\o l}}\rangle \simeq \frac{g_\varphi^2 v^{'2}Y_1^{(4)2}\mu^2}{1024\pi \, m_{\chi_I}^2(4m_{\chi_I}^2-m_\varphi^2)^2},
\end{align}
where $\langle\cdots\rangle$ represents the thermally averaged \cite{Gondolo:1990dk} and $v_{\mathrm{M\o l}}$ is the M\o ller velocity of DM.

Notice that even though $\varphi$ is neutral and lighter than $\chi$ components, it cannot be a DM candidate because it is not stable. In Figure \ref{fig:DMrelic}, the mass of $\varphi$ is smaller than that of $\chi_I$. 

\begin{figure}[htbp]
\includegraphics[width=80mm]{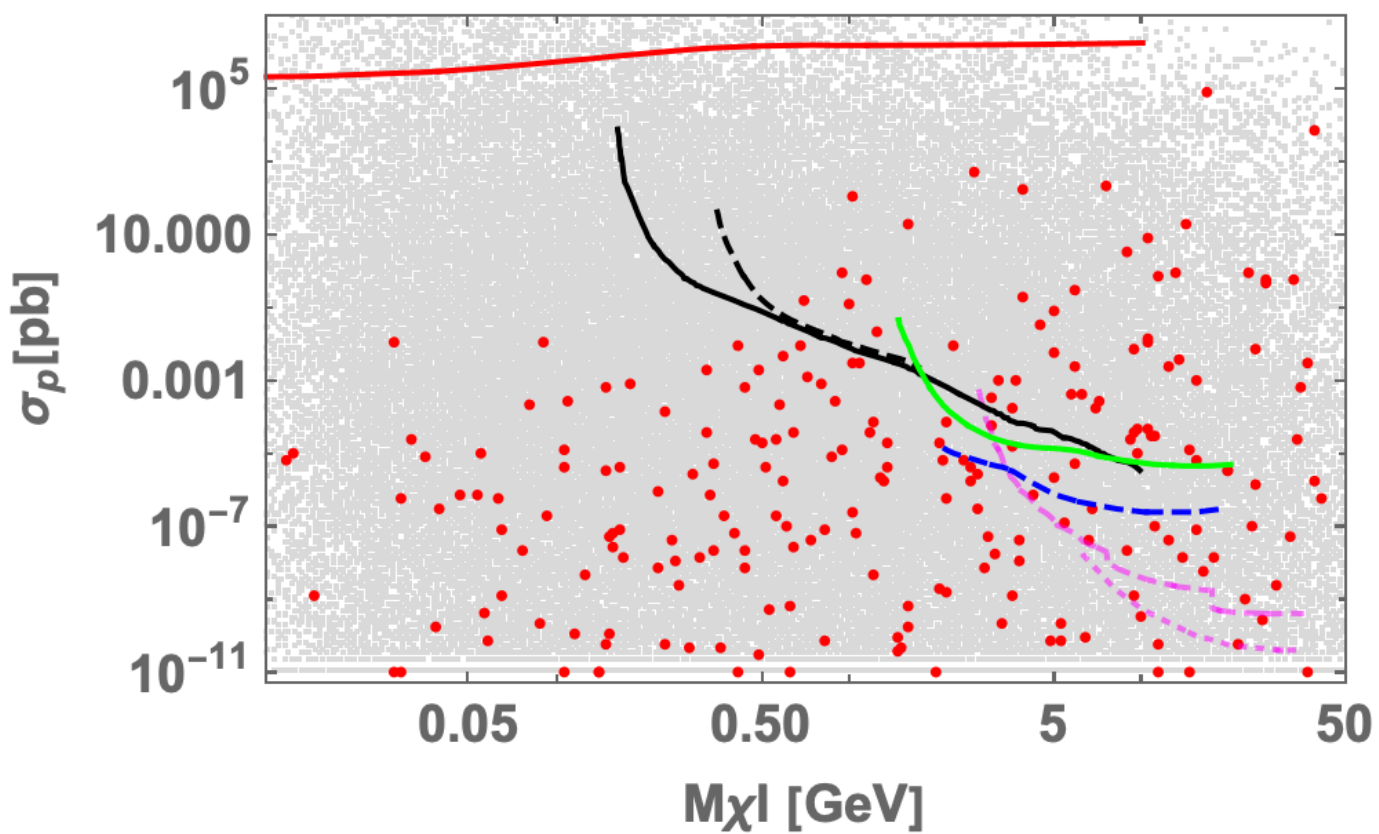}
   \caption{The spin-independent dark matter-proton cross section associated with constraints by the direct detection. Lines represent the constraints of direct detections. See the text for details of them.}
   \label{fig:DMdetection}
\end{figure}

In Figure \ref{fig:DMdetection}, constraint for spin-independent cross section of DM-proton interaction for direct detection is shown with current limits. In the figure, constraints are represented for cosmic-ray accelerated DM in red straight line, Xenon1T light DM search in magenta dashed line \cite{XENON:2019gfn}, Xenon1T in magenta dotted line \cite{XENON:2018voc, XENON:2019rxp}, DarkSide-50 in blue dashed line \cite{DarkSide:2018ppu, DarkSide:2018bpj, DarkSide:2018kuk}, CDMSlite in green dotted line \cite{SuperCDMS:2015eex}, CRESST-III in black dashed line \cite{CRESST:2017cdd} and in black straight line \cite{CRESST:2020}, respectively. 
As well as the Figure \ref{fig:DMrelic}, red and gray points correspond to parameter sets with which the relic abundance agrees with the constraint and is smaller than the constraints, respectively. 
DM $\chi$ does not directly couple to the standard model fields; thus, the only allowed interaction is through small $\chi_{I}$-Higgs mixing, $\sin{x}$ in Eq.(\ref{eq:parameters}). We can see there are enough parameter points not only in the small mass region where it is moderately bounded by the cosmic ray accelerated DM but also in the larger mass region than that where direct detection experiments seriously restrict it.
In the points which satisfy both relic abundance and direct detection constraints, the coupling $g_\varphi$ should not be too small. At the same time, the mixing angle $\sin{x}$ should be slight to avoid the direct detection constraints.

\subsection{Neutrino sector}
Similar to the $N_R$, we can derive the active neutrino mass matrix.

\subsubsection{Neutrino mass matrix}
The valid Lagrangian is now given in terms of $\psi_N$  as follows:
\begin{align}
&\frac1{\sqrt2} f_{ia} \bar \nu_{L_i} V_{N_{ak}} \psi_{N_k} (\eta_R-\eta_I)+{\rm h.c.} 
\equiv
F_{ik}  \bar \nu_{L_i}   \psi_{N_k} (\eta_R-\eta_I) +{\rm h.c.},
\end{align}
where $F_{ik}\equiv \sum_{a=1}^3 f_{ia}  V_{L_{ak}}/\sqrt2\equiv \alpha_\eta \sum_{a=1}^3 \tilde f_{ia}  V_{L_{ak}}/\sqrt2\equiv \alpha_\eta \tilde F$.
Then, the neutrino mass matrix is given at one-loop level as follows:
\begin{align}
m_{\nu_{ij}}&=\frac{\alpha_\eta^2}{(4\pi)^2}\sum_{k=1}^3
\tilde F_{ik} D_{N_k} \tilde F^T_{kj}
\left[
\frac{m_{\eta_R}^2}{m_{\eta_R}^2-D_{N_k}^2} \ln\left(\frac{m_{\eta_R}^2}{D_{N_k}^2}\right)
-
\frac{m_{\eta_I}^2}{m_{\eta_I}^2-D_{N_k}^2} \ln\left(\frac{m_{\eta_I}^2}{D_{N_k}^2}\right)
\right]\equiv \alpha_\eta^2 \tilde m_{\nu_{ij}}.
\end{align}
$m_\nu$ is diagonalzied by a unitary matrix $U_{\rm PMNS}$~\cite{Maki:1962mu}; $D_\nu=\alpha_\eta^2 \tilde D_\nu= U_{\rm PMNS}^T m_\nu U_{\rm PMNS}=\alpha_\eta^2 U_{\rm PMNS}^T \tilde m_\nu U_{\rm PMNS}$.
Then $\alpha_\eta$ is determined by
\begin{align}
(\mathrm{NH}):\  \alpha_\eta^4= \frac{|\Delta m_{\rm atm}^2|}{\tilde D_{\nu_3}^2-\tilde D_{\nu_1}^2},
\quad
(\mathrm{IH}):\  \alpha_\eta^4 = \frac{|\Delta m_{\rm atm}^2|}{\tilde D_{\nu_2}^2-\tilde D_{\nu_3}^2},
 \end{align}
where $\Delta m_{\rm atm}^2$ is atmospheric neutrino mass difference squares, and NH and IH represent the normal hierarchy and the inverted hierarchy, respectively. 
Subsequently, the solar mass different squares can be written in terms of $ \alpha_\eta$ as follows:
\begin{align}
\Delta m_{\rm sol}^2=  \alpha_\eta^4 ({\tilde D_{\nu_2}^2-\tilde D_{\nu_1}^2}),
 \end{align}
 which can be compared to the observed value.
 %
In our model, one finds $U_{\mathrm{PMNS}}=V_\nu$ since the charged-lepton is diagonal basis, and 
it is parametrized by three mixing angle $\theta_{ij} (i,j=1,2,3; i < j)$, one CP violating Dirac phase $\delta_{CP}$,
and two Majorana phases $\{\alpha_{21}, \alpha_{32}\}$ as follows:
\begin{equation}
U_{\mathrm{PMNS}} = 
\begin{pmatrix} c_{12} c_{13} & s_{12} c_{13} & s_{13} e^{-i \delta_{CP}} \\ 
-s_{12} c_{23} - c_{12} s_{23} s_{13} e^{i \delta_{CP}} & c_{12} c_{23} - s_{12} s_{23} s_{13} e^{i \delta_{CP}} & s_{23} c_{13} \\
s_{12} s_{23} - c_{12} c_{23} s_{13} e^{i \delta_{CP}} & -c_{12} s_{23} - s_{12} c_{23} s_{13} e^{i \delta_{CP}} & c_{23} c_{13} 
\end{pmatrix}
\begin{pmatrix} 1 & 0 & 0 \\ 0 & e^{i \frac{\alpha_{21}}{2}} & 0 \\ 0 & 0 & e^{i \frac{\alpha_{31}}{2}} \end{pmatrix},
\end{equation}
where $c_{ij}$ and $s_{ij}$ stands for $\cos \theta_{ij}$ and $\sin \theta_{ij}$ respectively. 
Then, each of mixing is given in terms of the component of $U_{\mathrm{PMNS}}$ as follows:
\begin{align}
\sin^2\theta_{13}=|(U_{\mathrm{PMNS}})_{13}|^2,\quad 
\sin^2\theta_{23}=\frac{|(U_{\mathrm{PMNS}})_{23}|^2}{1-|(U_{\mathrm{PMNS}})_{13}|^2},\quad 
\sin^2\theta_{12}=\frac{|(U_{\mathrm{PMNS}})_{12}|^2}{1-|(U_{\mathrm{PMNS}})_{13}|^2}.
\end{align}
Also, we compute the Jarlskog invariant, $\delta_{CP}$ derived from PMNS matrix elements $U_{\alpha i}$:
\begin{equation}
J_{CP} = \text{Im} [U_{e1} U_{\mu 2} U_{e 2}^* U_{\mu 1}^*] = s_{23} c_{23} s_{12} c_{12} s_{13} c^2_{13} \sin \delta_{CP},
\end{equation}
and the Majorana phases are also estimated in terms of other invariants $I_1$ and $I_2$:
\begin{equation}
I_1 = \text{Im}[U^*_{e1} U_{e2}] = c_{12} s_{12} c_{13}^2 \sin \left( \frac{\alpha_{21}}{2} \right), \
I_2 = \text{Im}[U^*_{e1} U_{e3}] = c_{12} s_{13} c_{13} \sin \left( \frac{\alpha_{31}}{2} - \delta_{CP} \right).
\end{equation}
In addition, the effective mass for the neutrinoless double beta decay is given by
\begin{align}
\langle m_{ee}\rangle= |\alpha_\eta|^2 |\tilde D_{\nu_1} \cos^2\theta_{12} \cos^2\theta_{13}+\tilde D_{\nu_2} \sin^2\theta_{12} \cos^2\theta_{13}e^{i\alpha_{21}}+\tilde D_{\nu_3} \sin^2\theta_{13}e^{i(\alpha_{31}-2\delta_{CP})}|,
\end{align}
where its observed value could be measured by KamLAND-Zen in future~\cite{KamLAND-Zen:2016pfg}. 
We will adopt the neutrino experimental data at 3$\sigma$ interval in NuFit5.0~\cite{Esteban:2018azc} as follows:
\begin{align}
&{\rm NH}: \Delta m^2_{\rm atm}=[2.431, 2.598]\times 10^{-3}\ {\rm eV}^2,\
\Delta m^2_{\rm sol}=[6.82, 8.04]\times 10^{-5}\ {\rm eV}^2,\\
&\sin^2\theta_{13}=[0.02034, 0.02430],\ 
\sin^2\theta_{23}=[0.407, 0.618],\ 
\sin^2\theta_{12}=[0.269, 0.343],\nn\\
&{\rm IH}: \Delta m^2_{\rm atm}=[2.412, 2.583]\times 10^{-3}\ {\rm eV}^2,\
\Delta m^2_{\rm sol}=[6.82, 8.04]\times 10^{-5}\ {\rm eV}^2,\\
&\sin^2\theta_{13}=[0.02053, 0.02436],\ 
\sin^2\theta_{23}=[0.411, 0.621],\ 
\sin^2\theta_{12}=[0.269, 0.343].\nn
\end{align}

\subsubsection{Numerical analysis \label{sec:NA}}
Here, we present our numerical analysis applying the following ranges of input parameters,
\begin{align}
&\{|\alpha_N|,\ |\beta_N|,\ |\gamma_N|,\ |\delta_N|\}\in [0.0001,1],\ \{\tilde\beta_{\eta},\ \tilde\gamma_{\eta} \} \in [0.1,10],\ \alpha_{L}  \in [0.0001,1],\  \nn\\
&  M_{S_R} \in [10^2,10^{10}]{\rm GeV}, \quad m_\eta \in [10^5,10^6] \ {\rm GeV},
, \quad m_{\chi_R} \in [10^3,10^5] \ {\rm GeV},
\end{align}
where in our work, we run whole the fundamental region  of $\tau$, and fix $v'=10^5$ GeV.

{\it Normal hierarchy case}:
\begin{figure}[htbp]
 \begin{minipage}{0.32\hsize}
  \begin{center}
\includegraphics[width=49mm]{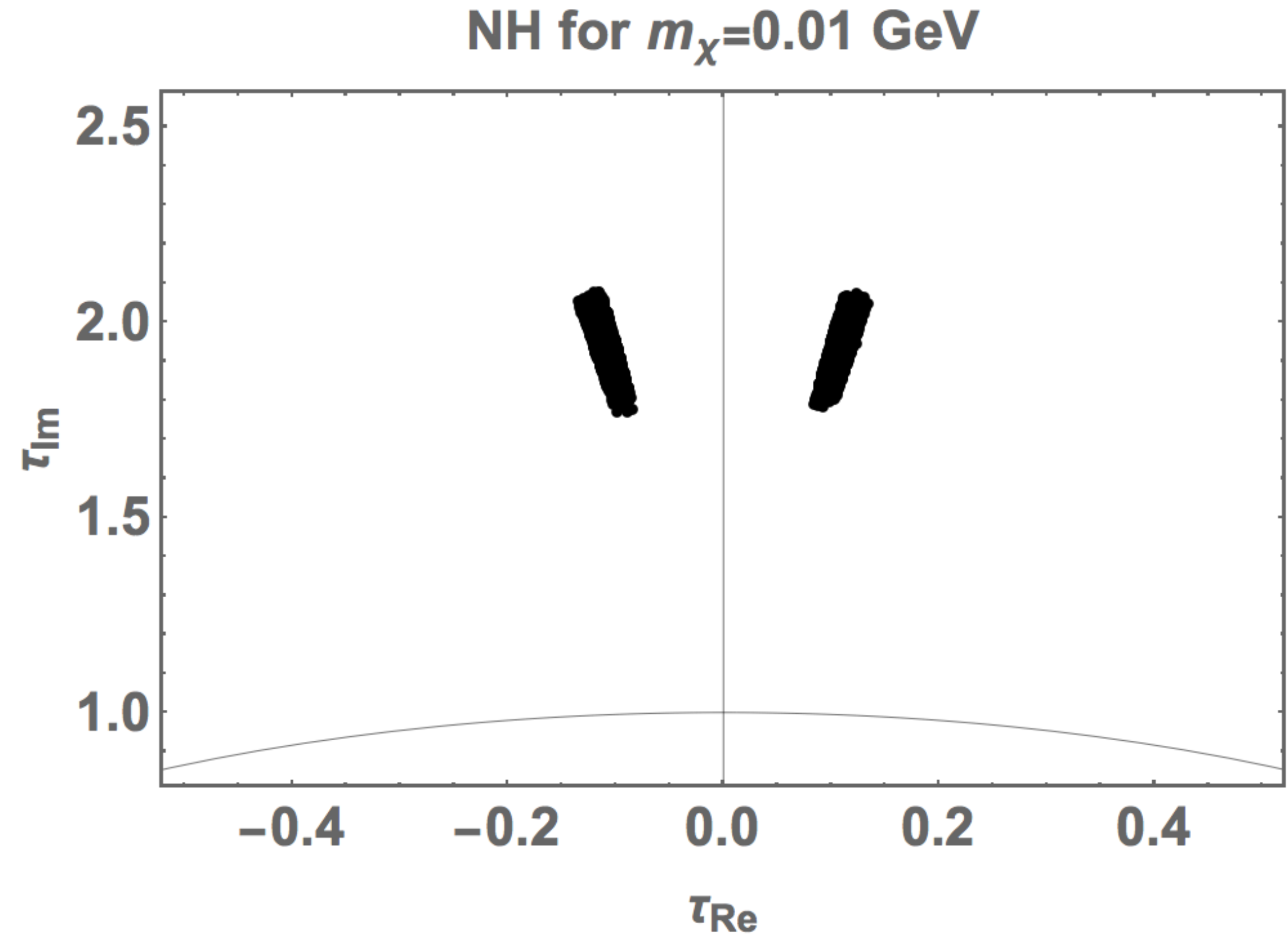}
  \end{center}
 \end{minipage}
 \begin{minipage}{0.32\hsize}
 \begin{center}
  \includegraphics[width=49mm]{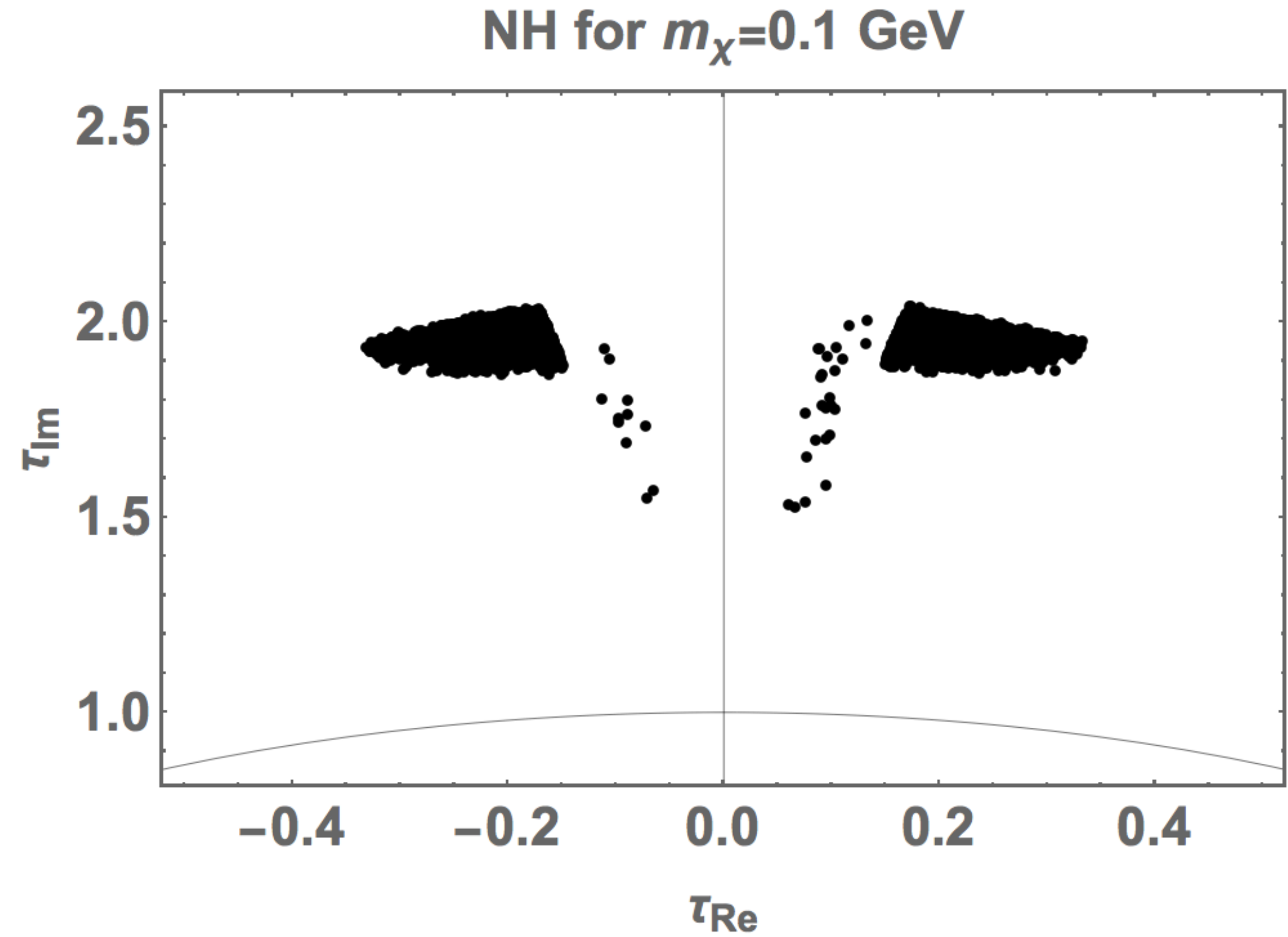}
 \end{center}
 \end{minipage}
 \\
 \begin{minipage}{0.32\hsize}
 \begin{center}
  \includegraphics[width=49mm]{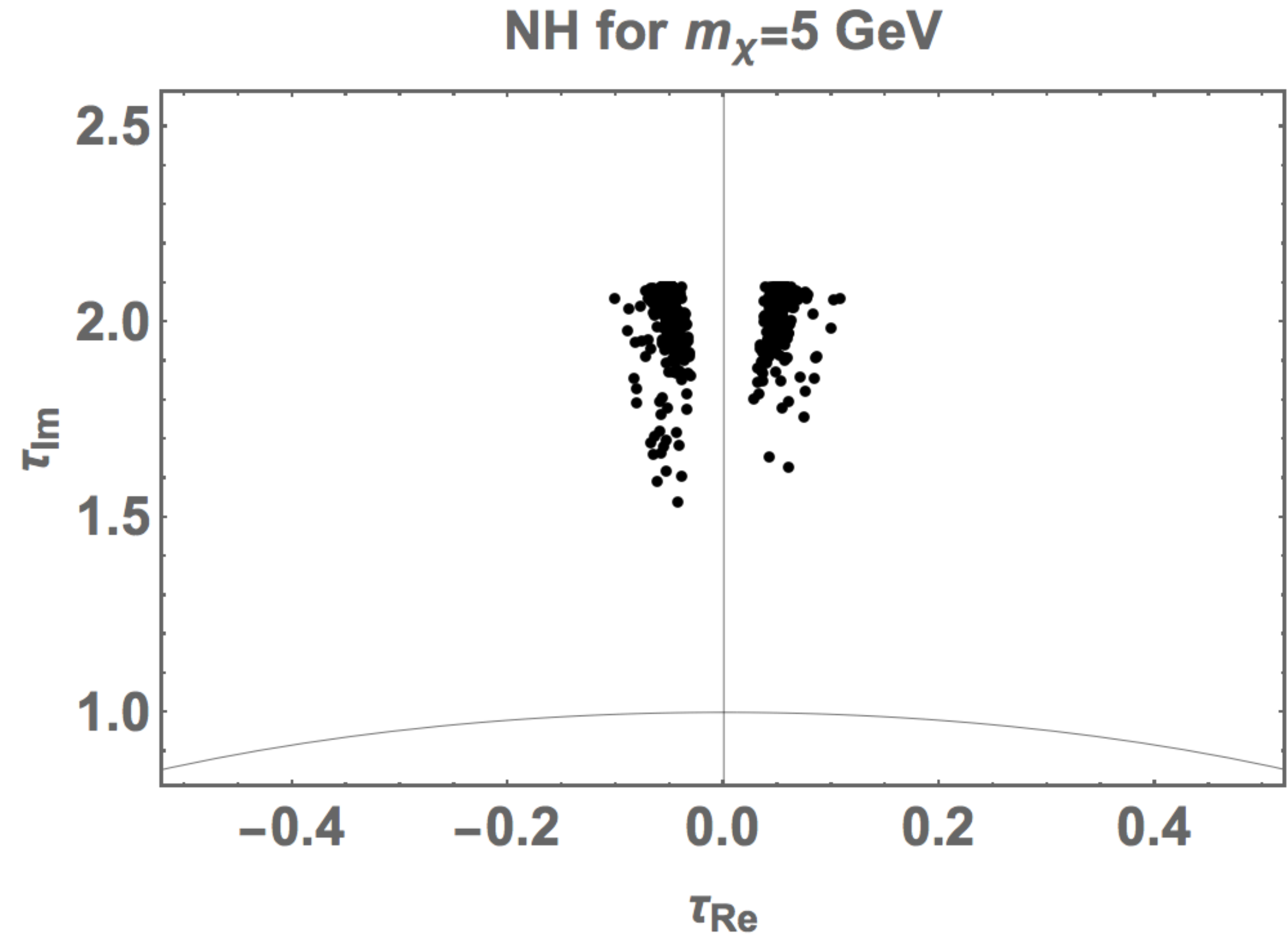}
 \end{center}
 \end{minipage}
   \begin{minipage}{0.32\hsize}
 \begin{center}
  \includegraphics[width=49mm]{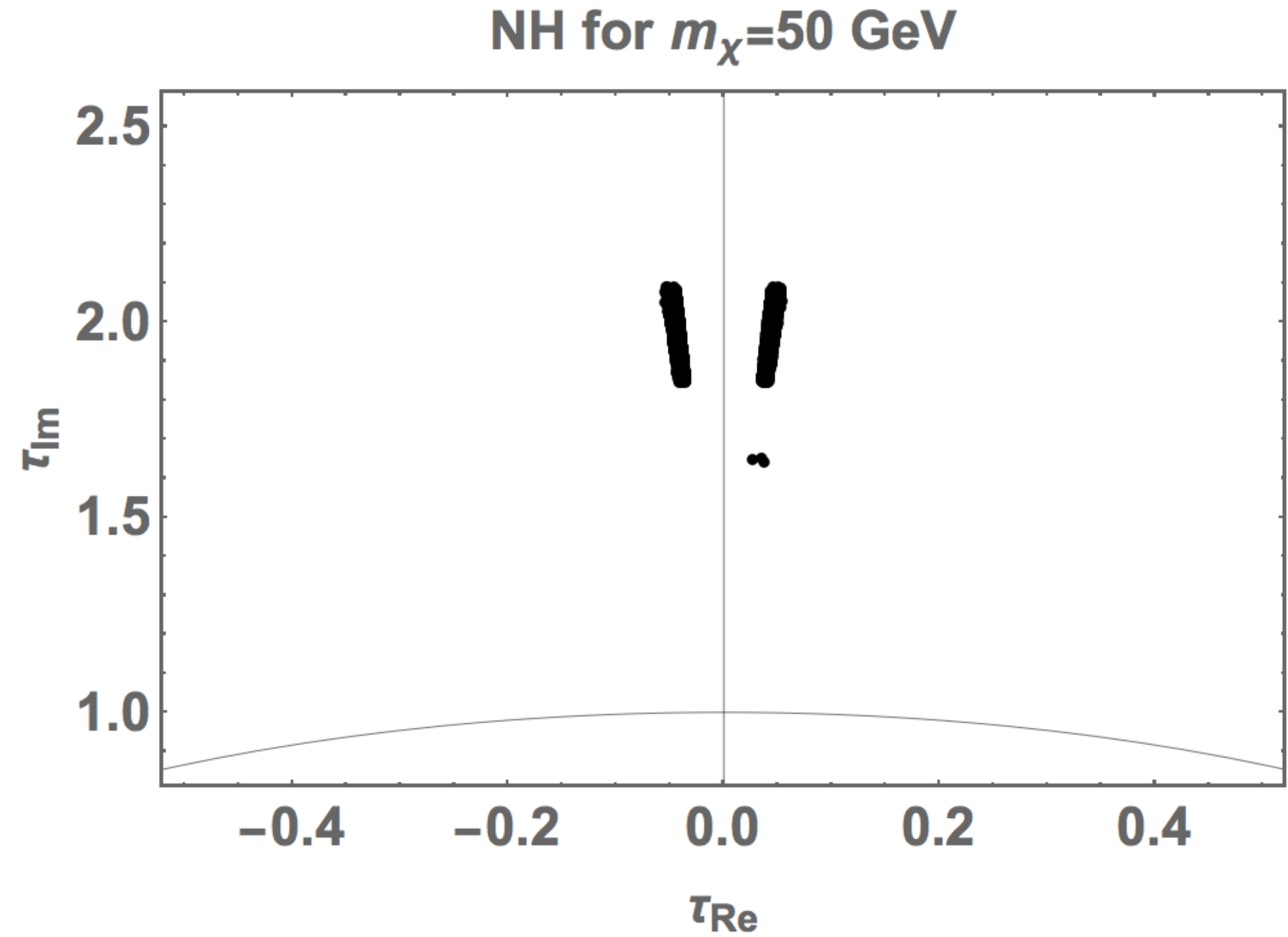}
 \end{center}
 \end{minipage}
 \caption{Scatter plots of Re$[\tau]\equiv \tau_{\rm Re}$ and  Im$[\tau]\equiv \tau_{\rm Im}$,
where the left up figure is the one for $m_\chi=0.01$ GeV, the right up one $m_\chi=0.1$ GeV, the left down one $m_\chi=5$ GeV, and the right down one $m_\chi=50$ GeV. }
   \label{fig:tau_nh}
\end{figure}
In Fig.~\ref{fig:tau_nh}, scatter plots of Re$[\tau]\equiv \tau_{\rm Re}$ and Im$[\tau]\equiv \tau_{\rm Im}$ are shown,
where the left up figure is the one for $m_\chi=0.01$ GeV, the right up one $m_\chi=0.1$ GeV, the left down one $m_\chi=5$ GeV, and the right down one $m_\chi=50$ GeV. 
Each of the allowed regions is at nearby 
$|\tau_{\rm Re}|=[0.07-0.5]$ and $\tau_{\rm Im}=[1.7-2.1]$ for  $m_\chi=0.01$ GeV,
$|\tau_{\rm Re}|=[0.05-0.35]$ and $\tau_{\rm Im}=[1.5-2.1]$ for  $m_\chi=0.1$ GeV,
$|\tau_{\rm Re}|=[0.02-0.1]$ and $\tau_{\rm Im}=[1.5-2.1]$ for  $m_\chi=5$ GeV, and
$|\tau_{\rm Re}|=[0.02-0.5]$ and $\tau_{\rm Im}=[1.6-2.1]$ for  $m_\chi=50$ GeV.

\begin{figure}[htbp]
 \begin{minipage}{0.32\hsize}
  \begin{center}
\includegraphics[width=49mm]{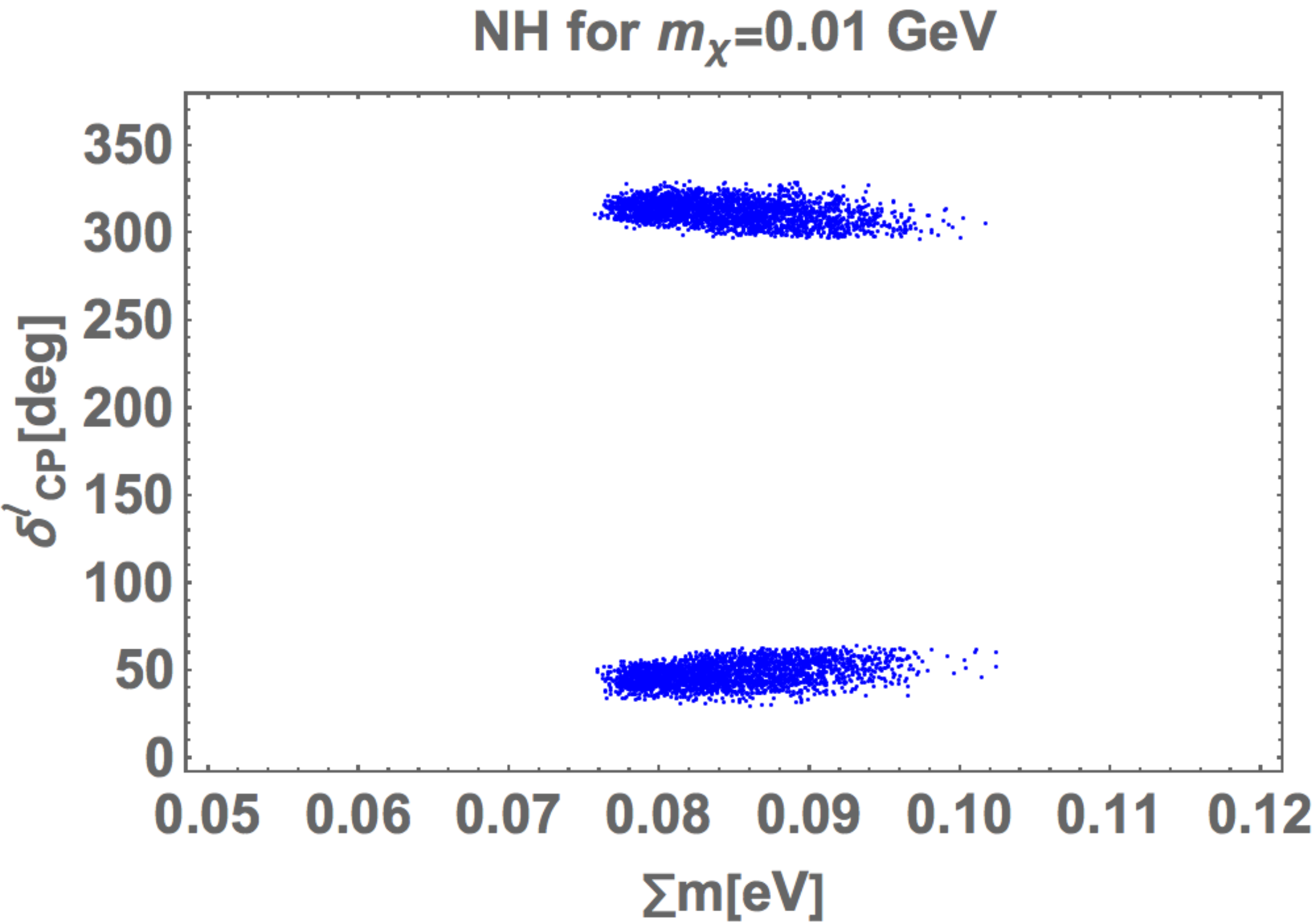}
  \end{center}
 \end{minipage}
 \begin{minipage}{0.32\hsize}
 \begin{center}
  \includegraphics[width=49mm]{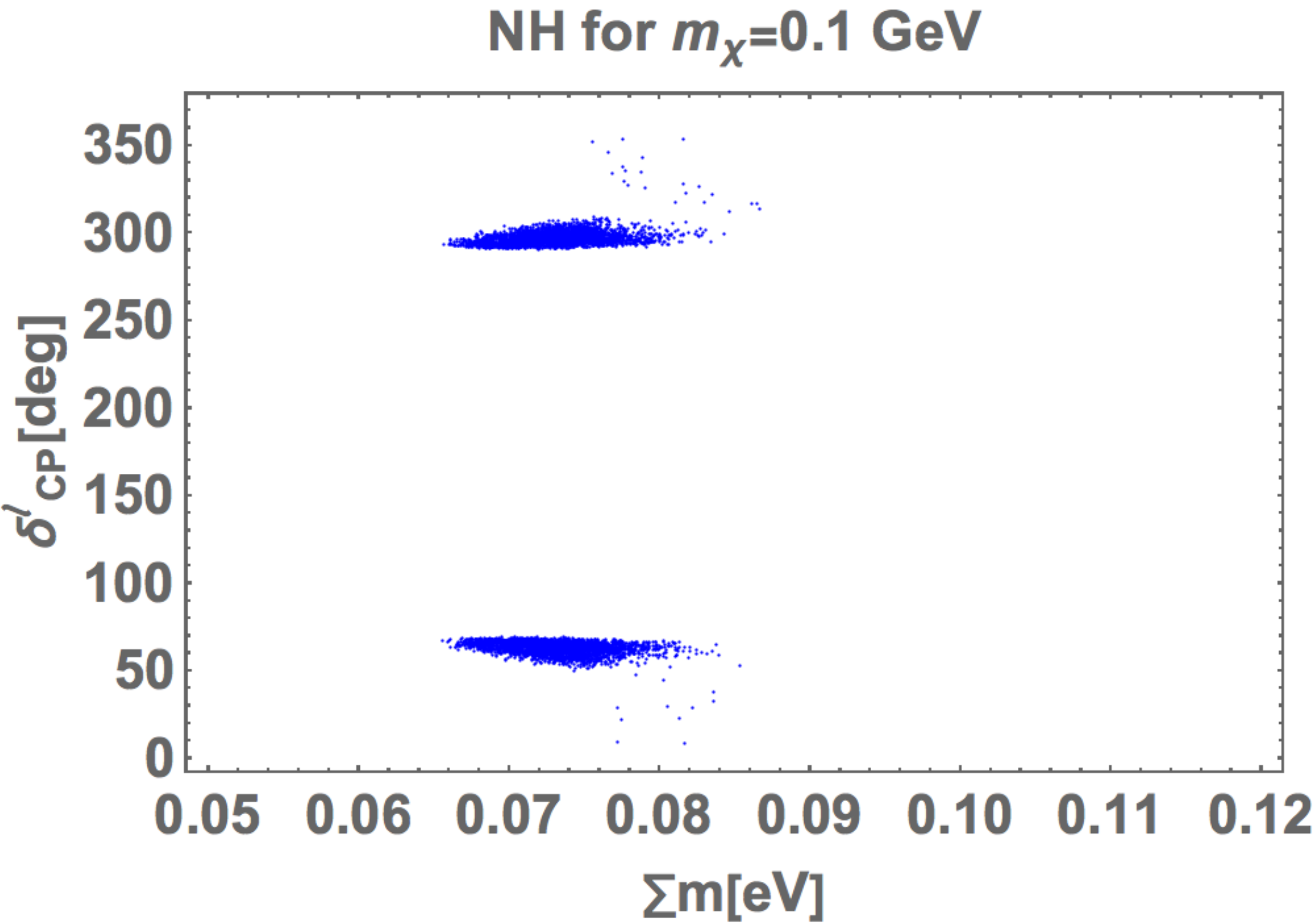}
 \end{center}
 \end{minipage}
 \\
 \begin{minipage}{0.32\hsize}
 \begin{center}
  \includegraphics[width=49mm]{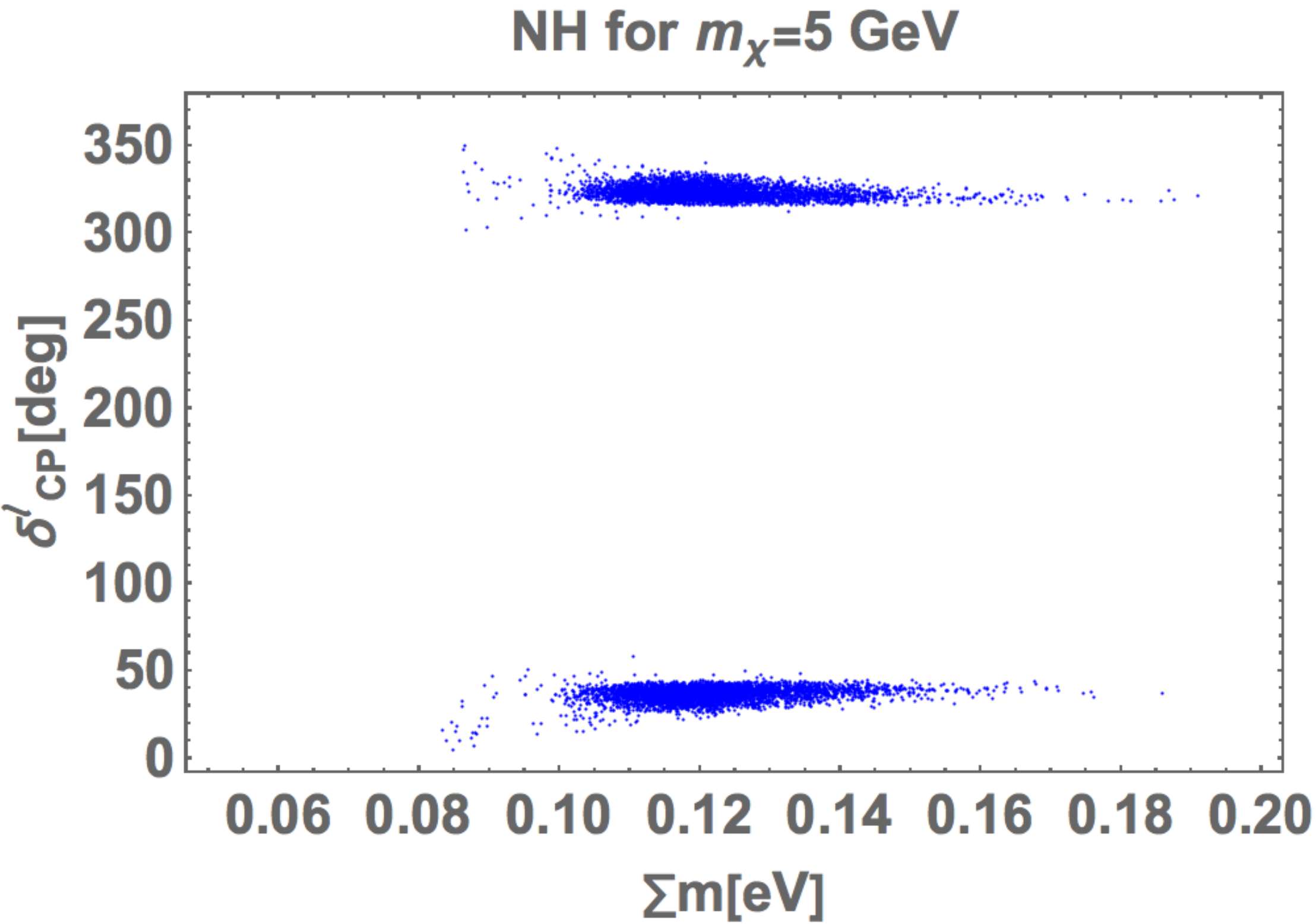}
 \end{center}
 \end{minipage}
   \begin{minipage}{0.32\hsize}
 \begin{center}
  \includegraphics[width=49mm]{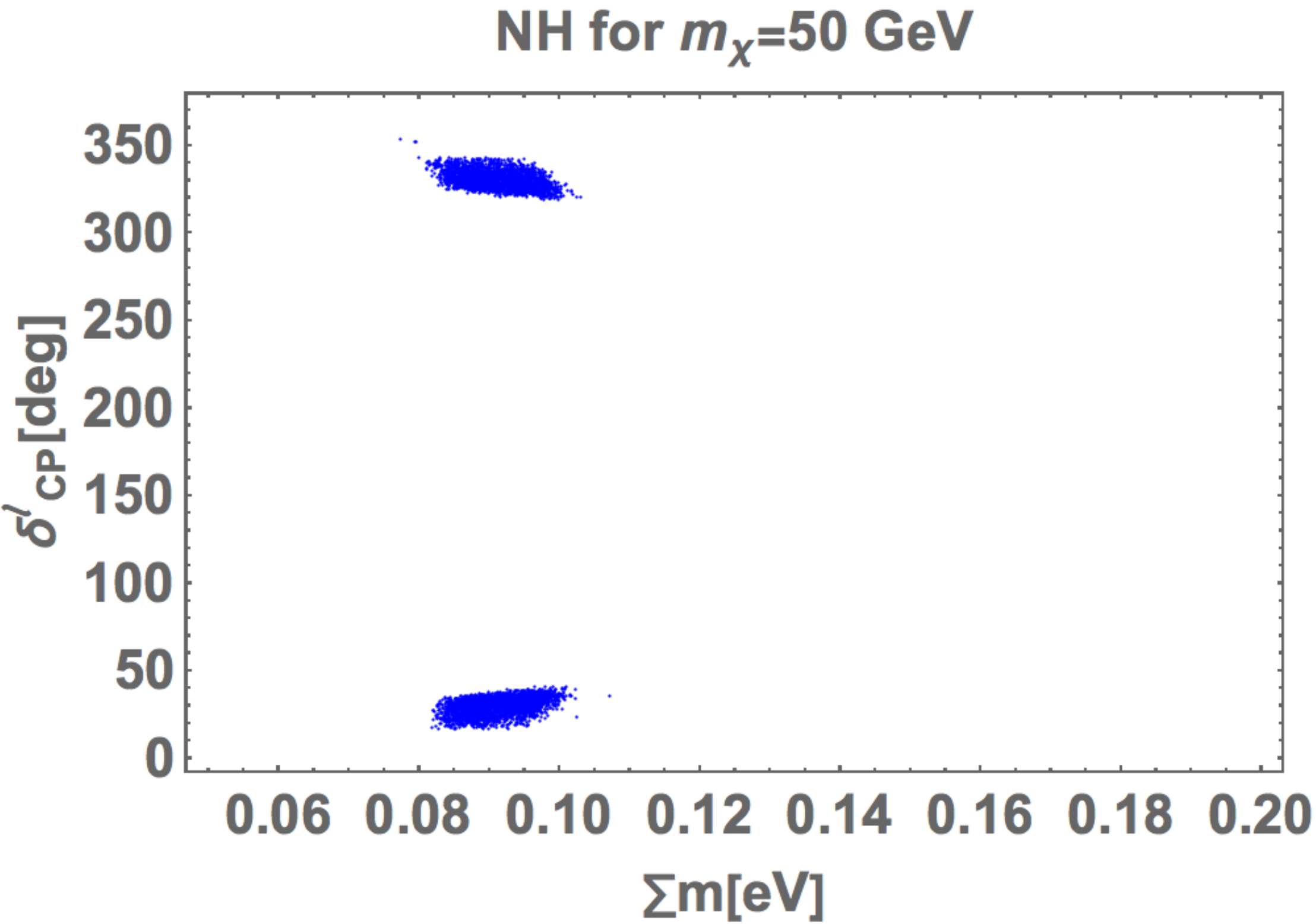}
 \end{center}
 \end{minipage}
 \caption{Scatter plots of sum of neutrino masses $\equiv\sum m$ eV and  $\delta^\ell_{\rm CP}$, where the legends are the same as Fig.~\ref{fig:tau_nh}.}
   \label{fig:sum-dcp_nh}
\end{figure}
In Fig.~\ref{fig:sum-dcp_nh}, scatter plots of sum of neutrino masses $\equiv\sum m$ eV and  $\delta^\ell_{\rm CP}$ are shown, where the legends are the same as Fig.~\ref{fig:tau_nh}.
Allowed regions are as follows:
$\sum m=[0.075-0.105]$ eV and $\delta^\ell_{\rm CP}=[30-60,300-330]$ for  $m_\chi=0.01$ GeV,
$\sum m=[0.065-0.085]$ eV and $\delta^\ell_{\rm CP}=[0-70,300-360]$ for  $m_\chi=0.1$ GeV,
$\sum m=[0.08 - 0.19]$ eV and $\delta^\ell_{\rm CP}=[0-60,300-360]$ for  $m_\chi=5$ GeV, and
$\sum m=[0.08-0.11]$ eV and $\delta^\ell_{\rm CP}=[30-50,320-350]$ for  $m_\chi=50$ GeV.

\begin{figure}[htbp]
 \begin{minipage}{0.32\hsize}
  \begin{center}
\includegraphics[width=49mm]{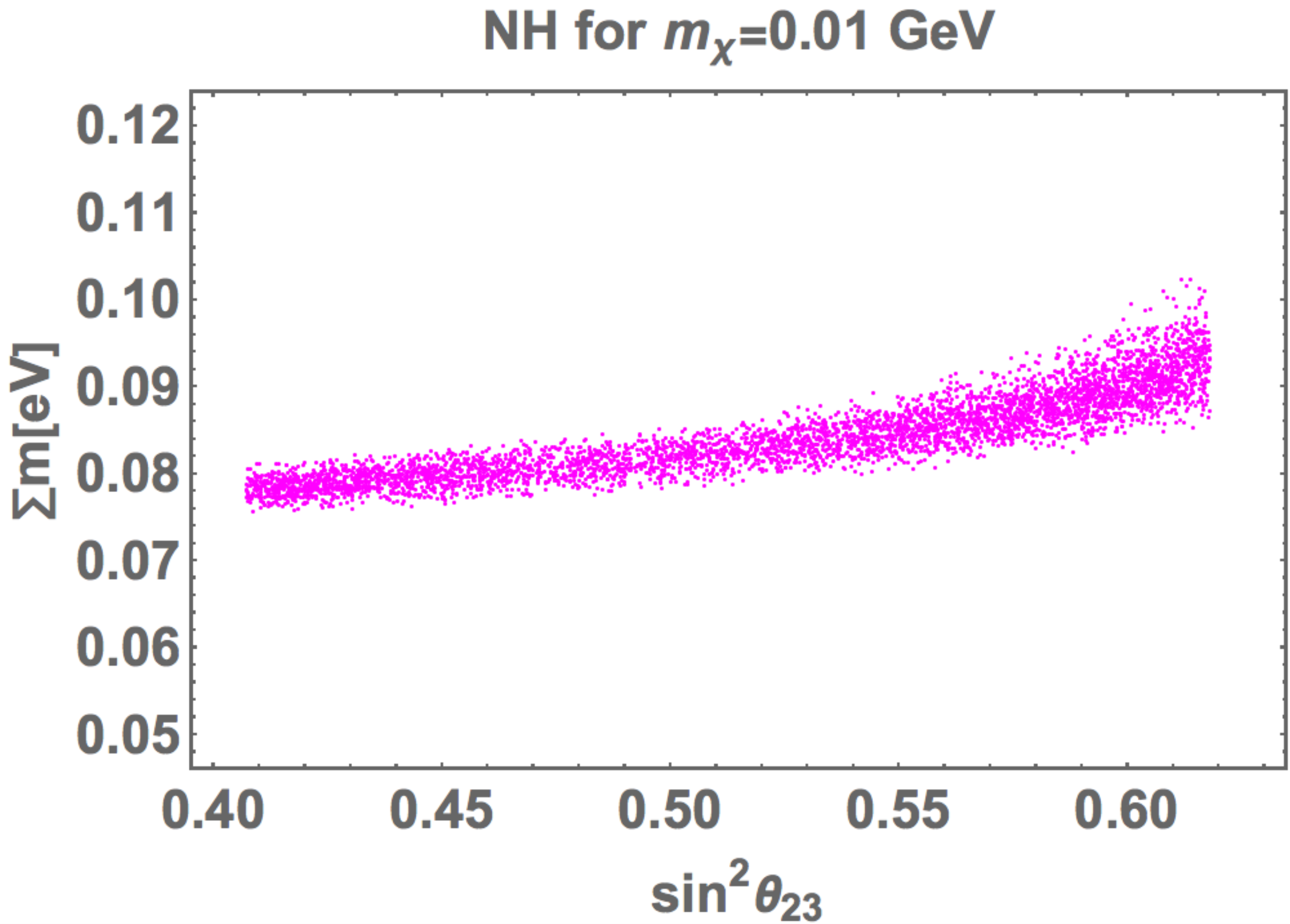}
  \end{center}
 \end{minipage}
 \begin{minipage}{0.32\hsize}
 \begin{center}
  \includegraphics[width=49mm]{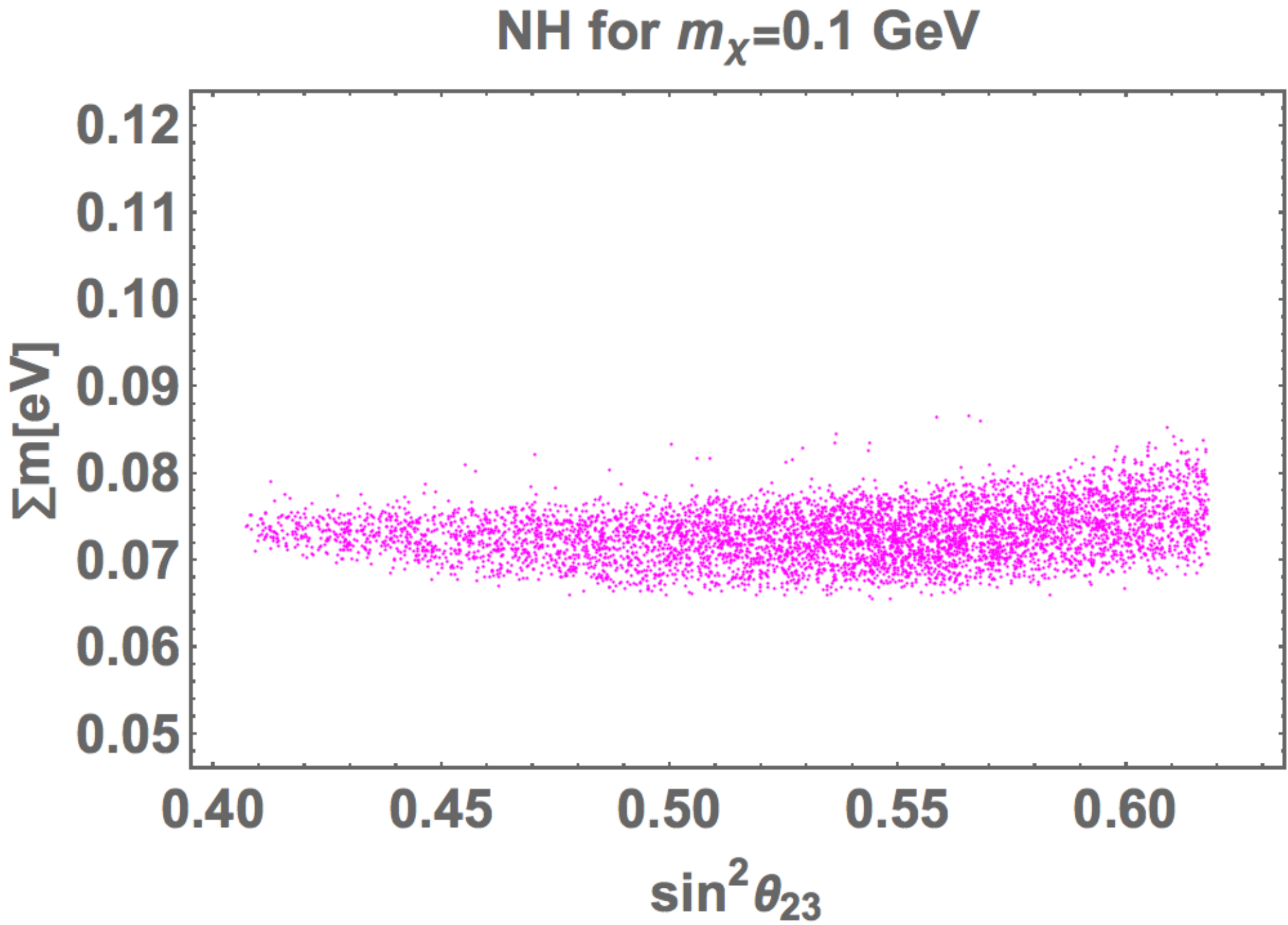}
 \end{center}
 \end{minipage}
 \\
 \begin{minipage}{0.32\hsize}
 \begin{center}
  \includegraphics[width=49mm]{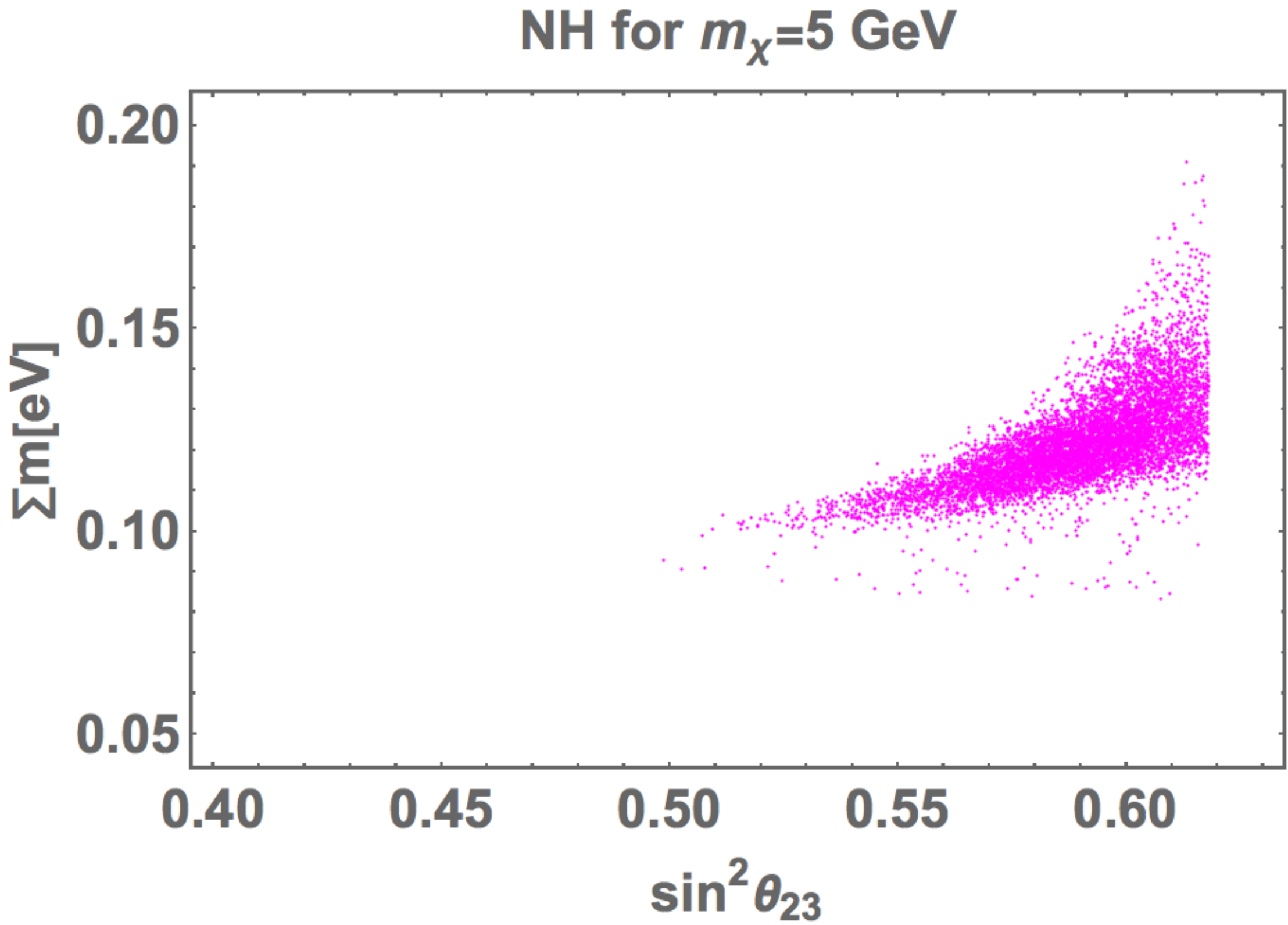}
 \end{center}
 \end{minipage}
   \begin{minipage}{0.32\hsize}
 \begin{center}
  \includegraphics[width=49mm]{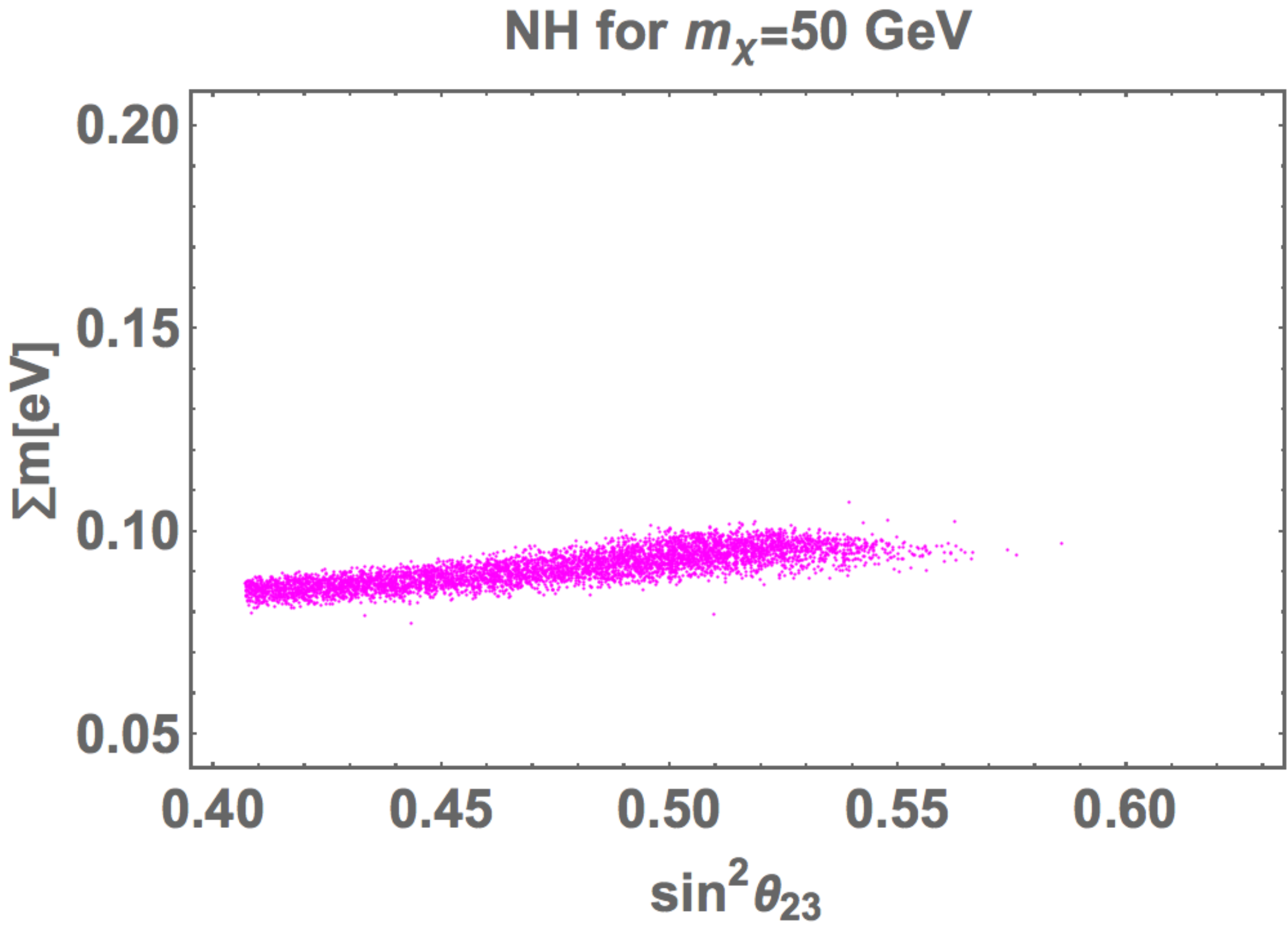}
 \end{center}
 \end{minipage}
 \caption{Scatter plots of $\sin^2\theta_{23}$ and  $\sum m$, where the legends are the same as Fig.~\ref{fig:tau_nh}.}
   \label{fig:s23-sum_nh}
\end{figure}
In Fig.~\ref{fig:s23-sum_nh}, scatter plots of $\sin^2\theta_{23}$ and  $\sum m$ are shown, where the legends are the same as Fig.~\ref{fig:tau_nh}.
Allowed regions are as follows:
$\sin^2\theta_{23}=[0.407-0.618]$  for  $m_\chi=0.01$ GeV,
$\sin^2\theta_{23}=[0.407-0.618]$  for  $m_\chi=0.1$ GeV,
$\sin^2\theta_{23}=[0.5-0.618]$  for  $m_\chi=5$ GeV, and
$\sin^2\theta_{23}=[0.407-0.58]$ for  $m_\chi=50$ GeV.

\begin{figure}[htbp]
 \begin{minipage}{0.32\hsize}
  \begin{center}
\includegraphics[width=49mm]{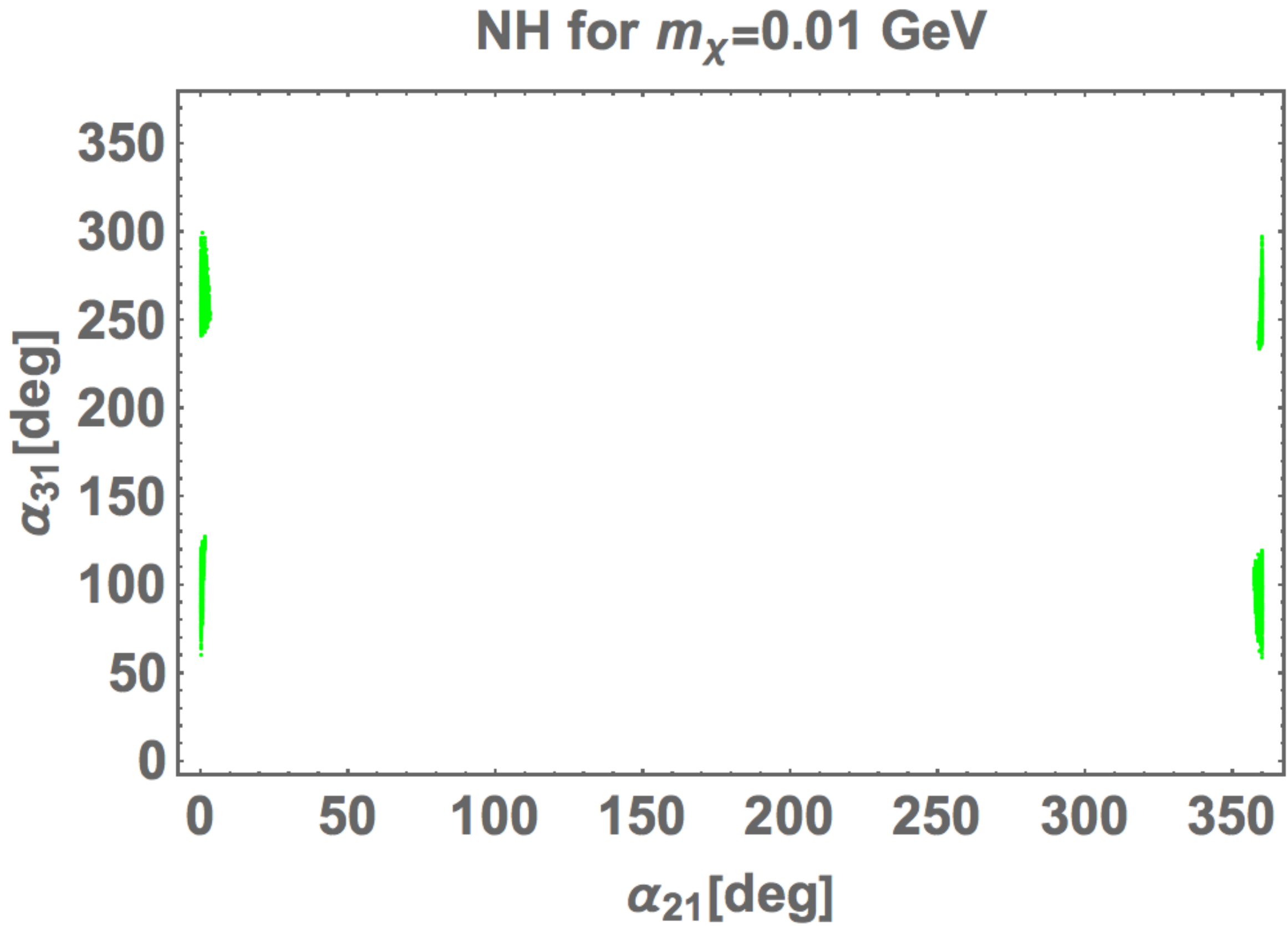}
  \end{center}
 \end{minipage}
 \begin{minipage}{0.32\hsize}
 \begin{center}
  \includegraphics[width=49mm]{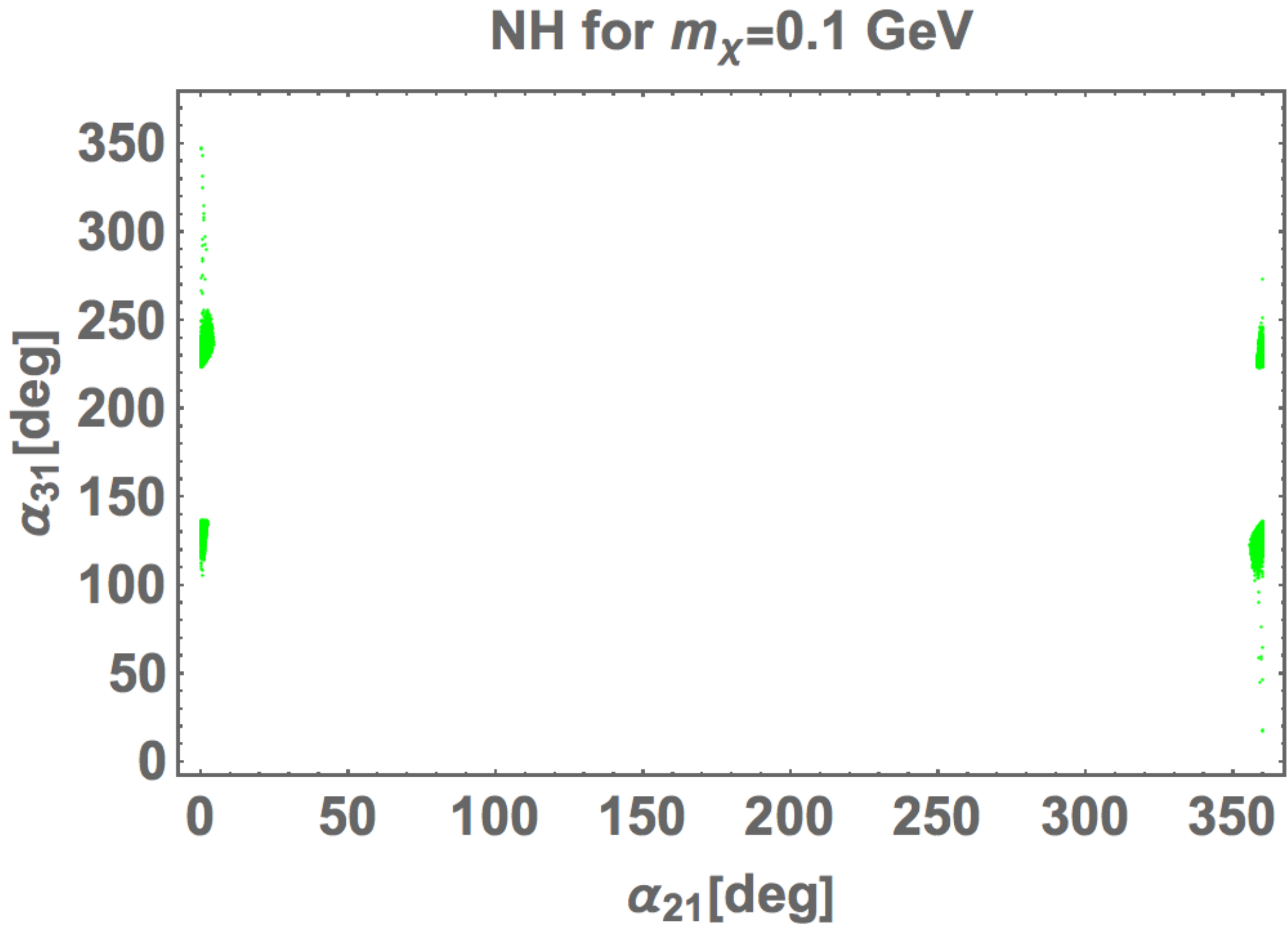}
 \end{center}
 \end{minipage}
 \\
 \begin{minipage}{0.32\hsize}
 \begin{center}
  \includegraphics[width=49mm]{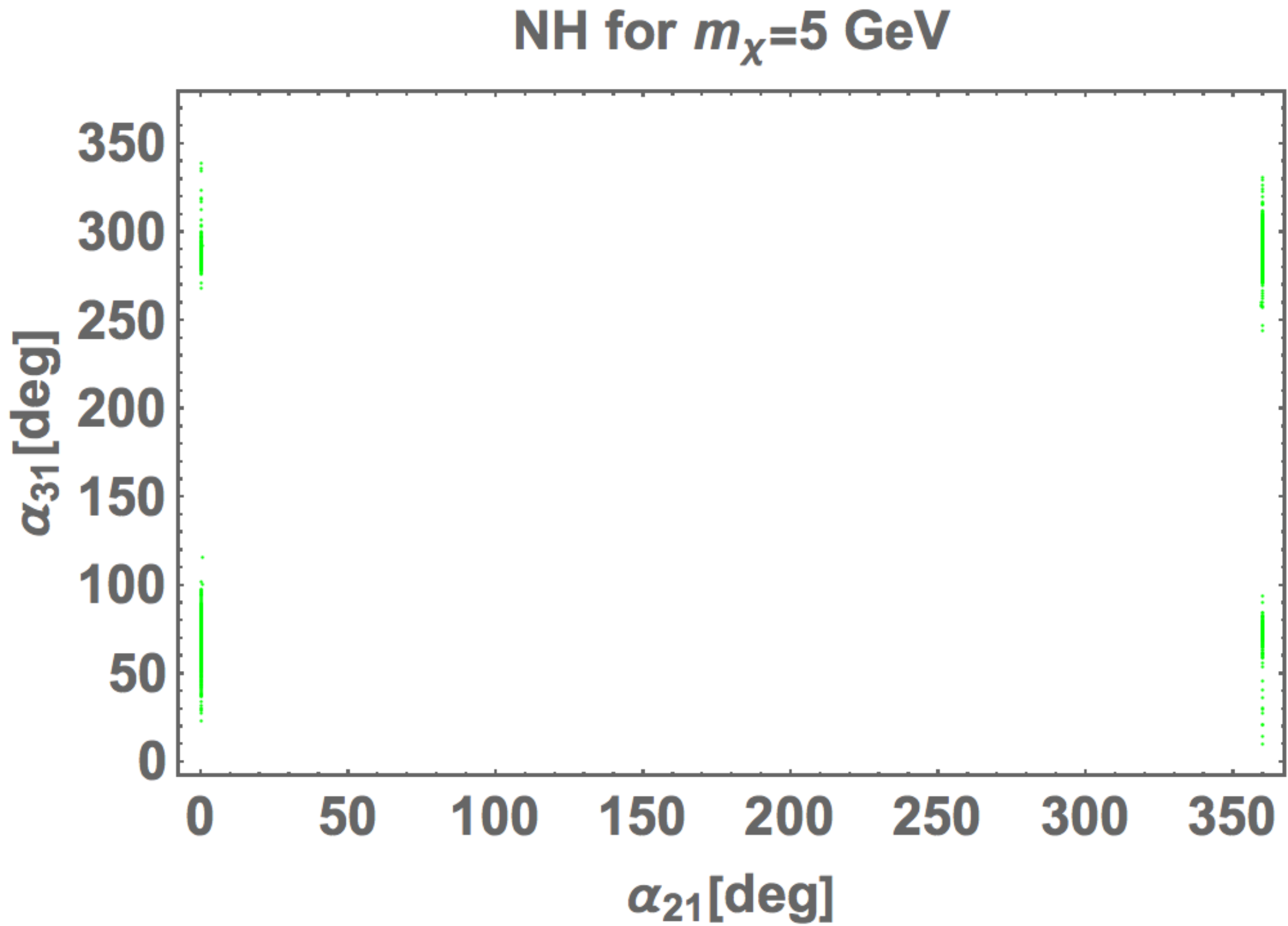}
 \end{center}
 \end{minipage}
   \begin{minipage}{0.32\hsize}
 \begin{center}
  \includegraphics[width=49mm]{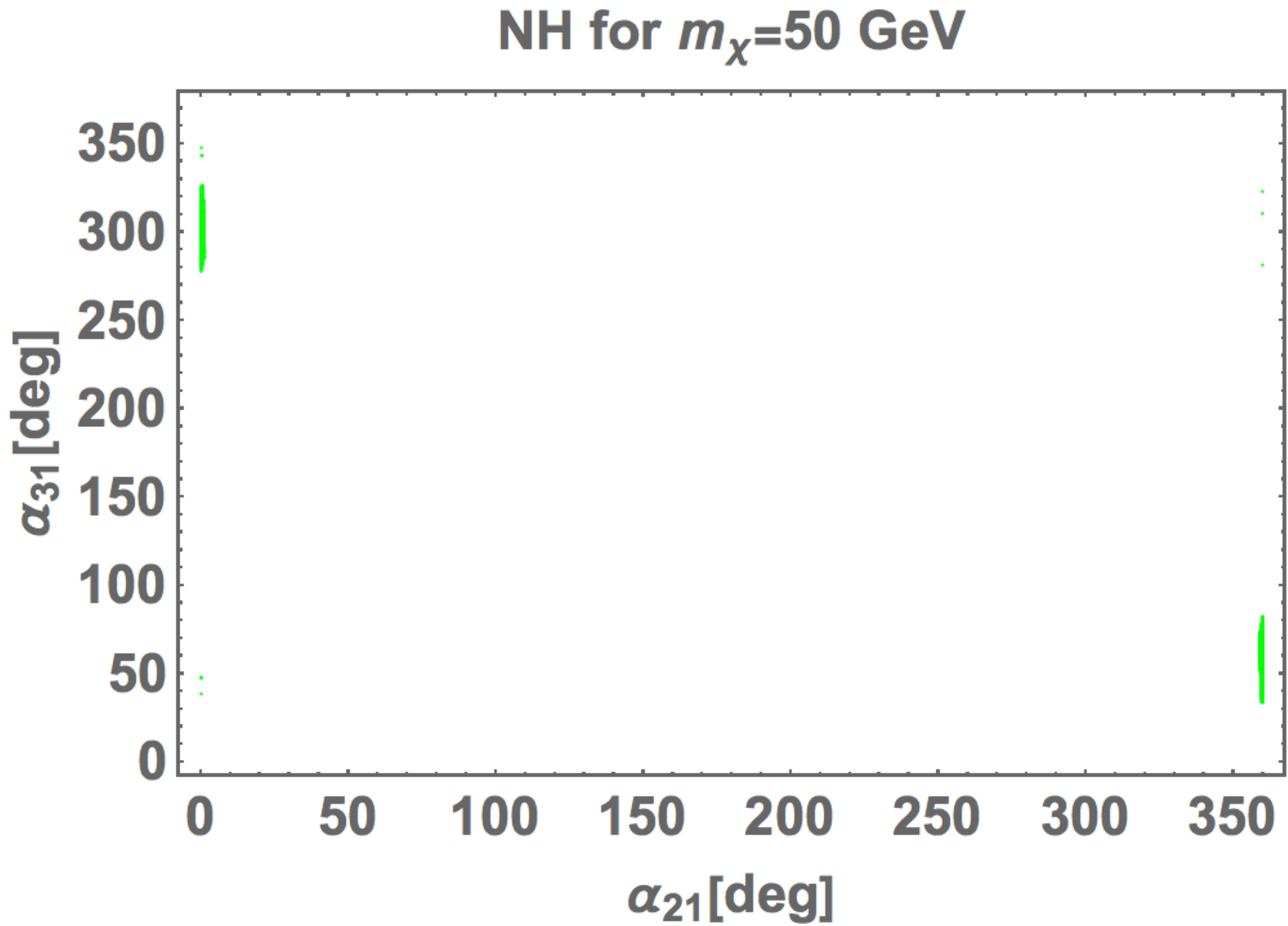}
 \end{center}
 \end{minipage}
 \caption{Scatter plots of $\alpha_{21}$ [deg] and $\alpha_{31}$ [deg], where the legends are the same as Fig.~\ref{fig:tau_nh}. }
   \label{fig:majo_nh}
\end{figure}
In Fig.~\ref{fig:majo_nh}, scatter plots of $\alpha_{21}$ [deg] and $\alpha_{31}$ [deg]  are shown, where the legends are the same as Fig.~\ref{fig:tau_nh}. 
Allowed regions are as follows:
$\alpha_{21}\sim0$  [deg] and $\alpha_{31}=[50-120, 240-300]$  [deg] for  $m_\chi=0.01$ GeV,
$\alpha_{21}\sim0$  [deg] and $\alpha_{31}=[110-130, 220-360]$  [deg] for  $m_\chi=0.1$ GeV,
$\alpha_{21}\sim0$  [deg] and $\alpha_{31}=[30-100, 250,340]$  [deg] for  $m_\chi=5$ GeV, and
$\alpha_{21}\sim0$  [deg] and $\alpha_{31}=[30-80, 280,350]$  [deg] for  $m_\chi=50$ GeV.

\begin{figure}[htbp]
 \begin{minipage}{0.32\hsize}
  \begin{center}
\includegraphics[width=49mm]{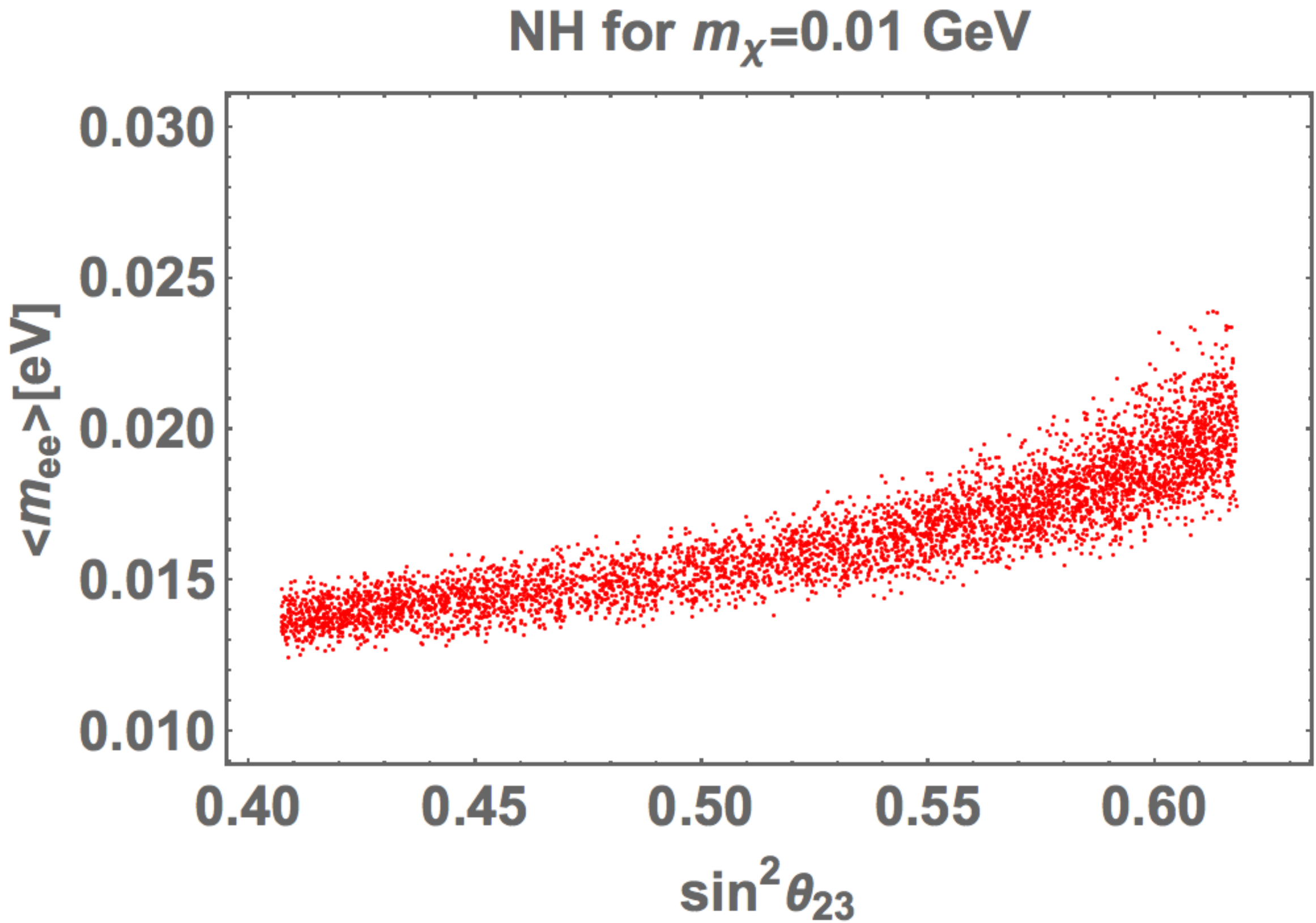}
  \end{center}
 \end{minipage}
 \begin{minipage}{0.32\hsize}
 \begin{center}
  \includegraphics[width=49mm]{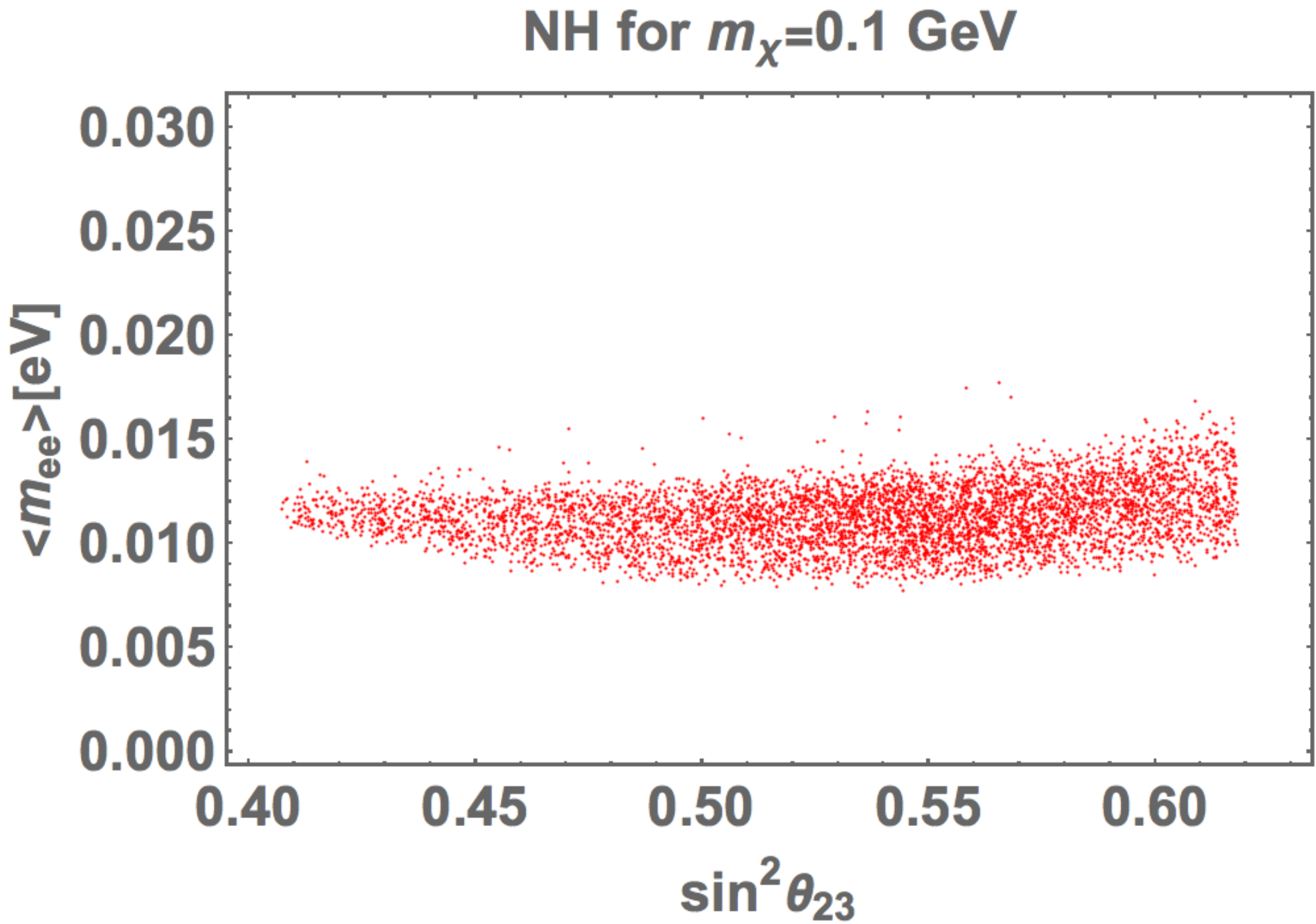}
 \end{center}
 \end{minipage}
 \\
 \begin{minipage}{0.32\hsize}
 \begin{center}
  \includegraphics[width=49mm]{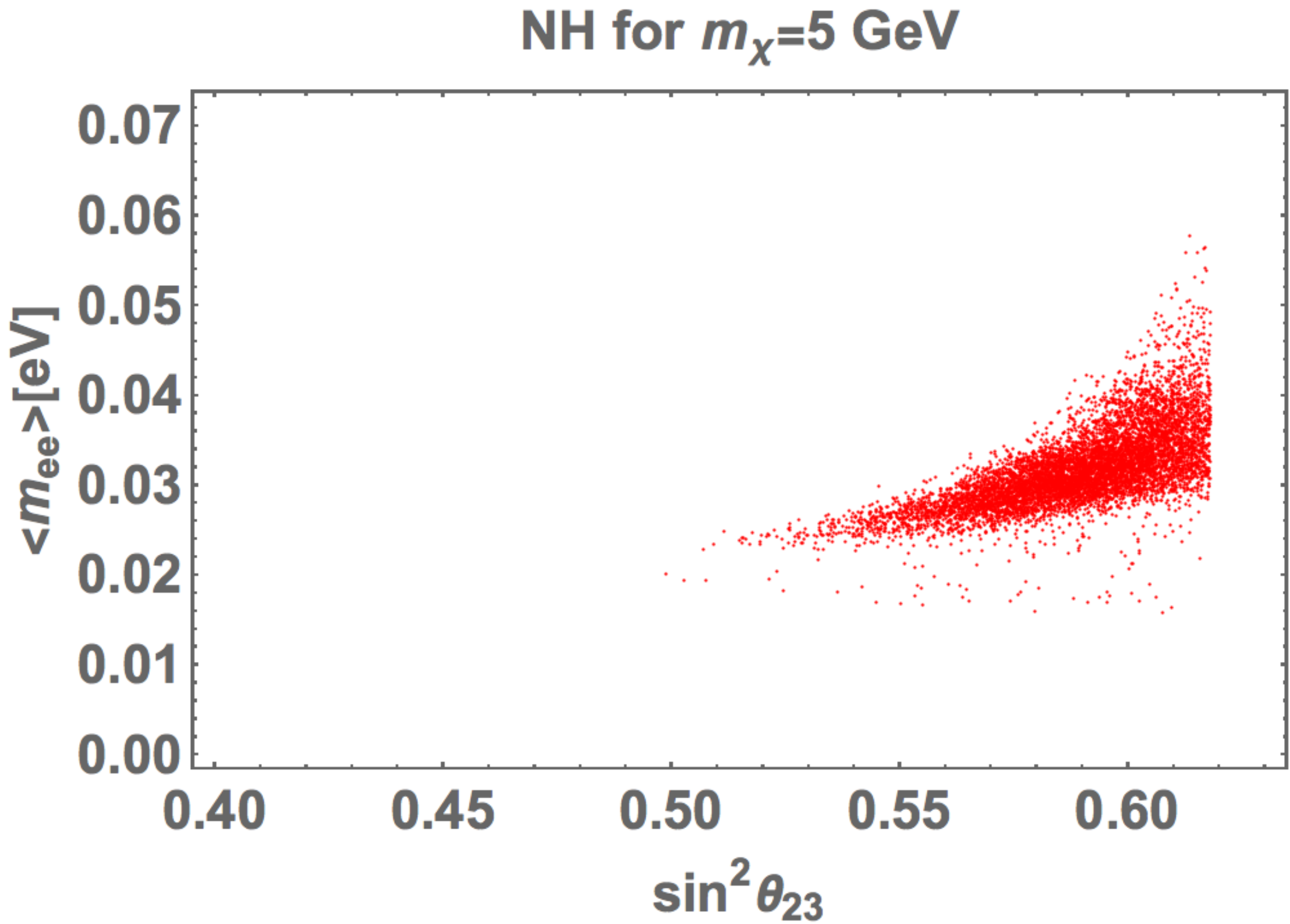}
 \end{center}
 \end{minipage}
   \begin{minipage}{0.32\hsize}
 \begin{center}
  \includegraphics[width=49mm]{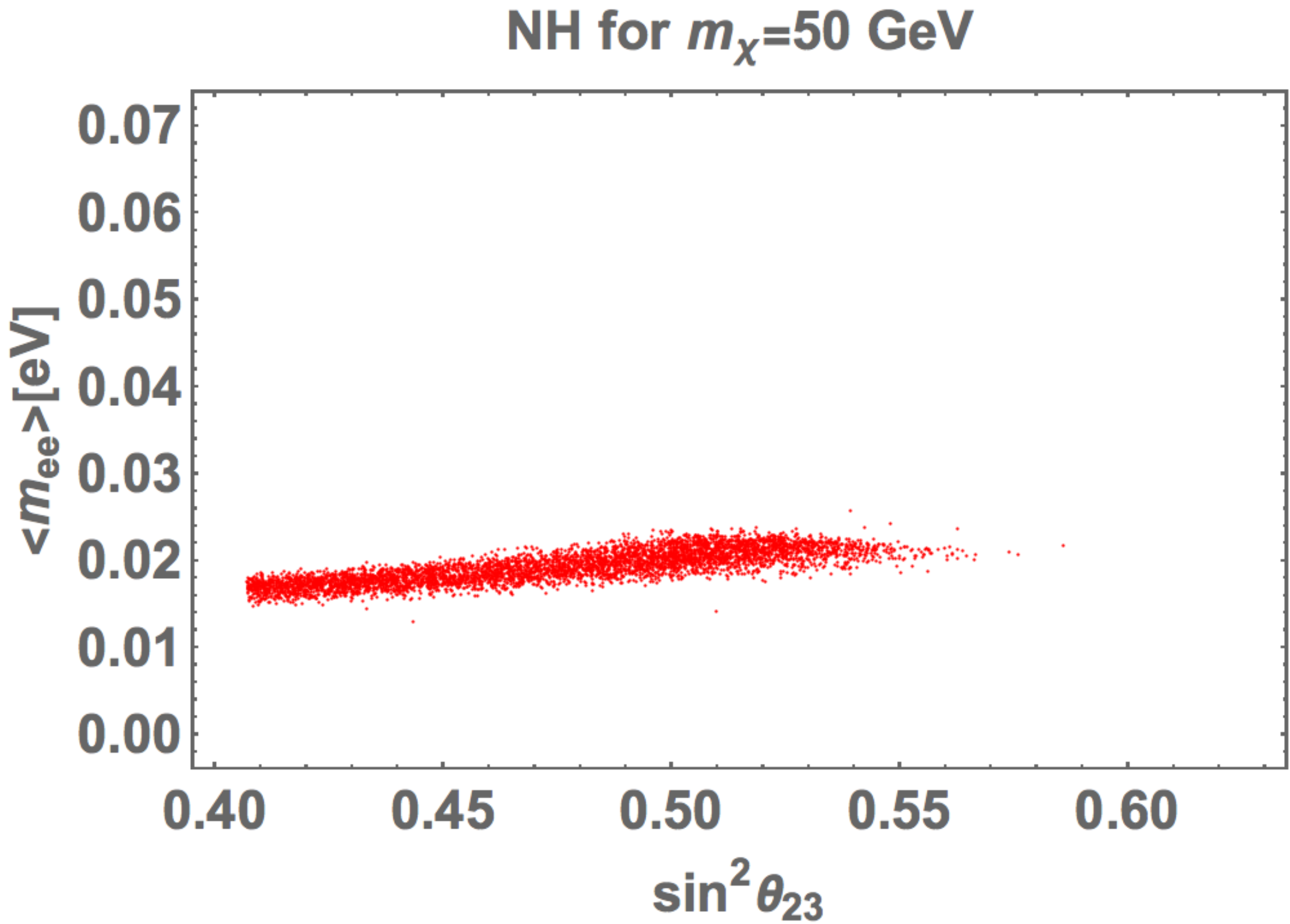}
 \end{center}
 \end{minipage}
 \caption{Scatter plots of $\sin^2\theta_{23}$ and  $\langle m_{ee} \rangle$, where the legends are the same as Fig.~\ref{fig:tau_nh}.}
   \label{fig:s23-mee_nh}
\end{figure}
In Fig.~\ref{fig:s23-mee_nh}, scatter plots of $\sin^2\theta_{23}$ and  $\langle m_{ee} \rangle$ are shown, where the legends are the same as Fig.~\ref{fig:tau_nh}.
Allowed regions are as follows:
$\langle m_{ee} \rangle = [0.013-0.024]$ eV for  $m_\chi=0.01$ GeV,
$\langle m_{ee} \rangle=[0.008-0.015]$ eV for  $m_\chi=0.1$ GeV,
$\langle m_{ee} \rangle=[0.015-0.058]$ eV for  $m_\chi=5$ GeV, and
$\langle m_{ee} \rangle=[0.016-0.021]$ eV for  $m_\chi=50$ GeV.

Here, we summarize some features in the case of NH.
In case of $m_\chi=5$ GeV, half of allowed region might be ruled out by the cosmological constraint $\sum m\le$ 0.12 eV and larger value of $\sin^2\theta_{23}$ is favored. While large value of $\sin^2\theta_{23}$ is disfavored in case of $m_\chi=50$ GeV.
For all the cases, $\alpha_{21}$ is almost zero, and $\delta_{CP}$ is localized at nearby 50 deg and 300 deg.

{\it Inverted hierarchy case}

\begin{figure}[htbp]
 \begin{minipage}{0.32\hsize}
  \begin{center}
\includegraphics[width=49mm]{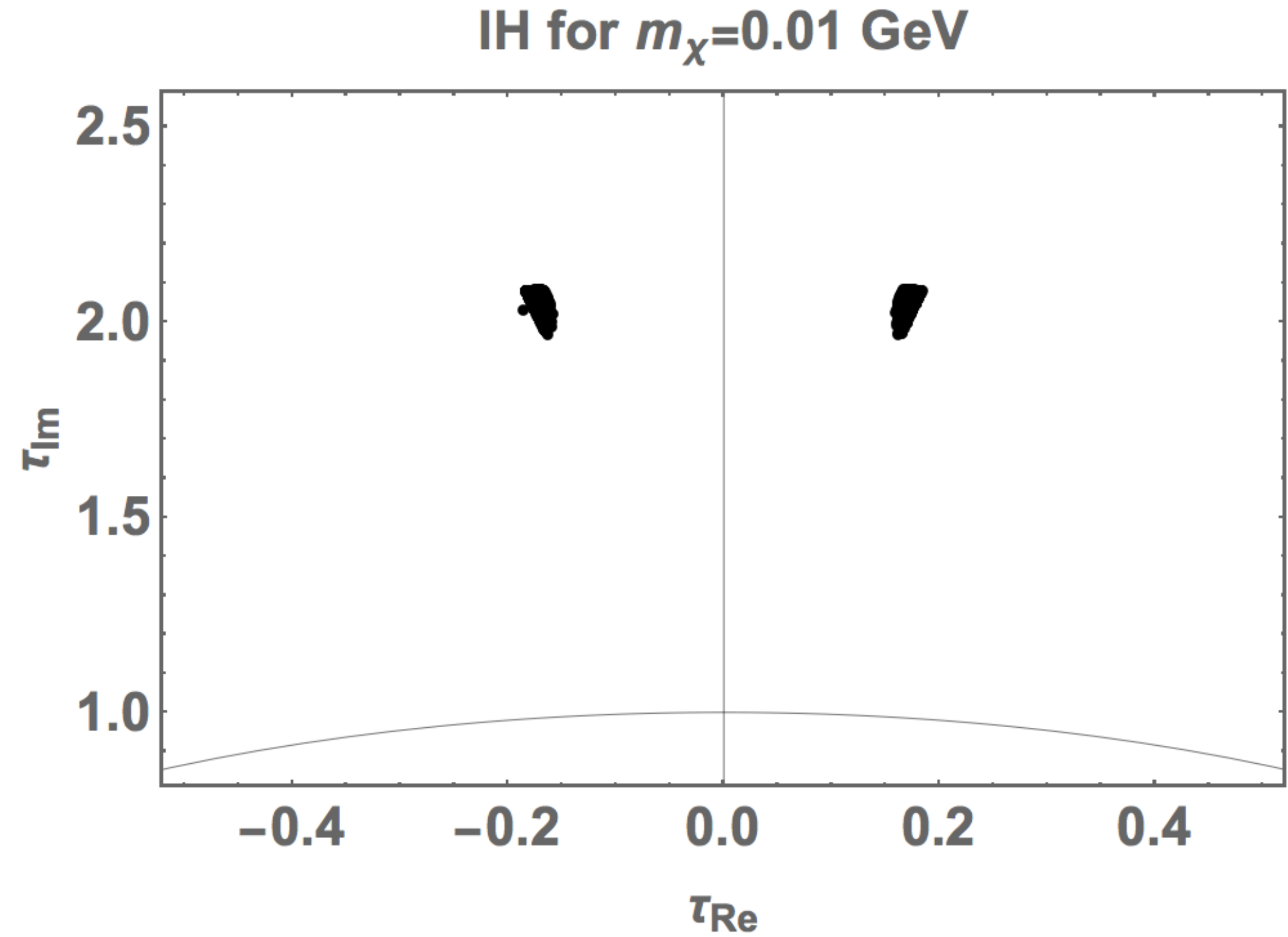}
  \end{center}
 \end{minipage}
 \begin{minipage}{0.32\hsize}
 \begin{center}
  \includegraphics[width=49mm]{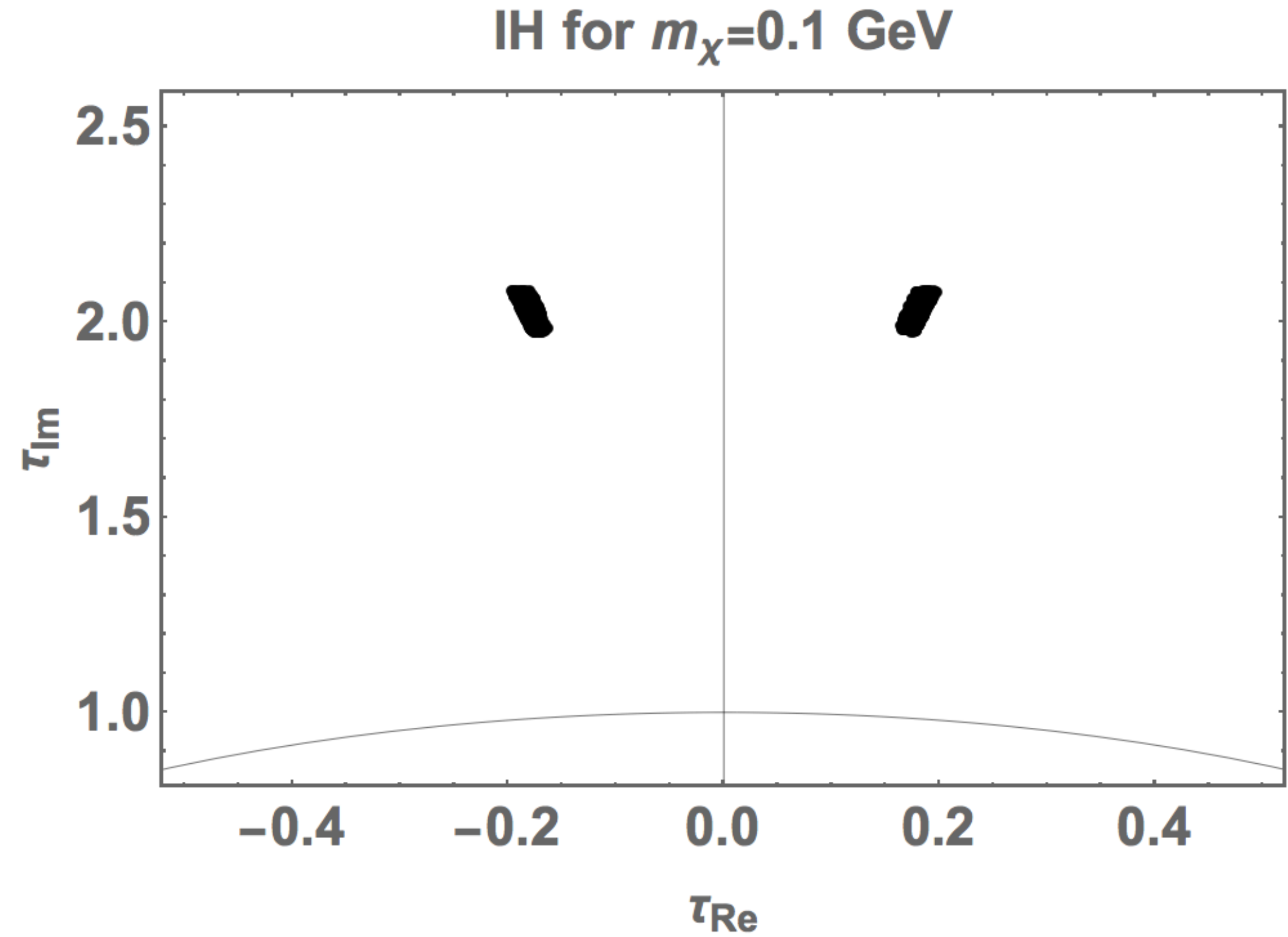}
 \end{center}
 \end{minipage}
 \\
 \begin{minipage}{0.32\hsize}
 \begin{center}
  \includegraphics[width=49mm]{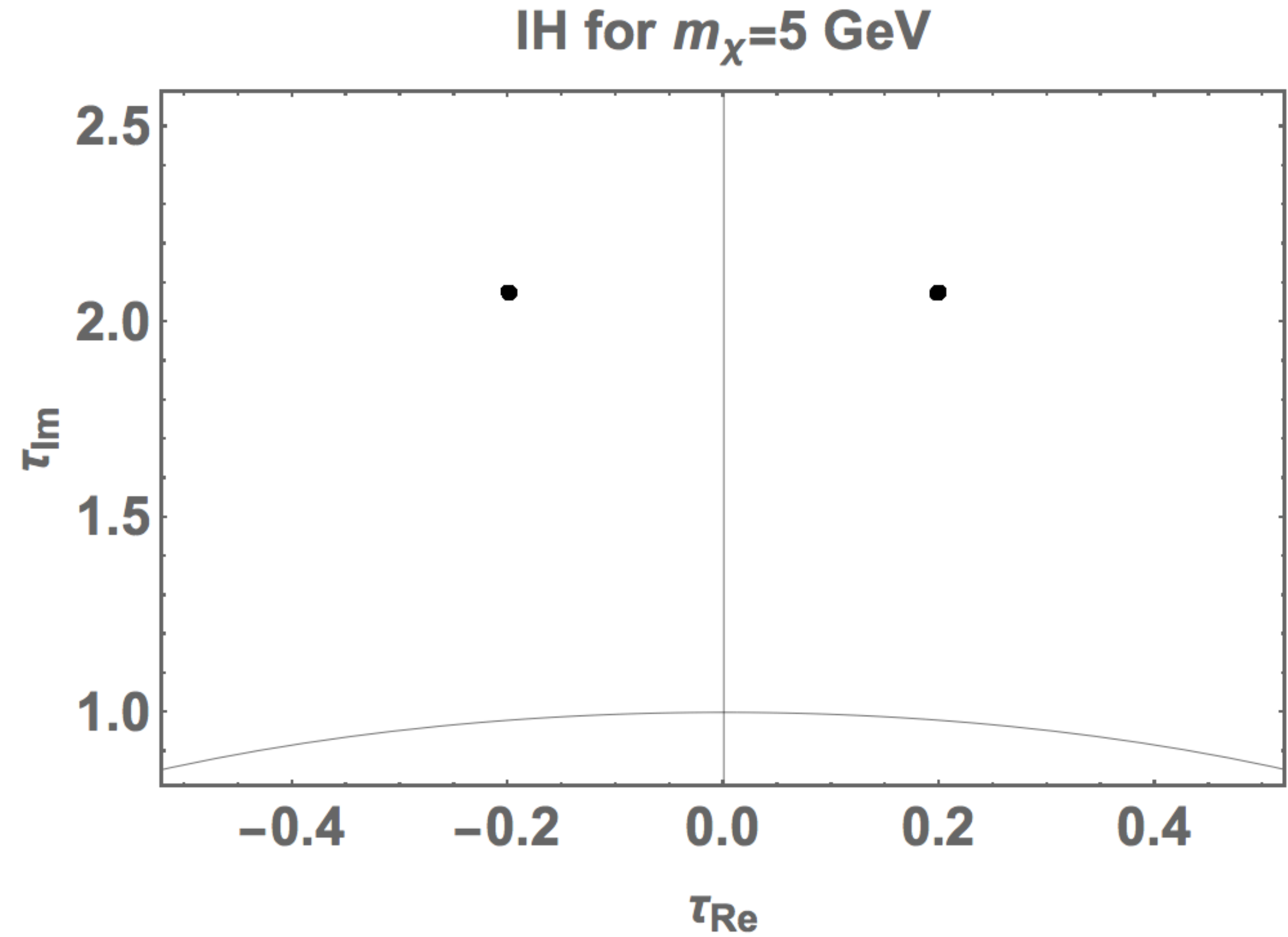}
 \end{center}
 \end{minipage}
   \begin{minipage}{0.32\hsize}
 \begin{center}
  \includegraphics[width=49mm]{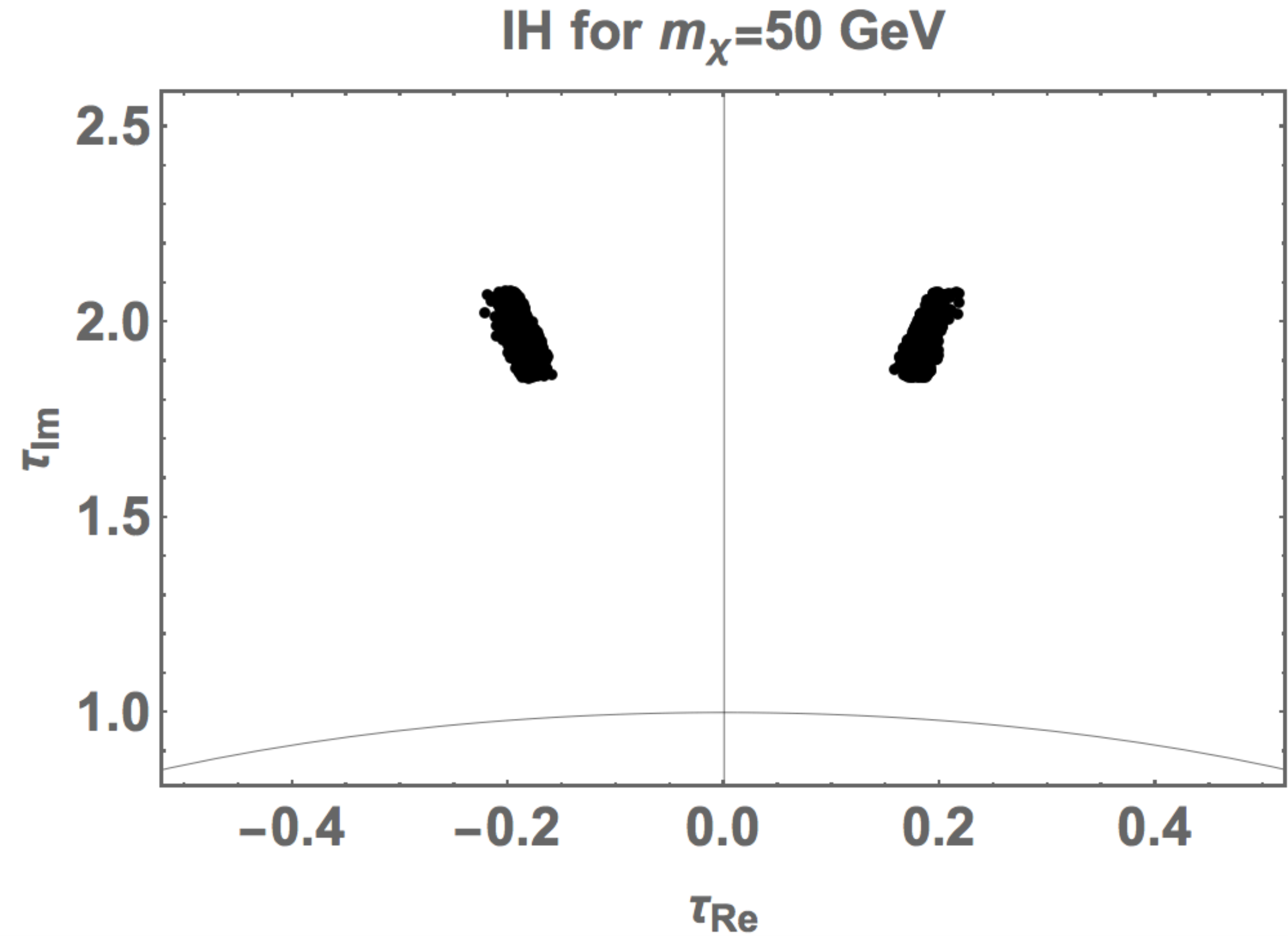}
 \end{center}
 \end{minipage}
 \caption{Scatter plots of Re$[\tau]\equiv \tau_{\rm Re}$ and  Im$[\tau]\equiv \tau_{\rm Im}$, where the legends are the same as Fig.~\ref{fig:tau_nh}.}
   \label{fig:tau_ih}
\end{figure}
In Fig.~\ref{fig:tau_ih}, scatter plots of Re$[\tau]\equiv \tau_{\rm Re}$ and  Im$[\tau]\equiv \tau_{\rm Im}$ are shown, where the legends are the same as Fig.~\ref{fig:tau_nh}.
Each of allowed regions is at nearby$|\tau_{\rm Re}=[0.19-0.21]$ and $\tau_{\rm Im}=[1.9,2.1]$ for $m_\chi=0.01$ GeV,
$|\tau_{\rm Re}=[0.19-0.21]$ and $\tau_{\rm Im}=[1.9,2.1]$ for  $m_\chi=0.1$ GeV,
$|\tau_{\rm Re}\sim0.2$ and $\tau_{\rm Im}=sim$ for  $m_\chi=5$ GeV, and
$|\tau_{\rm Re}|=[0.15-0.23]$ and $\tau_{\rm Im}=[1.85-2.1]$ for  $m_\chi=50$ GeV.

\begin{figure}[htbp]
 \begin{minipage}{0.32\hsize}
  \begin{center}
\includegraphics[width=49mm]{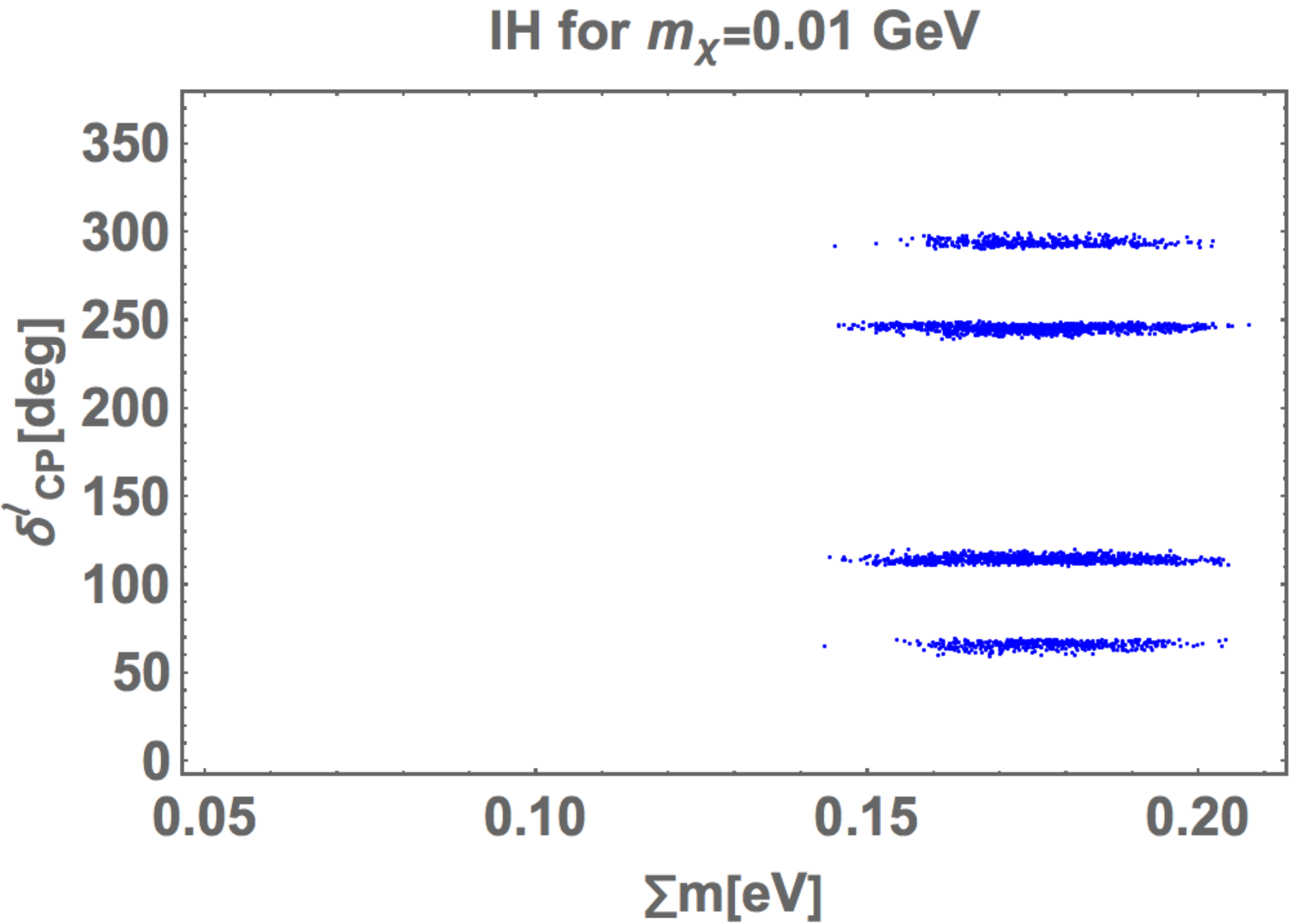}
  \end{center}
 \end{minipage}
 \begin{minipage}{0.32\hsize}
 \begin{center}
  \includegraphics[width=49mm]{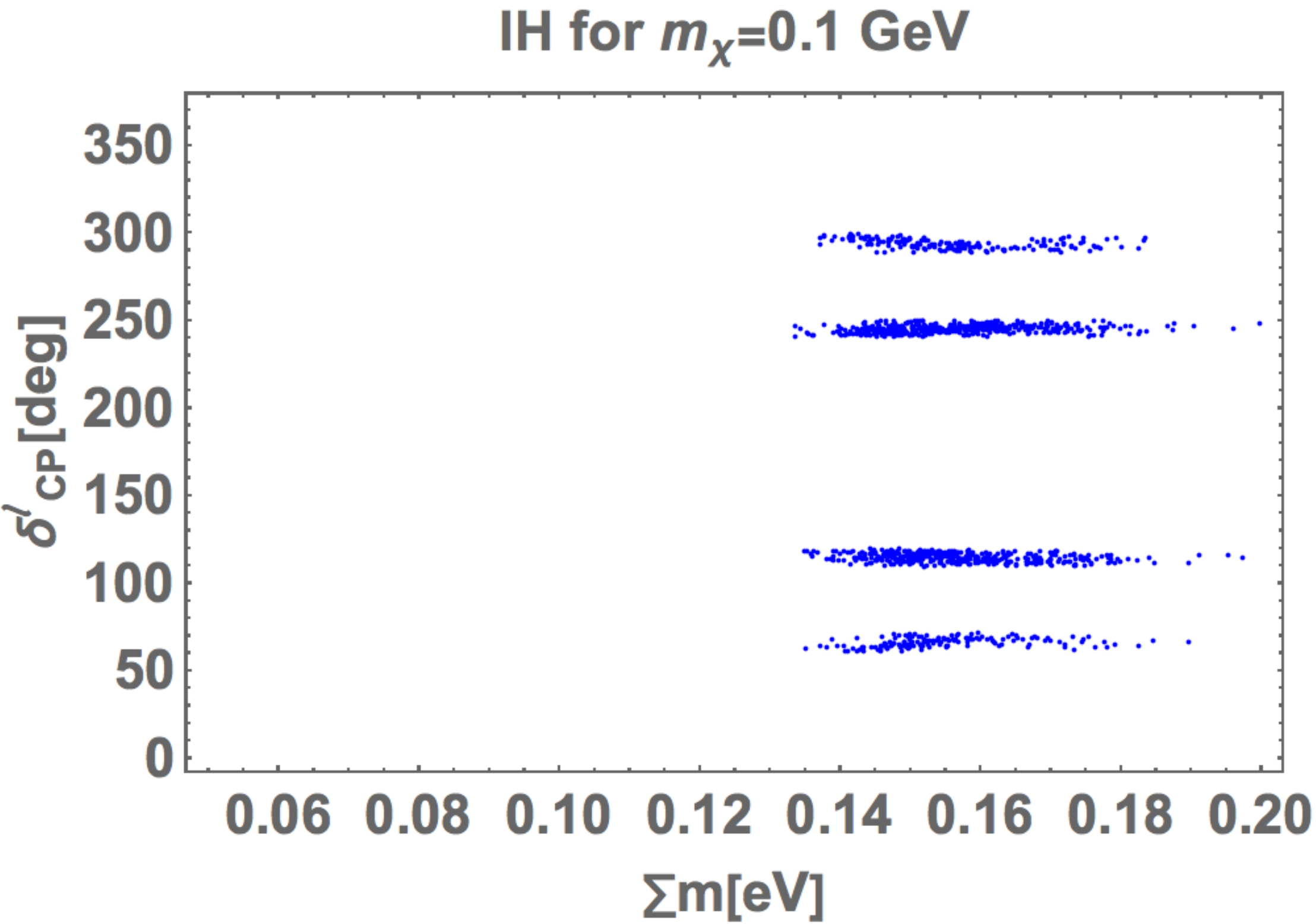}
 \end{center}
 \end{minipage}
 \\
 \begin{minipage}{0.32\hsize}
 \begin{center}
  \includegraphics[width=49mm]{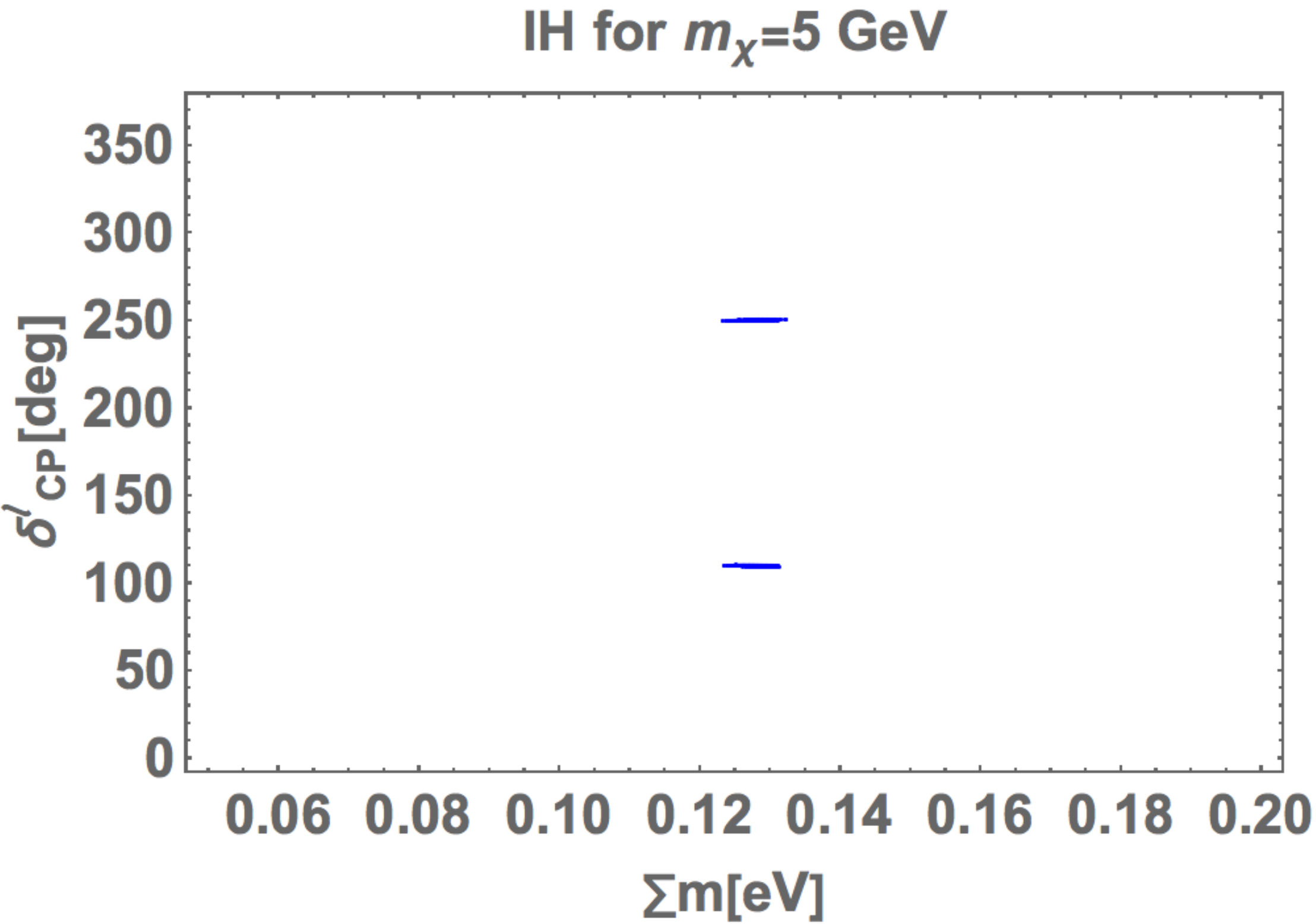}
 \end{center}
 \end{minipage}
   \begin{minipage}{0.32\hsize}
 \begin{center}
  \includegraphics[width=49mm]{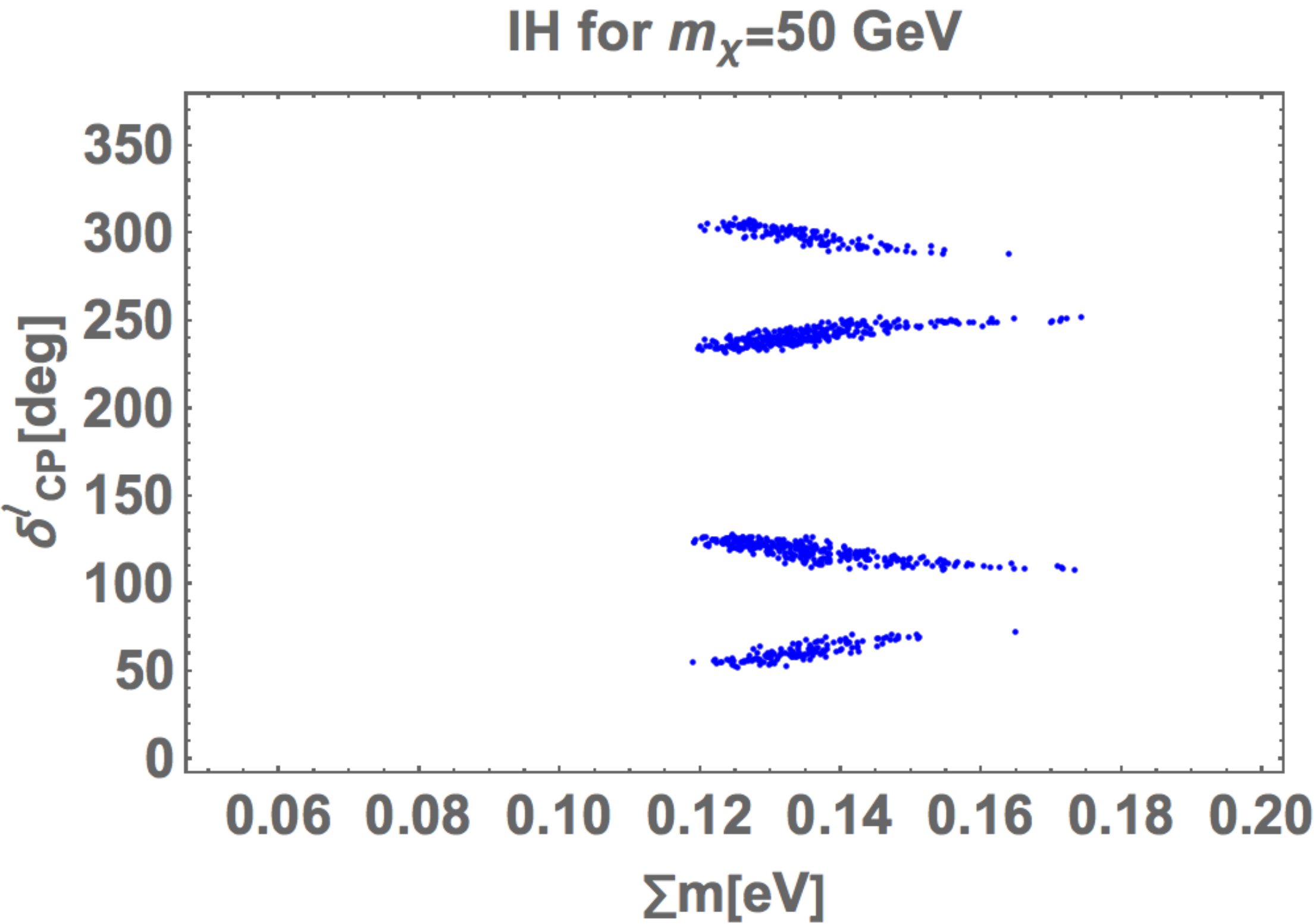}
 \end{center}
 \end{minipage}
 \caption{Scatter plots of sum of neutrino masses $\equiv\sum m$ eV and  $\delta^\ell_{\rm CP}$, where the legends are the same as Fig.~\ref{fig:tau_nh}.}
   \label{fig:sum-dcp_ih}
\end{figure}
In Fig.~\ref{fig:sum-dcp_ih}, scatter plots of sum of neutrino masses $\equiv\sum m$ eV and  $\delta^\ell_{\rm CP}$ are shown, where the legends are the same as Fig.~\ref{fig:tau_nh}.
Allowed regions are as follows:
$\sum m=[0.14-0.21]$ eV and $\delta^\ell_{\rm CP}\sim[70,110,250, 290]$ for  $m_\chi=0.01$ GeV,
$\sum m=[0.135-0.20]$ eV and $\delta^\ell_{\rm CP}\sim[70,120,250, 300]$ for  $m_\chi=0.1$ GeV,
$\sum m=[0.125 - 0.135]$ eV and $\delta^\ell_{\rm CP}=[110,250]$ for  $m_\chi=5$ GeV, and
$\sum m=[0.118-0.175]$ eV and $\delta^\ell_{\rm CP}=[50-70,100-130, 230-260, 290-310]$ for  $m_\chi=50$ GeV.

\begin{figure}[htbp]
 \begin{minipage}{0.32\hsize}
  \begin{center}
\includegraphics[width=49mm]{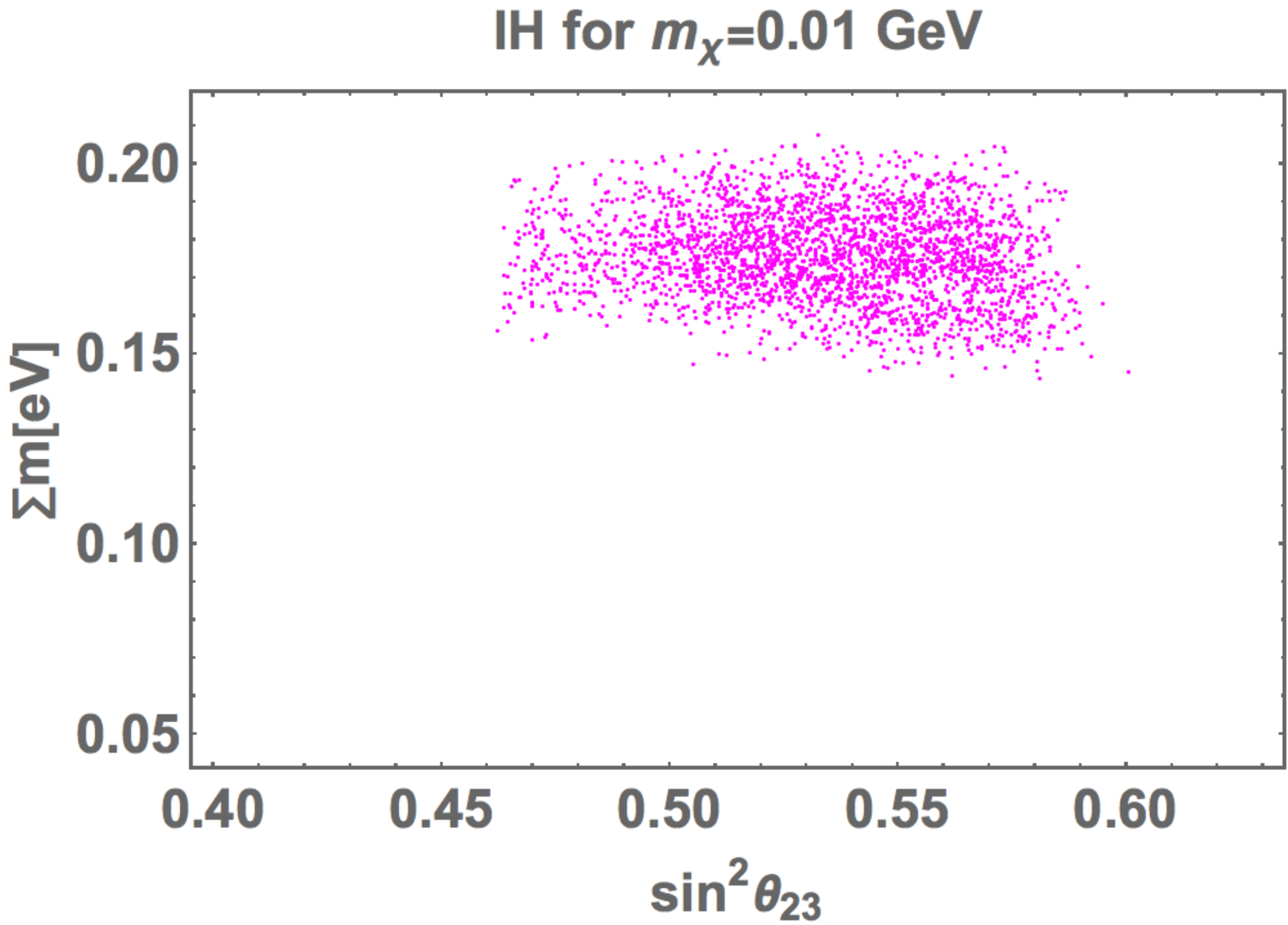}
  \end{center}
 \end{minipage}
 \begin{minipage}{0.32\hsize}
 \begin{center}
  \includegraphics[width=49mm]{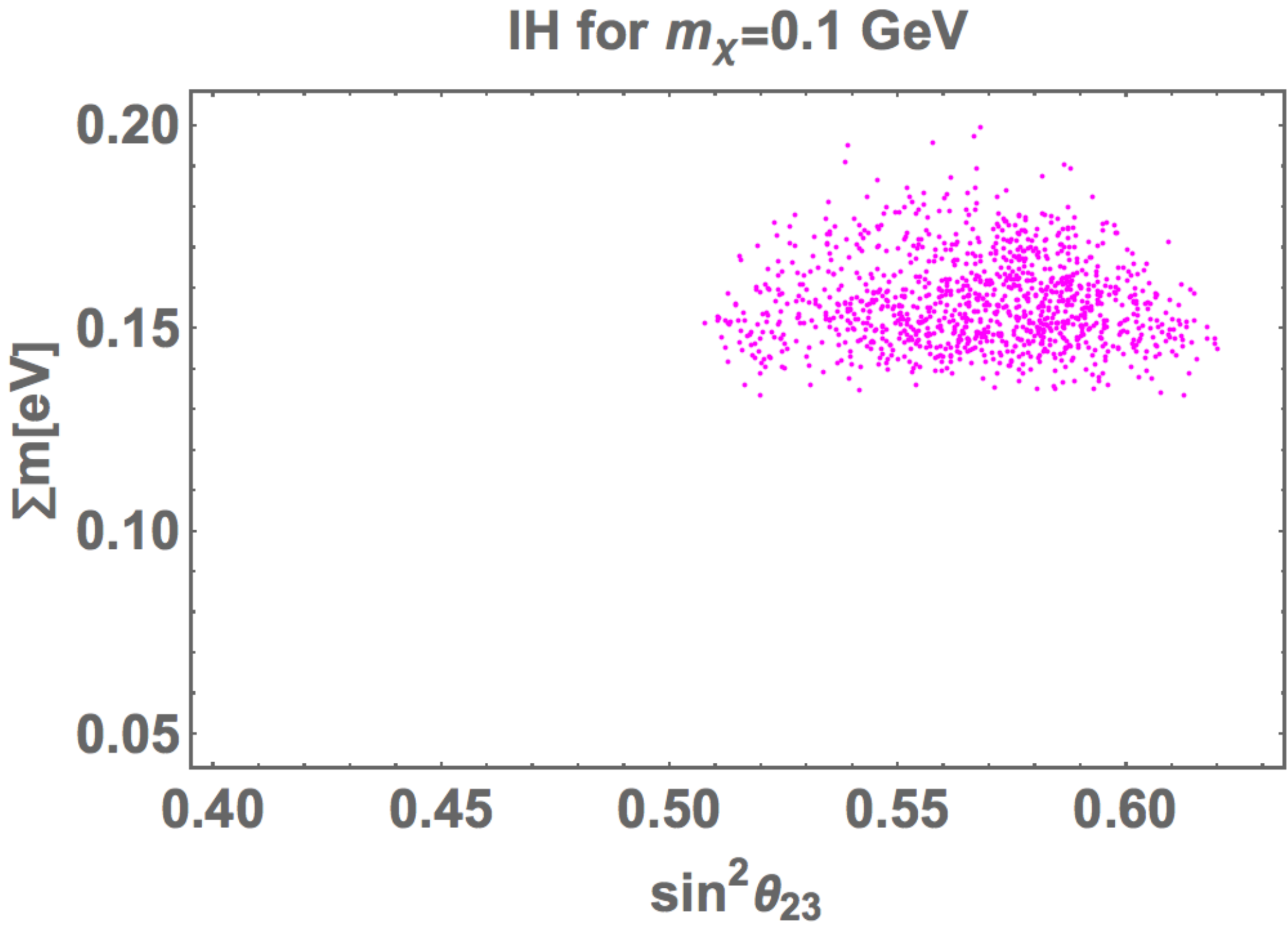}
 \end{center}
 \end{minipage}
 \\
 \begin{minipage}{0.32\hsize}
 \begin{center}
  \includegraphics[width=49mm]{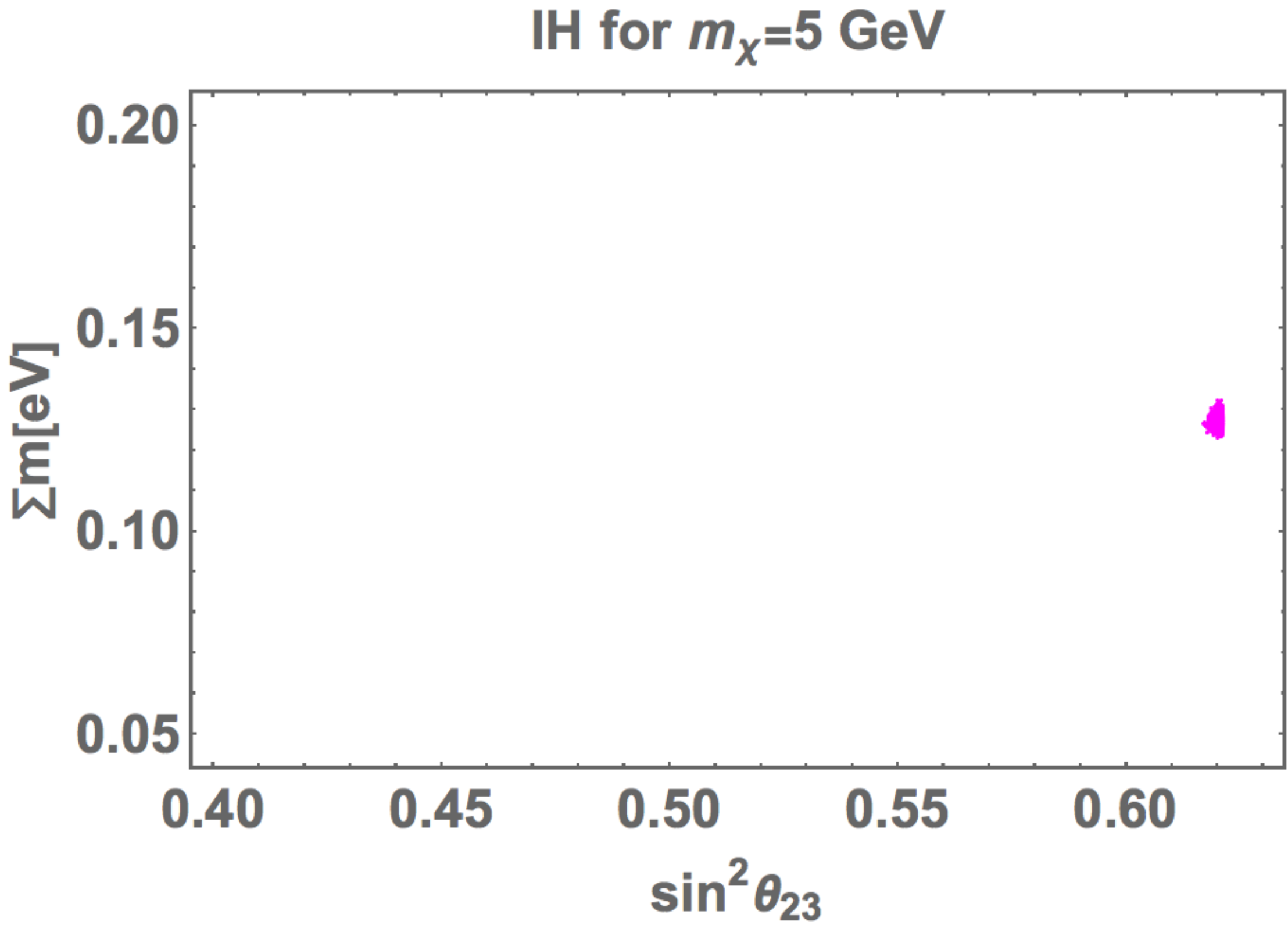}
 \end{center}
 \end{minipage}
   \begin{minipage}{0.32\hsize}
 \begin{center}
  \includegraphics[width=49mm]{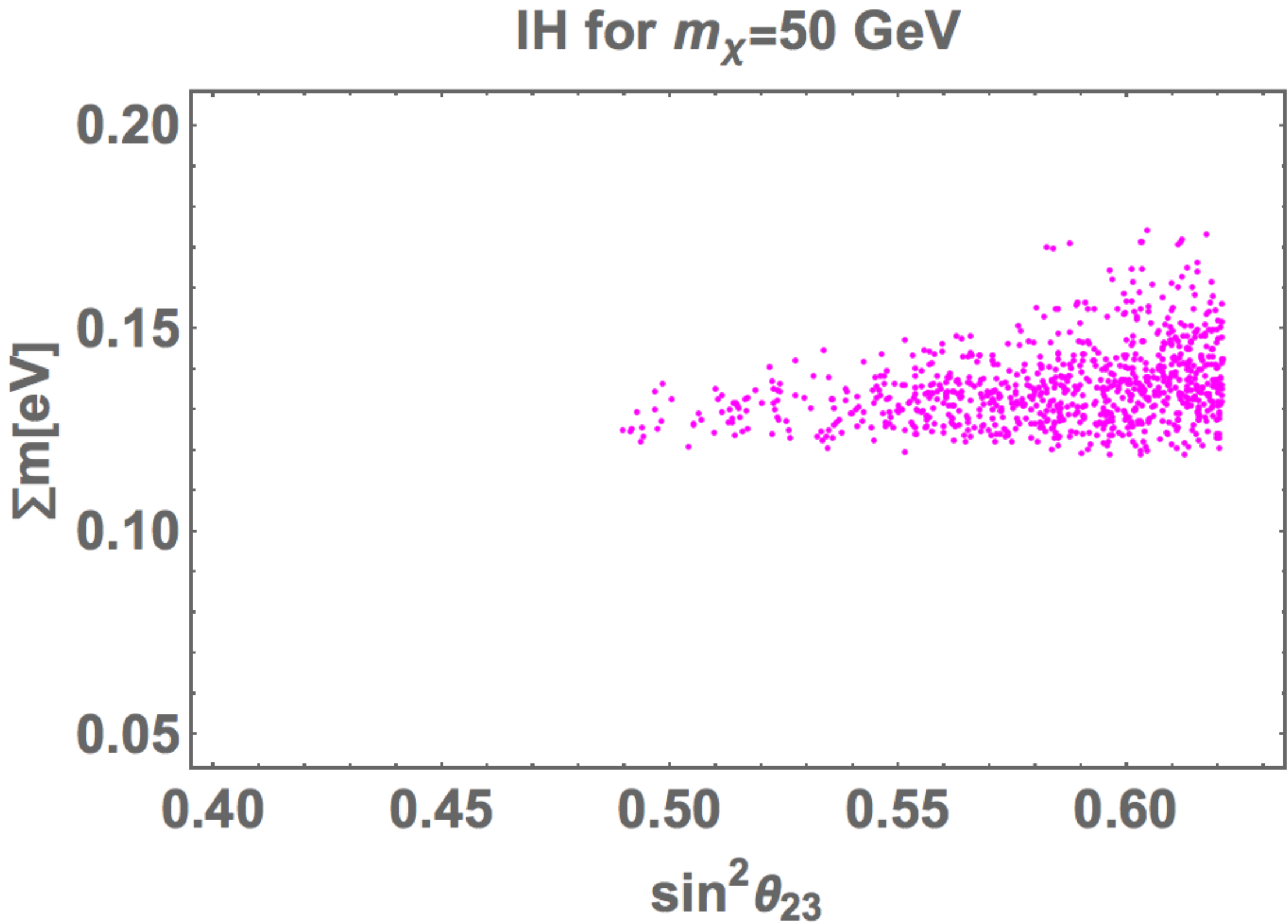}
 \end{center}
 \end{minipage}
 \caption{Scatter plots of $\sin^2\theta_{23}$ and  $\sum m$, where the legends are the same as Fig.~\ref{fig:tau_nh}.}
   \label{fig:s23-sum_ih}
\end{figure}
In Fig.~\ref{fig:s23-sum_ih}, scatter plots of $\sin^2\theta_{23}$ and  $\sum m$ are shown, where the legends are the same as Fig.~\ref{fig:tau_nh}.
Allowed regions are as follows:
$\sin^2\theta_{23}=[0.46-0.60]$  for  $m_\chi=0.01$ GeV,
$\sin^2\theta_{23}=[0.51-0.618]$  for  $m_\chi=0.1$ GeV,
$\sin^2\theta_{23}=[0.615-0.618]$  for  $m_\chi=5$ GeV, and
$\sin^2\theta_{23}=[0.48-0.618]$ for  $m_\chi=50$ GeV.

\begin{figure}[htbp]
 \begin{minipage}{0.32\hsize}
  \begin{center}
\includegraphics[width=49mm]{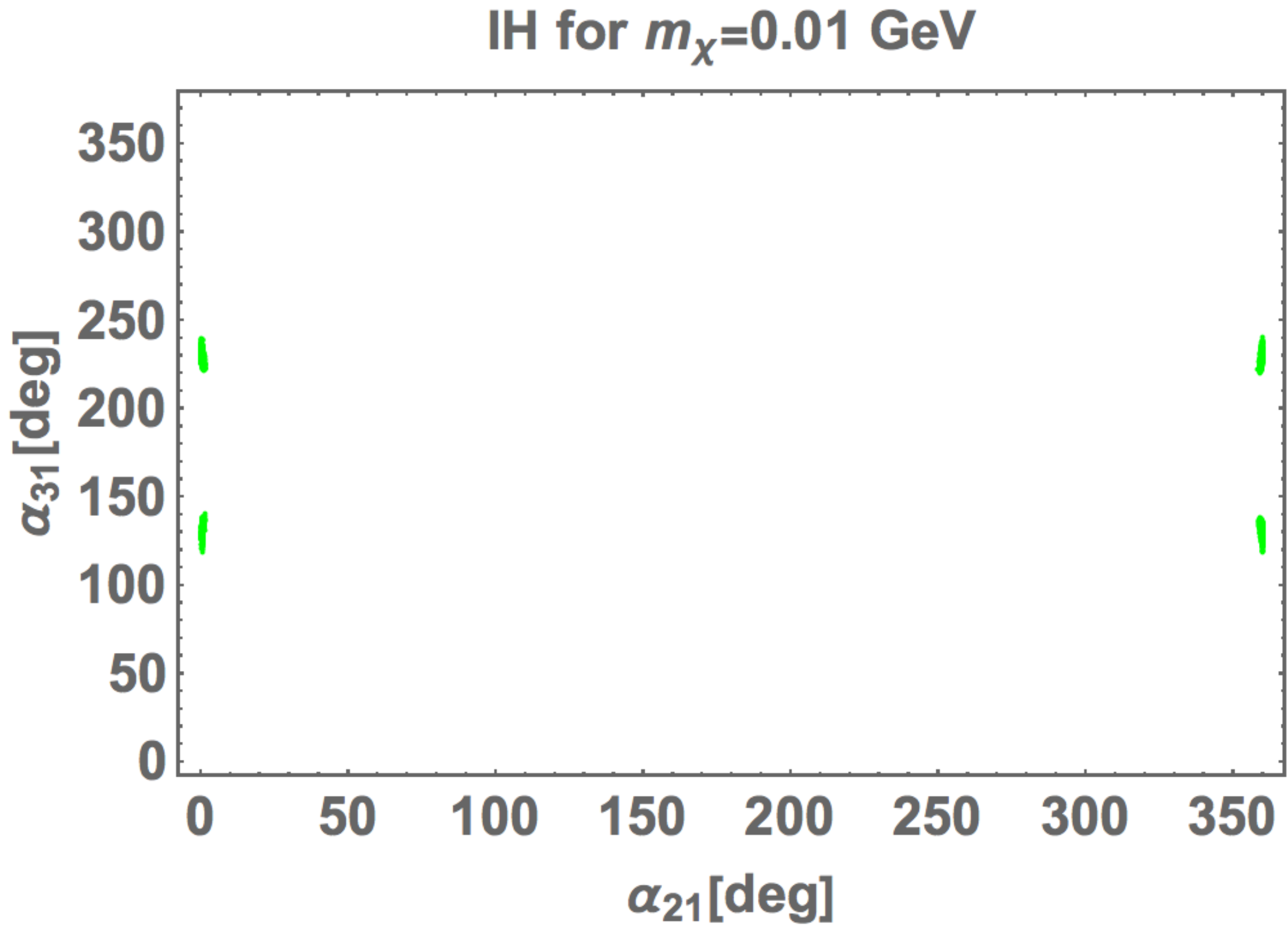}
  \end{center}
 \end{minipage}
 \begin{minipage}{0.32\hsize}
 \begin{center}
  \includegraphics[width=49mm]{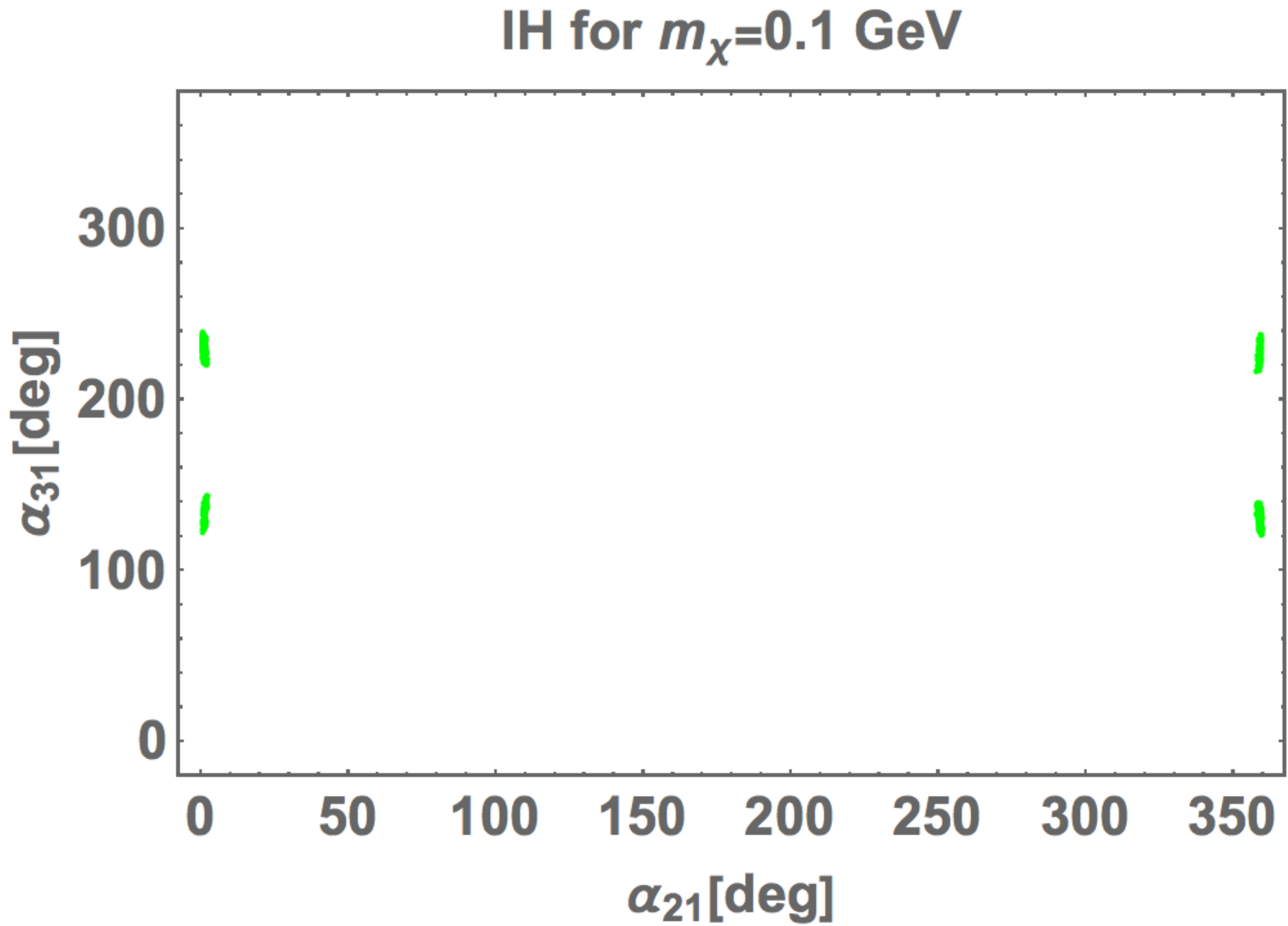}
 \end{center}
 \end{minipage}
 \\
 \begin{minipage}{0.32\hsize}
 \begin{center}
  \includegraphics[width=49mm]{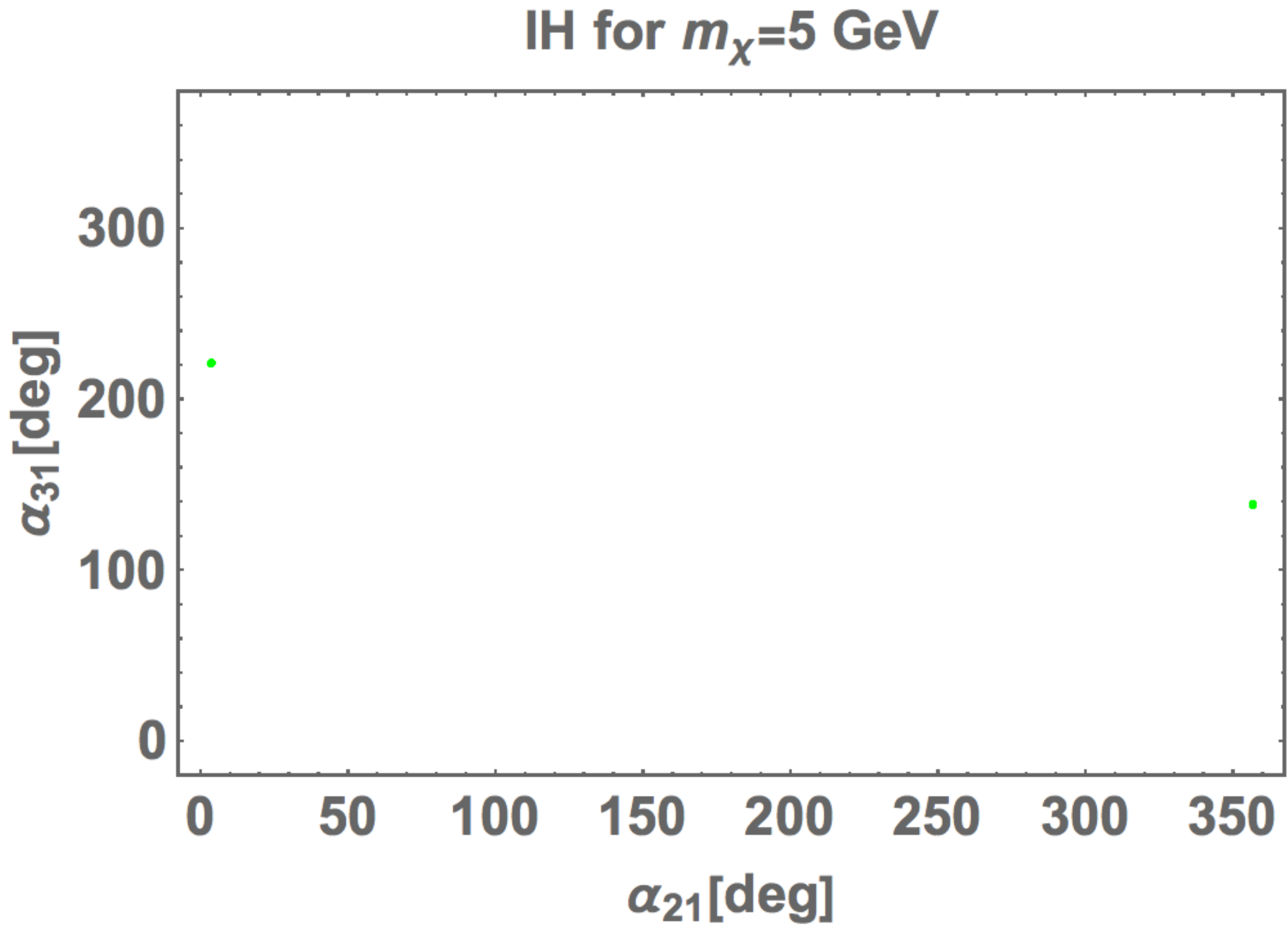}
 \end{center}
 \end{minipage}
   \begin{minipage}{0.32\hsize}
 \begin{center}
  \includegraphics[width=49mm]{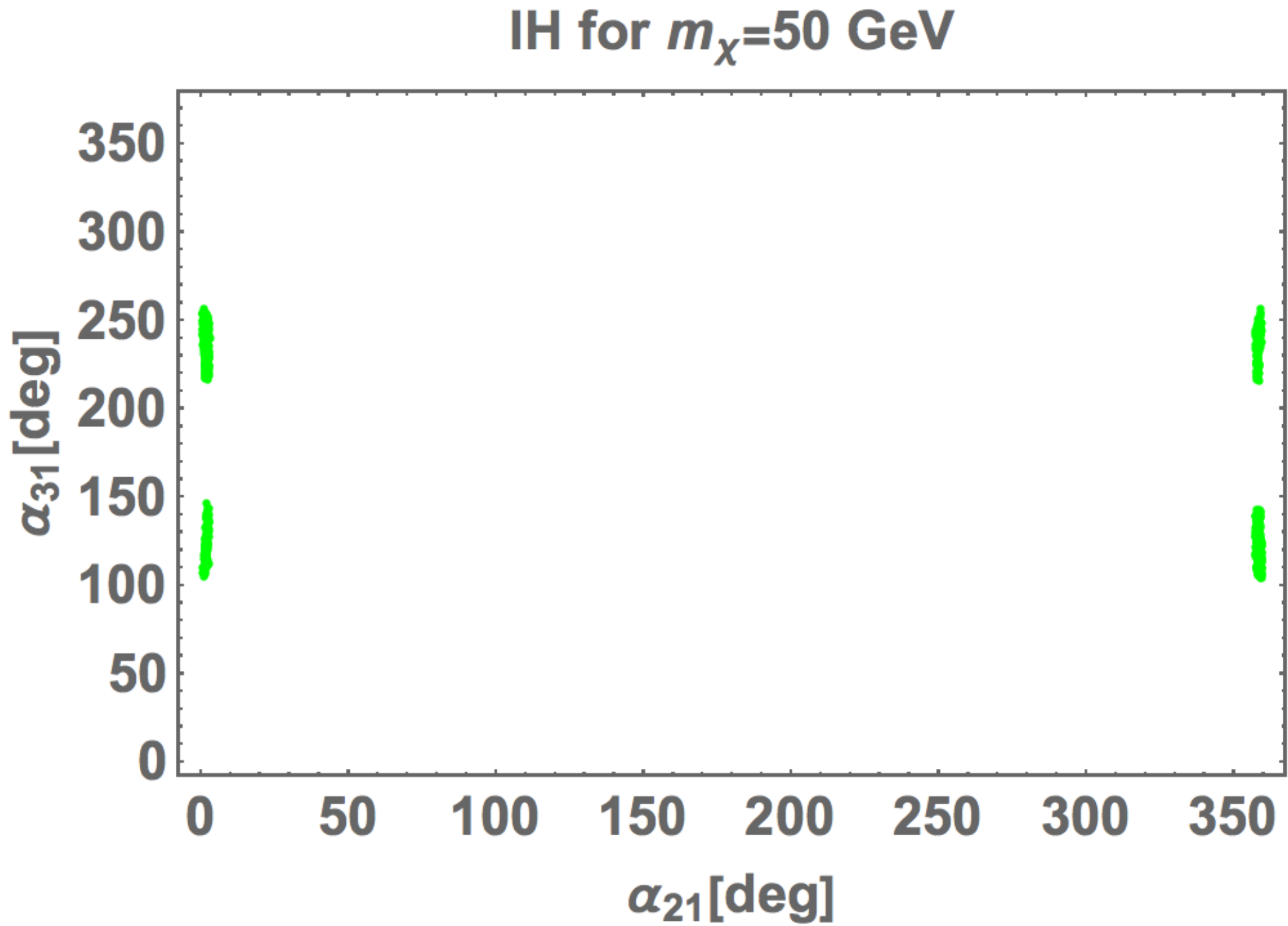}
 \end{center}
 \end{minipage}
 \caption{Scatter plots of $\alpha_{21}$ [deg] and $\alpha_{31}$ [deg], where the legends are the same as Fig.~\ref{fig:tau_nh}.}
   \label{fig:majo_ih}
\end{figure}
In Fig.~\ref{fig:majo_ih}, scatter plots of $\alpha_{21}$ [deg] and $\alpha_{31}$ [deg]  are shown, where the legends are the same as Fig.~\ref{fig:tau_nh}.
Allowed regions are as follows:
$\alpha_{21}\sim0$  [deg] and $\alpha_{31}=[120-140, 220-240]$  [deg] for  $m_\chi=0.01$ GeV,
$\alpha_{21}\sim0$  [deg] and $\alpha_{31}=[120-140, 220-240]$  [deg] for  $m_\chi=0.1$ GeV,
$\alpha_{21}\sim0$  [deg] and $\alpha_{31}=[140, 220]$  [deg] for  $m_\chi=5$ GeV, and
$\alpha_{21}\sim0$  [deg] and $\alpha_{31}=[100-140, 210-260]$  [deg] for  $m_\chi=50$ GeV.

\begin{figure}[htbp]
 \begin{minipage}{0.32\hsize}
  \begin{center}
\includegraphics[width=49mm]{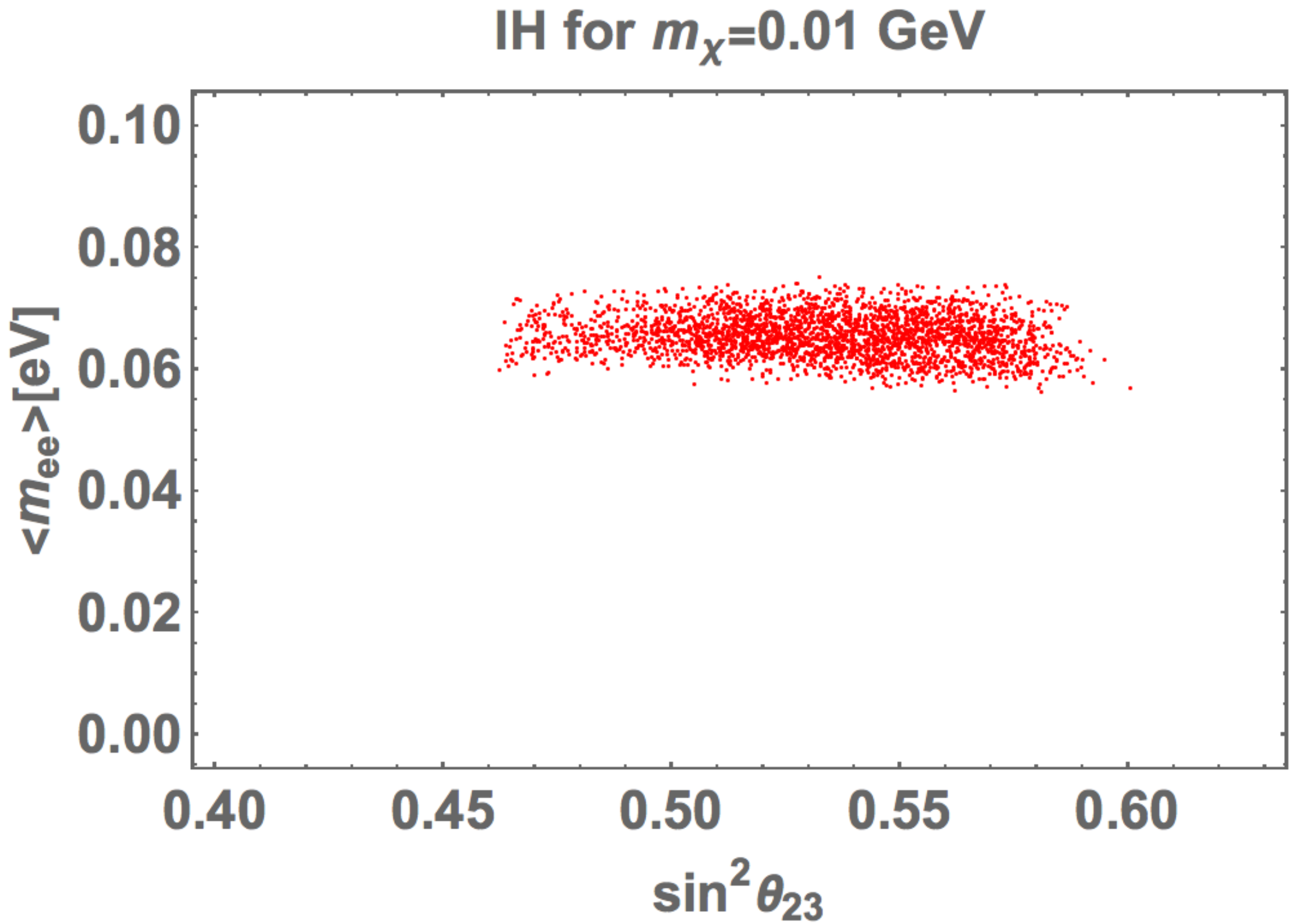}
  \end{center}
 \end{minipage}
 \begin{minipage}{0.32\hsize}
 \begin{center}
  \includegraphics[width=49mm]{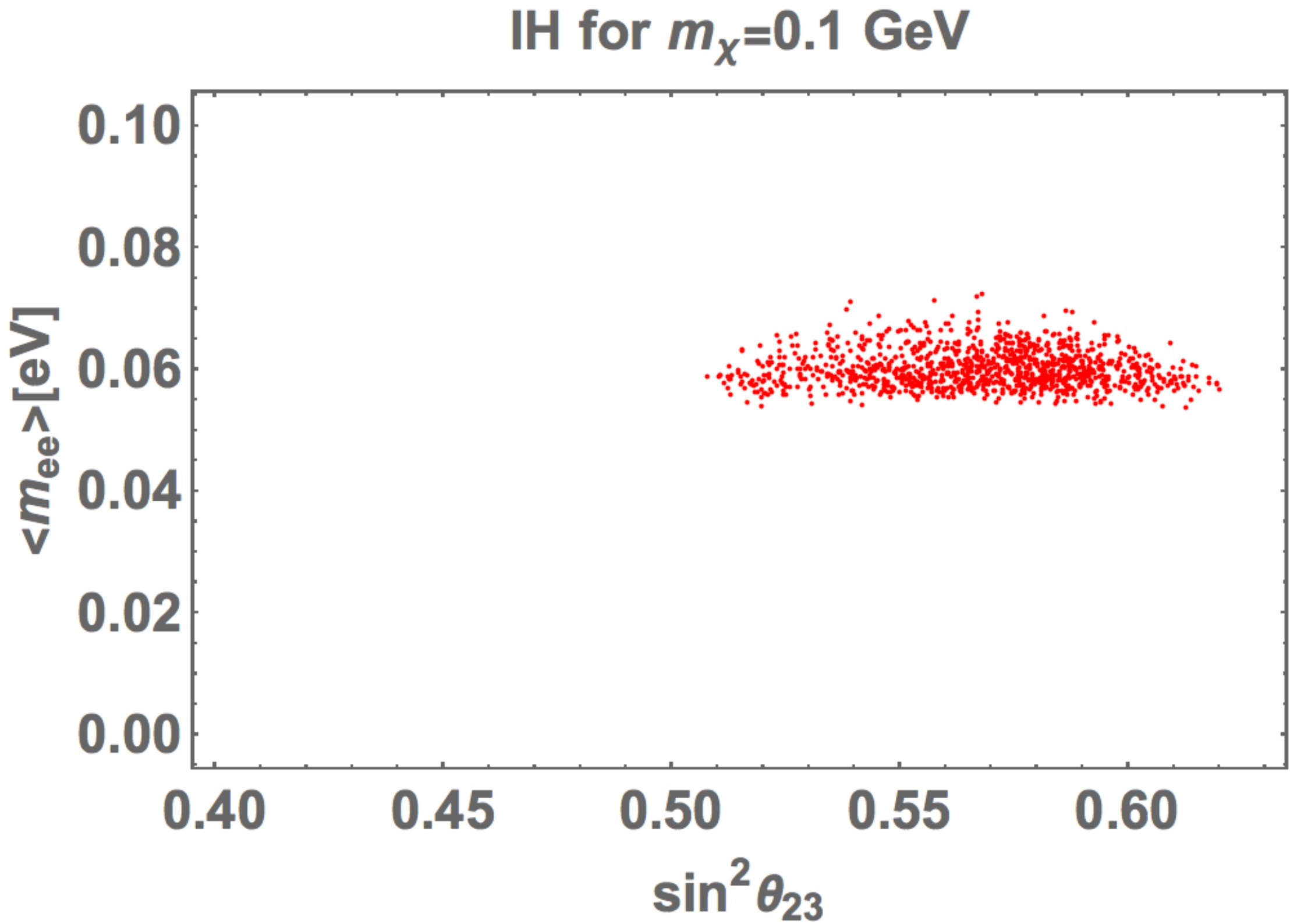}
 \end{center}
 \end{minipage}
 \\
 \begin{minipage}{0.32\hsize}
 \begin{center}
  \includegraphics[width=49mm]{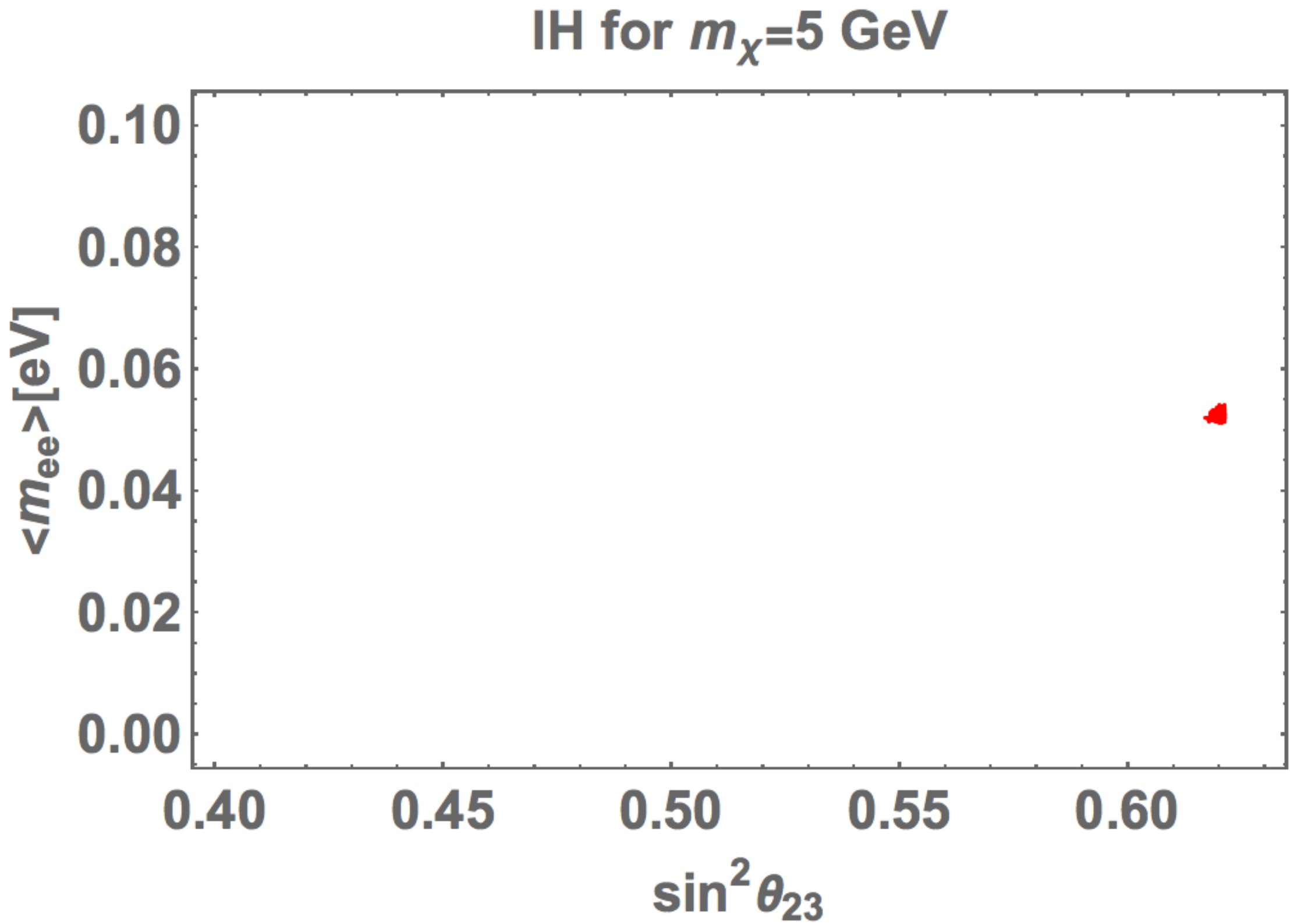}
 \end{center}
 \end{minipage}
   \begin{minipage}{0.32\hsize}
 \begin{center}
  \includegraphics[width=49mm]{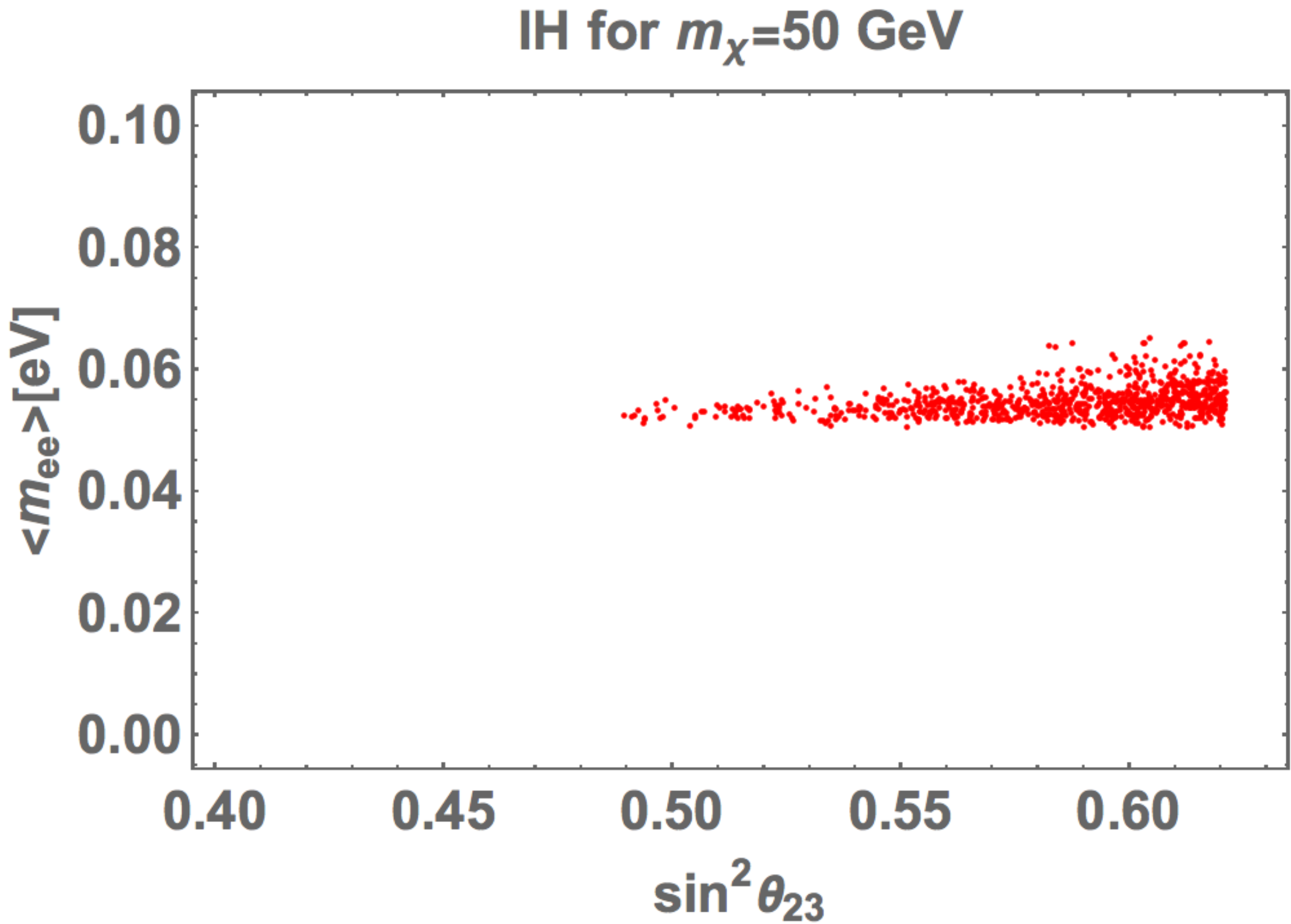}
 \end{center}
 \end{minipage}
 \caption{Scatter plots of $\sin^2\theta_{23}$ and  $\langle m_{ee} \rangle$, where the legends are the same as Fig.~\ref{fig:tau_nh}.}
   \label{fig:s23-mee_ih}
\end{figure}
In Fig.~\ref{fig:s23-mee_ih}, scatter plots of $\sin^2\theta_{23}$ and  $\langle m_{ee} \rangle$ are shown, where the legends are the same as Fig.~\ref{fig:tau_nh}.
Allowed regions are as follows:
$\langle m_{ee} \rangle = [0.055-0.07]$ eV for  $m_\chi=0.01$ GeV,
$\langle m_{ee} \rangle=[0.055-0.07]$ eV for  $m_\chi=0.1$ GeV,
$\langle m_{ee} \rangle\sim 0.053$ eV for  $m_\chi=5$ GeV, and
$\langle m_{ee} \rangle=[0.05-0.065]$ eV for  $m_\chi=50$ GeV.

Here, we summarize some features in the case of IH.
For all the cases, $\tau$ is localized at nearby $\tau=\pm0.2+2 i$.
In the cases of $m_\chi=5, 50$ GeV, there is a small allowed region to satisfy the cosmological constraint $\sum m\le$ 0.12 eV, while there is no allowed region for the other cases.
Larger value of $\sin^2\theta_{23}$ is favored except $m_\chi=0.01$ GeV. 
For all the cases, $\alpha_{21}$ is almost zero that is the same as the case of NH.

\section{Summary and Conclusions}
 We have proposed a radiative seesaw model with a light DM candidate under the modular $A_4$ and gauged $U(1)_{B-L}$ symmetries, in which the neutrino masses are generated via one-loop level, and we have a relatively light bosonic DM candidate $\sim 0.01-50$ GeV.
DM is stabilized by nonzero modular weight and a remnant $Z_2$ symmetry of $U(1)_{B-L}$ symmetry.
The naturalness of the tiny mass of DM has indirectly been realized by the radiatively induced mass of neutral fermions that interact with our DM candidate. Thus, DM mass must not exceed the mass of the neutral fermion masses, whose scale is assumed to be 50 GeV. 
 After checking there is a parameter space allowed by both the observations of relic density of DM and the direct detection constraints, taking four benchmark points of DM mass, we have also demonstrated the numerical analysis for the neutrino oscillation data in cases of NH and IH. We have found several features for both cases as follows.\\
{\it NH case}:
In case of $m_\chi=5$ GeV, half of allowed region might be ruled out by the cosmological constraint $\sum m\le$ 0.12 eV and larger value of $\sin^2\theta_{23}$ is favored. While large value of $\sin^2\theta_{23}$ is disfavored in case of $m_\chi=50$ GeV.
For all the cases, $\alpha_{21}$ is almost zero, and $\delta_{CP}$ is localized at nearby 50 deg and 300 deg.\\
{\it IH case}:
For all the cases, $\tau$ is localized at nearby $\tau=\pm0.2+2 i$.
In the cases of $m_\chi=5, 50$ GeV, there is a small allowed region to satisfy the cosmological constraint $\sum m\le$ 0.12 eV, while there is no allowed region for the other cases.
Larger value of $\sin^2\theta_{23}$ is favored except $m_\chi=0.01$ GeV. 
For all the cases, $\alpha_{21}$ is almost zero that is the same as the case of NH.


\begin{acknowledgements}
KIN was supported by JSPS Grant-in-Aid for Scientific Research (A) 18H03699, (C) 21K03562, (C) 21K03583, Okayama Foundation for Science and Technology, and Wesco Scientific Promotion Foundation.
This research was supported by an appointment to the JRG Program at the
APCTP through the Science and Technology Promotion Fund and Lottery Fund
of the Korean Government. This was also supported by the Korean Local
Governments - Gyeongsangbuk-do Province and Pohang City (H.O.).
H.O.~is sincerely grateful for the KIAS member. 
\end{acknowledgements}

\section*{Appendix}

Here, we show several properties of modular $A_4$ symmetry. 
In general, the modular group $\bar\Gamma$ is a group of linear fractional transformation
$\gamma$, acting on the modulus $\tau$ 
which belongs to the upper-half complex plane and transforms as
\begin{equation}\label{eq:tau-SL2Z}
\tau \longrightarrow \gamma\tau= \frac{a\tau + b}{c \tau + d}\ ,~~
{\rm where}~~ a,b,c,d \in \mathbb{Z}~~ {\rm and }~~ ad-bc=1, 
~~ {\rm Im} [\tau]>0 ~.
\end{equation}
This is isomorphic to $PSL(2,\mathbb{Z})=SL(2,\mathbb{Z})/\{I,-I\}$ transformation.
Then modular transformation is generated by two transformations $S$ and $T$ defined by:
\begin{eqnarray}
S:\tau \longrightarrow -\frac{1}{\tau}\ , \qquad\qquad
T:\tau \longrightarrow \tau + 1\ ,
\end{eqnarray}
and they satisfy the following algebraic relations, 
\begin{equation}
S^2 =\mathbb{I}\ , \qquad (ST)^3 =\mathbb{I}\ .
\end{equation}
More concretely, we can fix the basis of $S$ and $T$ as follows:
 \begin{align}
S=\frac13
 \begin{pmatrix}
 -1 & 2 & 2 \\
 -2 & -1 & 2 \\
 2 & 2 & -1 \\
 \end{pmatrix} ,\quad 
 T= 
 \begin{pmatrix}
 1 & 0 & 0 \\
0 & \omega & 0 \\
0 & 0 & \omega^2 \\
 \end{pmatrix} ,
 \end{align}
where $\omega\equiv e^{2\pi i/3}$.

Here we introduce the series of groups $\Gamma(N)~ (N=1,2,3,\dots)$ which are defined by
 \begin{align}
 \begin{aligned}
 \Gamma(N)= \left \{ 
 \begin{pmatrix}
 a & b \\
 c & d  
 \end{pmatrix} \in SL(2,\mathbb{Z})~ ,
 ~~
 \begin{pmatrix}
 a & b \\
 c & d  
 \end{pmatrix} =
 \begin{pmatrix}
 1 & 0 \\
 0 & 1  
 \end{pmatrix} ~~({\rm mod}~N) \right \}
 \end{aligned},
 \end{align}
and we define $\bar\Gamma(2)\equiv \Gamma(2)/\{I,-I\}$ for $N=2$.
Since the element $-I$ does not belong to $\Gamma(N)$
 for $N>2$ case, we have $\bar\Gamma(N)= \Gamma(N)$,
 that are infinite normal subgroup of $\bar \Gamma$ known as principal congruence subgroups.
  We thus obtain finite modular groups as the quotient groups defined by
  $\Gamma_N\equiv \bar \Gamma/\bar \Gamma(N)$.
For these finite groups $\Gamma_N$, $T^N=\mathbb{I}$ is imposed, and
the groups $\Gamma_N$ with $N=2,3,4$ and $5$ are isomorphic to
$S_3$, $A_4$, $S_4$ and $A_5$, respectively \cite{deAdelhartToorop:2011re}.

Modular forms of level $N$ are 
holomorphic functions $f(\tau)$ which are transformed under the action of $\Gamma(N)$ given by
\begin{equation}
f(\gamma\tau)= (c\tau+d)^k f(\tau)~, ~~ \gamma \in \Gamma(N)~ ,
\end{equation}
where $k$ is the so-called as the modular weight.

Under the modular transformation in Eq.(\ref{eq:tau-SL2Z}) in case of $A_4$ ($N=3$) modular group, a field $\phi^{(I)}$ is also transformed as 
\begin{equation}
\phi^{(I)} \to (c\tau+d)^{-k_I}\rho^{(I)}(\gamma)\phi^{(I)},
\end{equation}
where $-k_I$ is the modular weight and $\rho^{(I)}(\gamma)$ denotes a unitary representation matrix of $\gamma\in\Gamma(2)$ ($A_4$ reperesantation).
Thus Lagrangian, such as Yukawa terms, can be invariant if the sum of modular weight from fields and modular form in the corresponding term is zero (also invariant under $A_4$ and gauge symmetry).

The kinetic terms and quadratic terms of scalar fields can be written by 
\begin{equation}
\sum_I\frac{|\partial_\mu\phi^{(I)}|^2}{(-i\tau+i\bar{\tau})^{k_I}} ~, \quad \sum_I\frac{|\phi^{(I)}|^2}{(-i\tau+i\bar{\tau})^{k_I}} ~,
\label{kinetic}
\end{equation}
which is invariant under the modular transformation and overall factor is eventually absorbed by a field redefinition consistently.
Therefore the Lagrangian associated with these terms should be invariant under the modular symmetry.

The basis of modular forms with weight 2, $Y^{(2)}_3 = (y_{1},y_{2},y_{3})$, transforming
as a triplet of $A_4$ is written in terms of Dedekind eta-function $\eta(\tau)$ and its derivative \cite{Feruglio:2017spp}:
\begin{align} 
\label{eq:Y-A4}
y_{1}(\tau) &= \frac{i}{2\pi}\left( \frac{\eta'(\tau/3)}{\eta(\tau/3)} +\frac{\eta'((\tau +1)/3)}{\eta((\tau+1)/3)}  
+\frac{\eta'((\tau +2)/3)}{\eta((\tau+2)/3)} - \frac{27\eta'(3\tau)}{\eta(3\tau)} \right)\nn\\ 
&\simeq
1+12 q+36 q^2+12 q^3+\cdots,\\
y_{2}(\tau) &= \frac{-i}{\pi}\left( \frac{\eta'(\tau/3)}{\eta(\tau/3)} +\omega^2\frac{\eta'((\tau +1)/3)}{\eta((\tau+1)/3)}  
+\omega \frac{\eta'((\tau +2)/3)}{\eta((\tau+2)/3)} \right) , \label{eq:Yi} \nn\\ 
&\simeq
-6q^{1/3} (1+7 q+8 q^2+\cdots),\\
y_{3}(\tau) &= \frac{-i}{\pi}\left( \frac{\eta'(\tau/3)}{\eta(\tau/3)} +\omega\frac{\eta'((\tau +1)/3)}{\eta((\tau+1)/3)}  
+\omega^2 \frac{\eta'((\tau +2)/3)}{\eta((\tau+2)/3)} \right)\nn\\ 
&\simeq
-18q^{2/3} (1+2 q+5 q^2+\cdots),
\end{align}
where $q=e^{2\pi i \tau}$, and expansion form in terms of $q$ would sometimes be useful to have numerical analysis.

Then, we can construct the higher order of couplings $Y^{(4)}_1,Y^{(6)}_1, Y^{(6)}_3, Y^{(6)}_{3'}$ following the multiplication rules as follows:
\begin{align}
Y^{(4)}_1& = y^2_1+2 y_2 y_3, \ Y^{(6)}_1 = y^3_1+y^3_2 + y^3_3 -3 y_1y_2y_3, \
 \\
Y^{(6)}_3&\equiv (y'_1,y'_2,y'_3) = ( y^3_1+2y_1 y_2 y_3, y_1^2y_2+2 y^2_2 y_3, y^2_1 y_3+2 y^2_3 y_2),\\
Y^{(6)}_{3'}&\equiv (y''_1,y''_2,y''_3) = ( y^3_3+2y_1 y_2 y_3, y^2_3 y_1+2 y^2_1 y_2, y^2_3 y_2+2 y^2_2 y_1),
\end{align}
where the above relations are constructed by the multiplication rules under $A_4$ as shown below:
\begin{align}
\begin{pmatrix}
a_1\\
a_2\\
a_3
\end{pmatrix}_{\bf 3}
\otimes 
\begin{pmatrix}
b_1\\
b_2\\
b_3
\end{pmatrix}_{\bf 3'}
&=\left (a_1b_1+a_2b_3+a_3b_2\right )_{\bf 1} 
\oplus \left (a_3b_3+a_1b_2+a_2b_1\right )_{{\bf 1}'} \nonumber \\
& \oplus \left (a_2b_2+a_1b_3+a_3b_1\right )_{{\bf 1}''} \nonumber \\
&\oplus \frac13
\begin{pmatrix}
2a_1b_1-a_2b_3-a_3b_2 \\
2a_3b_3-a_1b_2-a_2b_1 \\
2a_2b_2-a_1b_3-a_3b_1
\end{pmatrix}_{{\bf 3}}
\oplus \frac12
\begin{pmatrix}
a_2b_3-a_3b_2 \\
a_1b_2-a_2b_1 \\
a_3b_1-a_1b_3
\end{pmatrix}_{{\bf 3'}\ } \ , \nonumber \\
\nonumber \\
{\bf 1} \otimes {\bf 1} = {\bf 1} \ , \quad &
{\bf 1'} \otimes {\bf 1'} = {\bf 1''} \ , \quad
{\bf 1''} \otimes {\bf 1''} = {\bf 1'} \ , \quad
{\bf 1'} \otimes {\bf 1''} = {\bf 1} \ .
\end{align}

\end{document}